\documentclass{jpp}

\usepackage{graphicx,amsmath}
\usepackage{natbib,xtab}
\usepackage[pdftex,colorlinks,citecolor=blue]{hyperref}

\ifCUPmtlplainloaded \else
  \checkfont{eurm10}
  \iffontfound
    \IfFileExists{upmath.sty}
      {\typeout{^^JFound AMS Euler Roman fonts on the system,
                   using the 'upmath' package.^^J}%
       \usepackage{upmath}}
      {\typeout{^^JFound AMS Euler Roman fonts on the system, but you
                   dont seem to have the}%
       \typeout{'upmath' package installed. JPP.cls can take advantage
                 of these fonts, if you use 'upmath' package.^^J}%
       \providecommand\upi{\pi}%
      }
  \else
    \providecommand\upi{\pi}%
  \fi
\fi

\ifCUPmtlplainloaded \else
  \checkfont{msam10}
  \iffontfound
    \IfFileExists{amssymb.sty}
      {\typeout{^^JFound AMS Symbol fonts on the system, using the
                'amssymb' package.^^J}%
       \usepackage{amssymb}%
         
       \let\ge=\geqslant  
      }{}
  \fi
\fi

\ifCUPmtlplainloaded \else
  \IfFileExists{amsbsy.sty}
    {\typeout{^^JFound the 'amsbsy' package on the system, using it.^^J}%
     \usepackage{amsbsy}}
    {\providecommand\boldsymbol[1]{\mbox{\boldmath $##1$}}}
\fi

\newcommand{\imag}{{\rm i}}
\providecommand\upi{\pi}
\newcommand{\pD}[2]{\frac{\partial #2}{\partial #1}}
\newcommand{\pDD}[2]{\frac{\partial^2 #2}{\partial #1^2}}
\newcommand{\D}[2]{\frac{{\rm d} #2}{{\rm d} #1}}
\newcommand{\DD}[2]{\frac{{\rm d}^2 #2}{{\rm d} #1^2}}
\newcommand\bs[1]{\boldsymbol{#1}}
\newcommand\bb[1]{\mbox{\boldmath{$#1$}}}
\newcommand\grad{\bb{\nabla}}
\newcommand\bcdot{\,\bb{\cdot}\,}
\newcommand\btimes{\,\bb{\times}\,}
\newcommand{\mc}[1]{\mathcal{#1}}

\newcommand{\msf}[1]{\mathsfi{#1}}
\newcommand{\msb}[1]{\mathsfbi{#1}}
\newcommand{\gas}{\!\bs{R}_s}
\newcommand{\ggs}{{\negthickspace\!\bs{R}_s}}
\newcommand{\eb}{\hat{\bb{b}}}
\newcommand{\ez}{\hat{\bb{e}}_z}
\newcommand{\ex}{\hat{\bb{e}}_x}
\newcommand{\ey}{\hat{\bb{e}}_y}

\newsavebox{\astrutbox}
\sbox{\astrutbox}{\rule[-5pt]{0pt}{20pt}}

\title[]{Linear Vlasov theory of a magnetised, \\ thermally stratified atmosphere}

\author[R.~Xu \& M.~W.~Kunz]%
{R.\ns X\ls U$^1$ \and M.\ns W.\ns K\ls U\ls N\ls Z\ls$^{1,2}$\thanks{Email address for correspondence: mkunz@princeton.edu}
}

\affiliation{$^1$Department of Astrophysical Sciences, Princeton University, Peyton Hall, Princeton, NJ 08544, USA\\[\affilskip]
$^2$Princeton Plasma Physics Laboratory, P.O.~Box 451, Princeton, NJ 08543, USA}

\pubyear{2016}
\volume{}
\pagerange{}
\date{20 Jun 2016}
\begin{document}

\maketitle

%
%
\begin{abstract}
The stability of a collisionless, magnetised plasma to local convective disturbances is examined, with a focus on kinetic and finite-Larmor-radius effects. Specific application is made to the outskirts of galaxy clusters, which contain hot and tenuous plasma whose temperature increases in the direction of gravity. At long wavelengths (the ``drift-kinetic'' limit), we obtain the kinetic version of the magnetothermal instability (MTI) and its Alfv\'{e}nic counterpart (Alfv\'{e}nic MTI), which were previously discovered and analysed using a magnetofluid (i.e.~Braginskii) description. At sub-ion-Larmor scales, we discover an overstability driven by the electron temperature gradient of kinetic-Alfv\'{e}n drift waves -- the {\em electron MTI} (eMTI) -- whose growth rate is even larger than the standard MTI. At intermediate scales, we find that ion finite-Larmor-radius effects tend to stabilise the plasma. We discuss the physical interpretation of these instabilities in detail, and compare them both with previous work on magnetised convection in a collisional plasma and with temperature-gradient-driven drift-wave instabilities well-known to the magnetic-confinement-fusion community. The implications of having both fluid and kinetic scales simultaneously driven unstable by the same temperature gradient are briefly discussed.
\end{abstract} 


%
%
%
\section{Introduction}\label{sec:introduction}

The problem of convective stability has a long history, one that has enjoyed over a century's worth of attention by the fluids community in the study of atmospheric dynamics, oceanography, geophysics, and stellar atmospheres ({\it ``Strahlungsgleichgewicht''}, to use the term coined by \citet{schwarzschild06}). The study of convective stability in magnetised {\em plasmas}, on the other hand, is just over 50 years old, with rigorous astrophysical inquiry emerging only in the last decade and a half. Without this context, it may come as a surprise to find a contemporary paper on something as pedestrian as the linear Vlasov theory of a magnetised, thermally stratified atmosphere. But, as we demonstrate here, such surprise is not warranted -- there are new and novel findings to be had. This is despite a large body of work in the magnetic-confinement-fusion literature on instability and anomalous transport in magnetised, thermally stratified plasmas \citep[e.g.][]{rs61,mikhailovskii62,crs67,cks91,horton99,dimits00,dorland00}, as well as two encyclop\ae dic texts by \citet{mikhailovskii74,mikhailovskii92} on electromagnetic instabilities in inhomogeneous plasmas.

The culprit is that prior work on the stability of magnetised, inhomogeneous plasmas has focused almost entirely on low-$\beta$ plasmas, that is, those whose thermal energy is strongly subdominant to the energy of the magnetic field that threads them. This is only natural. Energetically dominant magnetic fields are routinely employed to confine terrestrial plasmas, both for experiments in basic plasma physics and in efforts to produce a self-sustaining and controlled nuclear fusion reaction, and it is largely in these contexts that studies of plasma stability have been carried out. But most astrophysical plasmas are high-$\beta$, with the magnetic field energetically subdominant not only to the thermal energy but, often, to the bulk kinetic energy as well. While it may be tempting to ignore the field altogether in this case due to its dynamical weakness, one must exercise great care. Since magnetic fields serve as conduits of heat and momentum transport in magnetised plasmas, it matters where and how effectively the field is carried around by the plasma.

This paper adds to the study of magnetised, thermally stratified, high-$\beta$ plasmas that are commonplace in astrophysical environments such as galaxy clusters and certain classes of black-hole accretion flows. Here, we take a kinetic Vlasov (i.e.~collisionless) approach to study linear stability, both complementing and extending recent work on the topic that alternatively used a magnetofluid ({\it viz.}, Braginskii) treatment. In the next section (\S\ref{sec:prereqs}), we provide a brief review of what is known from these previous fluid analyses about the linear convective stability of a collisional magnetised plasma. We also highlight some additional physics that is brought into play by adopting a kinetic approach, including Landau resonances, collisionless wave damping, and finite-Larmor-radius effects. The calculation proper begins in Section \ref{sec:vlasov}, in which the basic equations are given (\S\ref{sec:equations}), the equilibrium state is constructed (\S\ref{sec:equilibrium}), and the equilibrium particle trajectories are elucidated (\S\S\ref{sec:drifts}, \ref{sec:trajectories}). In Section \ref{sec:linear} we derive the linear theory governing small-amplitude perturbations about this equilibrium and specialize it for a quasi-Maxwell-Boltzmann equilibrium distribution function. Certain technical aspects of the derivation are relegated to Appendices \ref{app:derivation} and \ref{app:integrals}. The heart of the paper is Section \ref{sec:results}, in which the linear theory is used to assess convective stability across a wide range of perturbation wavelengths and frequencies. The physical interpretation of instability in each parameter regime is discussed in detail. Connection is made to drift-wave instabilities driven by ion- and electron-temperature gradients that are well known in the magnetic-confinement-fusion community. A supplementary appendix is provided in which the nonlinear gyrokinetic theory of a thermally stratified atmosphere is systematically derived (Appendix \ref{app:gk}). Finally, a summary of our findings is given in Section \ref{sec:summary}.

\section{Prerequisites}\label{sec:prereqs}

\subsection{Convective stability of a collisional magnetised plasma}

Before embarking on a detailed kinetic analysis, it seems prudent to briefly review what is known about the convective stability of a {\em collisional} magnetised plasma, that is, one in which the collisional mean free path $\lambda_{mfp}$ is intermediate between the Larmor radius $\rho_s$ of plasma species $s$ and the scales of interest on which convection sets in (e.g., the thermal-pressure scale height $H$). Even in this case, the physics is surprisingly different than applies to hydrodynamic (or even ideal magnetohydrodynamic (MHD)) convection, for which the entropy gradient is the discriminating quantity for stability \citep{schwarzschild06}. 

The culprit is that frozen-in magnetic-field lines serve as conduits along which heat and/or (parallel) momentum are transported from one tethered fluid element to the next, the rate of transport increasing the more aligned the magnetic field becomes with the temperature and/or velocity gradients. This is a consequence of the small Larmor radius: the magnetic field interferes strongly with motions across itself, and so collisional transport occurs most readily along magnetic lines of force \citep{braginskii65}. For sufficiently weak collisions or on sufficiently small lengthscales such that $k^2_\parallel \lambda_{mfp} H \gg 1$, where $k_\parallel$ is the wavenumber along the magnetic field, field lines tend towards isotherms and/or isotachs.

This has profound implications for the convective stability of such plasmas \citep{balbus00,balbus01}. Upward and downward fluid displacements that carry magnetic-field lines with them tend not to be isentropic but rather isothermal, since they remain magnetically and thus thermodynamically connected to conditions at their native altitude. For atmospheres in which the temperature increases in the direction of gravity, this implies that an upwardly (downwardly) displaced fluid element is always warmer (cooler) than the surroundings it is passing through. As it rises (sinks), the frozen-in field lines become ever more parallel to the temperature gradient, with the rate of entropy transfer increasing with separation. The result is instability, the temperature gradient taking precedence over the entropy gradient.

To offer a concrete example of this process, let us consider a non-rotating, vertically ($z$) stratified atmosphere in hydrostatic equilibrium, threaded by a uniform, horizontal ($x$) magnetic field, subthermal in its strength. For simplicity, we take the gravitational acceleration $\bb{g} = -g\ez$ to be constant and assume that sound waves propagate fast enough to ensure near incompressibility (i.e.~the Boussinesq approximation). We also neglect any dynamical effect of the magnetic field; its only role is to channel the flow of entropy between magnetically tethered fluid elements. Under these conditions, a vertical displacement $\xi_z$ of a fluid element excites a linear perturbation in the temperature $\delta T$ according to
\begin{equation}\label{eqn:prelim1}
\left( \pD{t}{} + \omega_{cond} \right) \frac{\delta T}{T} = - \left( \frac{N^2}{g} \pD{t}{} + \omega_{cond} \D{z}{\ln T} \right) \xi_z ,
\end{equation}
where $\omega_{cond}$ is the (field-aligned, or ``parallel'') conduction frequency and $N$ is the Brunt-V\"{a}is\"{a}l\"{a} frequency \citep{vaisala25,brunt27}. (This follows from the energy equation.) When conduction is negligible ($\omega_{cond} \rightarrow 0$), we find $\delta T / T = -(N^2/g) \xi_z$. In an atmosphere with an upwardly increasing entropy profile ($N^2 > 0$), an upwardly displaced fluid element ($\xi_z > 0$) adiabatically cools and falls back down towards its equilibrium position. Oscillations ensue with frequency $N$, {\em viz.}, $\partial^2\xi_z / \partial t^2 = -N^2 \xi_z$. In contrast, when conduction is rapid ($\omega_{cond} \gg \partial / \partial t$), the perturbed temperature satisfies $\Delta T/T \doteq \delta T / T + \xi_z \,{\rm d}\ln T/{\rm d}z = 0$; i.e.~the Lagrangian change in a fluid element's temperature vanishes. Isentropic displacements are precluded, and the assessment of stability demands a comparison between the (unchanged) temperature of the fluid element and that of its new surroundings. For ${\rm d}\ln T/{\rm d}z < 0$, an upwardly (downwardly) displaced fluid element is too hot (cool), and the displacement grows exponentially: the $z$-component of the force equation becomes $\partial^2 \xi_z / \partial t^2 = g | {\rm d} \ln T / {\rm d} z| \, \xi_z$. This is the simplest version of the {\it magnetothermal instability} (MTI).

The dynamics illustrated here are complicated further by the field-aligned collisional transport of parallel momentum, which performs the dual role of viscously damping those motions which change the magnetic-field strength (at a rate $\omega_{visc}$) and of coupling Alfv\'{e}n and slow modes so that slow-mode perturbations excite a buoyantly unstable Alfv\'{e}nic response \citep{kunz11}. Restoring the magnetic tension and the associated Alfv\'{e}n frequency $\omega_A$ in our presentation, vertical displacements with spatial dependence $\exp(\imag k_x x + \imag k_y y + \imag k_z z)$ in the limit of rapid field-aligned conduction satisfy
\begin{equation}\label{eqn:prelim2}
\biggl( \pDD{t}{} + \omega^2_A \biggr) \Biggl( \pDD{t}{} + \omega_{visc} \frac{k^2_\perp}{k^2} \pD{t}{} + \omega^2_A + g \D{z}{\ln T} \frac{\mc{K}}{k^2} \Biggr) \xi_z = - \omega_{visc} \, g \D{z}{\ln T}\frac{k^2_y}{k^2}  \pD{t}{\xi_z} ,
\end{equation}
where $k^2_\perp \doteq k^2_y + k^2_z$ is square of the wavenumber perpendicular to the mean magnetic field and $\mc{K} \doteq k^2_x + k^2_y$ (cf.~equation (38) of \citet{kunz11}). The first term in parentheses represents Alfv\'{e}n waves that are polarized with the perturbed magnetic field oriented along the $y$-axis. In the absence of parallel viscosity, they are unaffected by buoyancy. The second term in (\ref{eqn:prelim2}) represents slow modes, which are viscously damped at the rate $\omega_{visc}(k^2_\perp/k^2) $ and subject to the MTI when ${\rm d}\ln T/{\rm d}z < 0$. For $k_y \ne 0$, the right-hand side of (\ref{eqn:prelim2}) is non-zero, and parallel viscosity couples the Alfv\'{e}n and slow modes; slow-mode perturbations induce an Alfv\'{e}nic response. When the rate of transfer of parallel momentum $\omega_{visc}$ is much faster than the dynamical timescale, the slow mode is rapidly damped; in this limit, (\ref{eqn:prelim2}) becomes
\begin{equation}\label{eqn:prelim3}
\biggl( \pDD{t}{} + \omega^2_A \biggr) \pD{t}{\xi_z} \simeq - g \D{z}{\ln T} \frac{k^2_y}{k^2_\perp} \pD{t}{\xi_z} .
\end{equation}
If the temperature increases in the direction of gravity (${\rm d}\ln T / {\rm d} z < 0$), slow-mode perturbations excite a buoyantly unstable Alfv\'{e}nic response (``Alfv\'{e}nic MTI'', to borrow the term used by \citet{kunz11}), an effect otherwise absent in an inviscid fluid. By damping motions along field lines, parallel viscosity effectively reorients magnetic-field fluctuations to be nearly perpendicular to the background magnetic field (via flux freezing). These modes therefore display characteristics of both slow and Alfv\'{e}n modes: they have density and temperature perturbations, and therefore are subject to buoyancy forces, but their velocity and magnetic-field perturbations are predominantly polarized across the mean field. As long as field-aligned conduction is rapid, regardless of whether or not parallel viscosity is effective, the maximum growth rate is $\sqrt{-g \,{\rm d}\ln T / {\rm d}z}$.

This concludes our recapitulation of convective instability in a collisional magnetised plasma. We now train our focus on collisionless plasmas.\footnote{A reader educated in buoyancy instabilities in weakly collisional, magnetised plasmas will note a glaring omission from the preceding five paragraphs -- the heat-flux-driven buoyancy instability (HBI), discovered by \citet[][see also \citet{kunz11} and \citet{lk12}]{quataert08} and studied using dedicated numerical simulations by \citet{pq08}, \citet{parrish09}, \citet{bogdanovic09}, \citet{mccourt11}, \citet{mikellides11}, \citet{parrish12}, \citet{kunz12}, and \citet{avara13}. In this paper, we concentrate solely on equilibrium atmospheres whose (uniform) magnetic field is oriented perpendicular to gravity, a situation stable to the HBI. The reason we do so is quite simple -- without collisions, any equilibrium field-aligned temperature gradient in a magnetized plasma will be wiped out on a sound-crossing time of the atmosphere by free-streaming particles. As this timescale is generally comparable to or even smaller than the growth times of the HBI and the other instabilities investigated herein, this does not constitute a reasonable background state about which one may perform a linear stability analysis. The collisional case skirts this issue by having large-scale field-aligned temperature gradients relaxed on a diffusive timescale, which is generally longer than the instabilities of interest. One may even construct equilibrium states in this case by ensuring that any background collisional heat flux is divergence free, or simply by balancing the downward field-aligned transport of heat with losses from radiative cooling.}

\subsection{Collisionless physics in stratified atmospheres}

The main difference between a collisional magnetised plasma comprising a stratified atmosphere and its collisionless counterpart is the ability of particles in the latter case to resonate with emergent waves in the system. For low-frequency, long-wavelength fluctuations -- those with frequencies $\omega$ much smaller than the Larmor frequency $\Omega_s$ and wavenumbers $k_\parallel$ much larger than the inverse Larmor radius $\rho^{-1}_s$ ($s$ denotes the species index) -- the primary resonance is the Landau resonance, $v_\parallel = \omega / k_\parallel$. In the presence of an electric field, an inhomogeneous magnetic field, and a gravitational field, the parallel velocity $v_\parallel$ of a particle with mass $m_s$, charge $q_s$, and magnetic moment $\mu_s$ evolves according to
\begin{equation}\label{eqn:vprl}
m_s \D{t}{v_\parallel} = q_s E_\parallel - \mu_s \nabla_\parallel B + m_s g_\parallel ,
\end{equation}
where the subscript $\parallel$ on each field denotes the vector component along the local magnetic-field direction $\eb$. Landau-resonant particles interacting with the parallel electric field $E_\parallel$ result in collisionless damping of the electric-field fluctuations \citep{landau46}. A similar effect is caused by the second term in (\ref{eqn:vprl}), which goes by the name transit-time or \citet{barnes66} damping. Particles that are almost at rest with respect to the slow magnetosonic wave are subject to the action of the mirror force associated with the magnetic compressions in the wave. Since, for a monotonically decreasing distribution function ($\partial f / \partial v_\parallel < 0$), there are more particles with $v_\parallel < \omega / k_\parallel$ than with $v_\parallel > \omega / k_\parallel$, the energy exchange between resonant particles and the wave leads to a net gain (loss) of energy by the particles (wave). This is true even for non-oscillatory disturbances (which will constitute the majority of the instabilities investigating in this paper): particles with $v_\parallel \sim 0$ are resonant with the zero-frequency wave and extract energy through betatron acceleration. Put differently, the only way to maintain perpendicular pressure balance for such a disturbance is to increase the energy of the resonant particles at the expense of the wave energy. The fewer particles there are at small parallel velocity to be resonant, the higher the damping rate (i.e.~energy transfer) needs to be to maintain pressure balance. In both cases -- oscillatory waves and aperiodic modes -- the result is wave damping. The final term in (\ref{eqn:vprl}) represents the parallel acceleration of particles once the magnetic field acquires a component along the direction of gravity: $g_\parallel = \bb{g}\bcdot\eb =  - g b_z$. In that case, the plasma streams downwards along the perturbed magnetic-field lines, with Landau-resonant particles transferring the free energy stored in the vertical magnetic-field fluctuations to the particle distribution function. As with Landau and Barnes damping, the result is collisionless damping of electromagnetic fluctuations.

There are two further differences between the physics investigated in this paper and that which was the focus of previous (fluid) studies. First, the ions and electrons need not remain in thermal equilibrium as the plasma is perturbed. In standard fluid (collisional) treatments of the MTI, the electrons are responsible for ensuring isothermal displacements, while the ions, with their larger mass, are responsible for the buoyant response. The two are connected via thermal equilibration. In the collisionless case, this connection is lost. Second, we account for finite-Larmor-radius (FLR) effects. The consideration of FLR effects does not, of course, require specialization to a collisionless plasma, but our kinetic treatment does afford a systematic and rigorous way of deriving their impact on plasma stability. In particular, we will show that fluctuations with $k_\perp \rho_i \sim k_\parallel H \gg 1$ are unstable to a kinetic-Alfv\'{e}n version of the MTI, in which the ions are convected across the field lines as the electrons remain isothermal along them.

\vspace{0.1in}
These prerequisites fulfilled, our analysis begins.

\section{Vlasov description of a magnetised thermally stratified atmosphere}\label{sec:vlasov}

\subsection{Basic equations}\label{sec:equations}

We start with the kinetic Vlasov equation
\begin{equation}\label{eqn:vlasov}
\D{t}{f_s} \doteq \pD{t}{f_s} + \bb{v}\bcdot\grad f_s + \left[ \frac{q_s}{m_s} \left( \bb{E} + \frac{\bb{v}}{c}\btimes\bb{B} \right) + \bb{g} \right] \bcdot \pD{\bb{v}}{f_s} = 0 ,
\end{equation}
which governs the phase-space evolution of the distribution function of species $s$, $f_s = f_s(t,\bb{r},\bb{v})$. Our notation is standard: $q_s$ is the charge and $m_s$ is the mass of species $s$, $\bb{E}$ is the electric field, $\bb{B}$ is the magnetic field, and $\bb{g}$ is the gravitational acceleration. We take the latter to lie along the $-\ez$ direction of an $(x,y,z)$ Cartesian coordinate system, $\bb{g} = -g \ez$ with $g = {\rm const} > 0$. In the non-relativistic limit, the electromagnetic fields satisfy the plasma quasineutrality constraint (which follows from the Poisson equation to lowest order in $k^2 \lambda^2_D$, where $\lambda_D$ is the Debye length),
\begin{equation}\label{eqn:quasineutrality}
0 = \sum_s q_s n_s \doteq \sum_s q_s \int{\rm d}^3\bb{v}\,f_s,
\end{equation}
and the pre-Maxwell version of Amp\`{e}re's law,
\begin{equation}\label{eqn:ampere}
\grad\btimes\bb{B} = \frac{4\upi}{c} \bb{j} \doteq \frac{4\upi}{c} \sum_s q_s \int{\rm d}^3\bb{v}\,\bb{v} f_s.
\end{equation}
The magnetic field evolves according to Faraday's law of induction,
\begin{equation}\label{eqn:faraday}
\pD{t}{\bb{B}} = -c \grad\btimes\bb{E} ,
\end{equation}
while satisfying the solenoidality constraint $\grad\bcdot\bb{B} = 0$. We neglect particle-particle collisions in this paper; the linear theory of a weakly collisional, magnetised, thermally stratified atmosphere, with which we make frequent contact, is presented in \citet{kunz11}.

For future reference, we define the pressure tensor of species $s$,
\begin{equation}\label{eqn:ptensor}
\msb{P}_s \doteq m_s \int{\rm d}^3\bb{v}\,( \bb{v} - \bb{u}_s ) ( \bb{v} - \bb{u}_s ) f_s,
\end{equation}
as well as its mean velocity,
\begin{equation}\label{eqn:u}
\bb{u}_s \doteq \frac{1}{n_s} \int{\rm d}^3\bb{v}\,\bb{v} f_s.
\end{equation}
The total thermal pressure $P_s = n_s T_s \doteq (1/3)\, {\rm tr}\msb{P}_s$, where the temperature $T_s$ is defined {\it in situ}. In many of the limiting cases we investigate, the pressure tensor is diagonal in a coordinate system defined by the directions perpendicular ($\perp$) and parallel ($\parallel$) to the magnetic-field direction $\eb \doteq \bb{B}/B$:
\begin{equation}\label{eqn:gyrotropicP}
\msb{P}_s \rightarrow P_{\perp s} ( \msb{I} - \eb\eb ) + P_{\parallel s} \eb \eb ,
\end{equation}
where $\msb{I}$ is the unit dyadic, and
\begin{gather}\label{eqn:pprp}
P_{\perp s} = n_s T_{\perp s} \doteq m_s \int{\rm d}^3\bb{v}\,  \frac{1}{2} |\bb{v}_\perp - \bb{u}_{\perp,s} |^2 f_s ,\\*
\label{eqn:pprl}
P_{\parallel s} = n_s T_{\parallel s} \doteq m_s \int{\rm d}^3\bb{v}\,(v_\parallel - u_{\parallel s})^2 f_s
\end{gather}
are the thermal pressures perpendicular and parallel to the magnetic-field direction, respectively, of species $s$; $T_{\perp s}$ ($T_{\parallel s}$) is the perpendicular (parallel) temperature. The total thermal pressure $P_s = (2/3) P_{\perp s} + (1/3) P_{\parallel s}$.

We will sometimes express the electric and magnetic fields in terms of scalar and vector potentials:
\begin{equation}\label{eqn:potentials}
\bb{E} = - \grad\varphi - \frac{1}{c} \pD{t}{\bb{A}} \quad {\rm and} \quad \bb{B} = \grad\btimes\bb{A} ,
\end{equation}
where $\grad\bcdot\bb{A} = 0$ (the Coulomb gauge). Working with potentials instead of fields is particularly useful when taking the gyrokinetic limit (\S\ref{sec:gyrokinetic}, Appendix \ref{app:gk}).

\subsection{Equilibrium state}\label{sec:equilibrium}

We seek equilibrium solutions to (\ref{eqn:vlasov})--(\ref{eqn:faraday}) in the presence of a uniform, stationary background magnetic field oriented perpendicular to gravity, $\bb{B}_0 = B_0 \ex$, whose magnitude $B_0$ is such that the Larmor radius of species $s$, $\rho_s \doteq v_{th\perp s} / \Omega_s$, where $v_{th\perp s} \doteq (2T_{\perp 0s}/m_s)^{1/2}$ is the (perpendicular) thermal speed and $\Omega_s = q_s B_0 / m_s c$ is the Larmor frequency, is much smaller than the thermal-pressure scale height $H$ of the equilibrium distribution function $f_{0s}$, i.e.~the plasma is ``magnetised''. In a collisionless plasma, this scale separation demands that $\bb{B}_0\bcdot\grad f_{0s} = 0$, or $\partial f_{0s} / \partial x = 0$. Physically, this is due to the fast streaming along the field lines of particles setting up thermal equilibrium (see \S\ref{sec:gkzeroth} for a proof). The cross-field drift of particles is instead constrained by the Larmor motion, and gradients in the field-perpendicular directions are thus allowed. In this paper, we assume $\partial f_{0s} / \partial y = 0$, and hence consider equilibrium distribution functions that are dependent only upon the vertical ($z$) coordinate.

Under these conditions, the first moment of (\ref{eqn:vlasov}) becomes
\begin{equation}\label{eqn:equilibriumforce}
\D{z}{\msf{P}^{(zz)}_{0s}} - q_s n_{0s} E_0 + m_s n_{0s} g = 0 ,
\end{equation}
where $\msf{P}^{(zz)}_{0s}$ is the $zz$-component of the equilibrium pressure tensor (cf.~(\ref{eqn:ptensor})). Summing this equation over species and using quasineutrality (\ref{eqn:quasineutrality}), we find 
\begin{equation}\label{eqn:hydrostatic}
\sum_s \D{z}{\msf{P}^{(zz)}_{0s}} = - \sum_s m_s n_{0s} g \doteq - \varrho g < 0,
\end{equation}
a statement of hydrostatic equilibrium. That an equilibrium electrostatic field $\bb{E}_0 = E_0 \ez$ is in general required may be seen by multiplying (\ref{eqn:equilibriumforce}) by $q_s/m_s$, summing over species, and using quasineutrality (\ref{eqn:quasineutrality}) to find
\begin{equation}\label{eqn:equilibriumE}
E_0 = \sum_s \frac{q_s}{m_s} \D{z}{\msf{P}^{(zz)}_{0s}} \bigg/ \sum_s \frac{q^2_s n_{0s}}{m_s} .
\end{equation}
For this to vanish in a hydrogenic plasma with constant ion-to-electron temperature ratio, the thermal speed must be species-independent -- a rather limiting condition.

Up to this point, we have made no assumptions about $f_{0s}$ other than its lack of $y$ dependence. We now investigate what is allowed. A necessary and sufficient condition for a stationary solution to the Vlasov equation (\ref{eqn:vlasov}) is that $f_{0s}$ be a function only of the integrals of the motion. These may be identified by considering the Lagrangian
\begin{equation}\label{eqn:lagrangian}
\mc{L}(\bb{r},\bb{v}) = \frac{1}{2} m_s v^2 - q_s \left( \Phi_s - \frac{\bb{v}\bcdot\bb{A}_0}{c} \right) = \frac{1}{2} m_s v^2 - q_s \Phi_s - m_s  v_y \Omega_s z,
\end{equation}
where $\bb{A}_0 = -B_0 z\ey$ is the equilibrium vector potential and we have introduced the total (i.e.~electrostatic plus gravitational) potential $\Phi_s$, defined via
\begin{equation}\label{eqn:Phis}
\D{z}{\Phi_s} = - E_0 + \frac{m_s}{q_s} g = - \frac{T_{0s}}{q_s} \D{z}{\ln\msf{P}^{(zz)}_{0s}} > 0,
\end{equation}
the final equality following from (\ref{eqn:equilibriumforce}). Since the Lagrangian (\ref{eqn:lagrangian}) is independent of the $y$ coordinate, the $y$-component of the canonical momentum $\bb{\wp} = \partial\mc{L} / \partial \bb{v}$ is conserved: $\wp_y = m_s v_y - m_s \Omega_s z = {\rm const}$. This in turn implies that
\begin{equation}\label{eqn:Z}
\mc{Z}_s \doteq z - \frac{v_y}{\Omega_s} = {\rm const} .
\end{equation}
Because the Lagrangian is time-independent, the Hamiltonian
\begin{equation}\label{eqn:hamiltonian}
\mc{H}(\bb{r},\bb{\wp}) \doteq \bb{v} \bcdot \pD{\bb{v}}{\mc{L}} - \mc{L} = \frac{1}{2m_s} \Bigl[ \wp^2_x + ( \wp_y + m_s \Omega_s z )^2 + \wp^2_z \Bigr] + q_s \Phi_s
\end{equation}
is also conserved. Expressing $\mc{H}$ in terms of the particle velocity $\bb{v} = \partial\mc{H}/\partial\bb{\wp}$ yields the total (i.e.~kinetic plus potential) particle energy 
\begin{equation}\label{eqn:E}
\mc{E}_s \doteq \frac{1}{2} m_s v^2 + q_s \Phi_s ,
\end{equation}
which is an invariant of the motion. Finally, the Lagrangian is independent of $x$, and so $\wp_x = m_s v_x$ is conserved as well. The equilibrium distribution function for species $s$ may therefore be written
\begin{equation}\label{eqn:F0}
f_{0s} = F_{0s}(\mc{Z}_s,\mc{E}_s,v_x) .
\end{equation}
In this work, we specialize to isotropic equilibrium distribution functions, i.e.~$F_{0s} = F_{0s}(\mc{Z}_s,\mc{E}_s)$. In magnetised, high-beta plasmas, such as the intracluster medium, velocity-space anisotropy in the equilibrium distribution function can drive myriad linear instabilities at Larmor scales, all of which we seek to exclude from (at least this first pass at) the calculation.

Before proceeding any further, we forewarn the reader that, throughout the manuscript, the upper-case $F_{0s}$ refers to the equilibrium distribution function written as a function of the integrals of motion ($\mc{Z}_s,\mc{E}_s$), while the lower-case $f_{0s}$ denotes the equilibrium distribution function written as a function of the standard phase-space coordinates ($z,\bb{v}$). Special attention must be paid to what variable is held fixed under partial differentiation.

\subsection{Guiding-centre drifts and $\mu_s$ conservation}\label{sec:drifts}

A particular feature of our equilibrium is that it supports guiding-centre drifts, namely
\begin{equation}\label{eqn:vdrift}
\bb{v}_{ds} \doteq \frac{1}{\Omega_s} \left( \frac{q_s}{m_s} \bb{E}_0 + \bb{g} \right) \!\btimes \eb_0 = - \frac{c}{B_0} \D{z}{\Phi_s} \ey \doteq - v_{ds} \ey ,
\end{equation}
where we have used (\ref{eqn:Phis}) to obtain the second equality (note that $v_{ds} > 0$). The velocities of individual particles of species $s$ can thus be decomposed in terms of the parallel velocity $v_\parallel$, the perpendicular velocity $v_\perp$ relative to the guiding-centre drift $v_{ds}$, and the gyrophase angle $\vartheta$:
\begin{equation}\label{eqn:vvec}
\bb{v} = v_\parallel \ex - v_{ds} \ey + v_\perp \bigl( \cos\vartheta \ey + \sin\vartheta \ez \bigr) .
\end{equation}
Despite the presence of guiding-centre drifts in the equilibrium state, the plasma itself has no net momentum. This is because the equilibrium pressure gradients drive diamagnetic drifts $\propto$$\grad P_s \btimes \eb_0$, which precisely cancel the guiding-centre drifts when the plasma is in hydrostatic equilibrium. This is shown explicitly in Section \ref{sec:maxwellian}.

Expressed in terms of the velocity-space variables $(v_\parallel,v_\perp,\vartheta)$, the integrals of the motion are
\begin{subequations}\label{eqn:ZEv}
\begin{gather}\label{eqn:Z2}
\mc{Z}_s = z - \frac{v_\perp}{\Omega_s} \cos\vartheta + \frac{v_{ds}}{\Omega_s} = {\rm const.}, \\*
\label{eqn:energy}
\mc{E}_s = \frac{1}{2} m_s \bigl( v^2_\parallel + v^2_\perp \bigr) + q_s \Phi_s - m_s v_\perp v_{ds} \cos\vartheta + \frac{1}{2} m_s v^2_{ds} = {\rm const.},
\end{gather}
\end{subequations}
and $v_x = v_\parallel = {\rm const}$. The first two terms in (\ref{eqn:Z2}) are simply the projection of the particle guiding centre
\begin{equation}\label{eqn:R}
\bb{R}_s \doteq \bb{r} + \frac{(\bb{v}-\bb{v}_{ds})\btimes\ex}{\Omega_s}
\end{equation}
onto the $z$ axis. Differentiating (\ref{eqn:ZEv}), it is easy to show that the magnetic moment of a particle of species $s$, $\mu_s \doteq m_s v^2_\perp / 2 B_0$, is also conserved:
\begin{equation}\label{eqn:mu}
\D{t}{\mu_s} = \frac{1}{B_0} \biggl( \D{t}{\mc{E}_s} - \frac{m_s}{2} \D{t}{v^2_x} -  q_s \D{z}{\Phi_s} \D{t}{\mc{Z}_s} \biggr) = 0 .
\end{equation}
Were we to be considering anisotropic equilibrium distribution functions, $F_{0s}(\mc{Z}_s,\mc{E}_s,v_x)$ could equivalently be replaced with $F_{0s}(\mc{Z}_s,\mc{E}_s,\mu_s)$.

\subsection{Particle trajectories}\label{sec:trajectories}

Having established that the particles drift across magnetic-field lines while conserving $\mc{E}_s$, $\mc{Z}_s$, and $\mu_s$, we now determine their full phase-space trajectories. The equations of particle motion may be obtained from the characteristics of the steady-state Vlasov equation or, equivalently, from the Hamiltonian (\ref{eqn:hamiltonian}). For some set of phase-space coordinates $[\bb{r}'(t'),\bb{v}'(t')]$, they are
\begin{equation}\label{eqn:characteristics}
\D{t'}{\bb{r}'} = \bb{v'} \quad {\rm and} \quad \D{t'}{\bb{v}'} = \Omega_s \bigl( \bb{v}' \btimes \ex - v_{ds} \ez \bigr) .
\end{equation}
Given some data $[\bb{r}(t), \bb{v}(t)]$ at time $t = t' + \tau$, (\ref{eqn:characteristics}) may be straightforwardly integrated to yield the phase-space trajectories
\begin{subequations}\label{eqn:orbits}
\begin{gather}
\bb{r}' = \bb{r} - v_\parallel \tau \ex + v_{ds} \tau \ey - \frac{v_\perp}{\Omega_s} \bigl[ \sin(\vartheta + \Omega_s \tau) - \sin\vartheta \bigr] \ey + \frac{v_\perp}{\Omega_s} \bigl[ \cos(\vartheta + \Omega_s\tau) - \cos\vartheta \bigr] \ez , \\
\bb{v}' = v_\parallel \ex - v_{ds} \ey + v_\perp \cos(\vartheta + \Omega_s \tau) \ey + v_\perp \sin(\vartheta + \Omega_s \tau) \ez .
\end{gather}
\end{subequations}
The particles stream along field lines, $-\grad\Phi_s\btimes\eb_0$ drift across field lines, and execute Larmor gyrations about field lines. This information will be needed in Section \ref{sec:deltaf}, where we integrate the plasma response to electromagnetic fluctuations along these trajectories.

\section{Linear theory}\label{sec:linear}

In this Section, we consider the linear evolution of small perturbations to the equilibrium state. By considering the response of the distribution function to the fluctuating electromagnetic fields along the phase-space characteristics (\ref{eqn:orbits}), an integral relation is derived relating the current density to the fluctuating electric field (\S\ref{sec:deltaf}). This defines the conductivity tensor of the plasma and hence, in conjunction with Amp\`{e}re's law (\ref{eqn:ampere}), the dispersion properties of the linear waves. The procedure is a fairly standard one \citep[e.g.][]{mikhailovskii62,stix62}. In Sections \ref{sec:orbitintegral} and \ref{sec:maxwellian}, the conductivity tensor is calculated explicitly for an equilibrium plasma described by a Maxwell-Boltzmann distribution in which the Larmor radii of all species are much smaller than the equilibrium gradient lengthscales. Some of the more technical steps in the derivation of the linear theory can be found in Appendix \ref{app:derivation}.

\subsection{Perturbed distribution function and dispersion tensor}\label{sec:deltaf}

The calculation begins by decomposing the distribution function and electromagnetic fields into their equilibrium and fluctuating parts,
\begin{equation}\label{eqn:perturbations}
f_s = F_{0s}(\mc{Z}_s,\mc{E}_s) + \delta f_s(t,\bb{r},\bb{v}), \quad \bb{B} = \bb{B}_0 + \delta\bb{B}(t,\bb{r}), \quad \bb{E} = \bb{E}_0 + \delta\bb{E}(t,\bb{r}) ,
\end{equation}
and substituting them into (\ref{eqn:vlasov})--(\ref{eqn:faraday}). Retaining terms up to first order in the perturbation amplitudes, the linearised version of the Vlasov equation (\ref{eqn:vlasov}) may be written as
\begin{equation}\label{eqn:linearVlasov1}
\frac{{\rm D}}{{\rm D}t} \delta f_s(t,\bb{r},\bb{v}) = - \frac{q_s}{m_s} \left[ \delta\bb{E}(t,\bb{r}) + \frac{\bb{v}}{c} \btimes \delta\bb{B}(t,\bb{r}) \right]\! \bcdot \pD{\bb{v}}{F_{0s}} ,
\end{equation}
where ${\rm D}/{\rm D}t$ is the time derivative taken along the unperturbed orbits given by (\ref{eqn:orbits}). Equation (\ref{eqn:linearVlasov1}) is solved by inverting the differential operator on the left-hand side and integrating along the phase-space trajectory $[\bb{r}'(t'),\bb{v}'(t')]$ terminating at $[\bb{r}(t), \bb{v}(t)]$ (see (\ref{eqn:orbits})):
\begin{equation}\label{eqn:linearVlasov2}
\delta f_s(t,\bb{r},\bb{v}) = - \frac{q_s}{m_s} \int_{-\infty}^t {\rm d}t' \left[ \delta\bb{E}(t',\bb{r}') + \frac{\bb{v}'}{c}\btimes\delta\bb{B}(t',\bb{r}') \right]\! \bcdot \!\left. \pD{\bb{v}}{F_{0s}} \right|_{\bs{r}',\bs{v}'} .
\end{equation}
In order to ease the notation, we henceforth drop the `0' subscript on equilibrium quantities; because we are constructing a linear theory, this omission should cause no confusion.

To proceed, we assume that the equilibrium varies slowly on the scale of the fluctuations and adopt WKBJ plane-wave solutions:
\begin{subequations}
\begin{align}
\delta f_s(t,\bb{r},\bb{v}) &= \delta f_s(\bb{v}) \exp(-\imag\omega t + \imag \bb{k}\bcdot\bb{r}) ,\\*
\delta\bb{B}(t,\bb{r}) &= \delta\bb{B} \exp(-\imag\omega t + \imag\bb{k}\bcdot\bb{r}) ,\\*
\delta\bb{E}(t,\bb{r}) &= \delta\bb{E} \exp(-\imag\omega t + \imag\bb{k}\bcdot\bb{r} ) ,
\end{align}
\end{subequations}
where $\omega$ is the (complex) frequency and $\bb{k}$ is the wavevector. For such disturbances, Faraday's law (\ref{eqn:faraday}) is simply
\begin{equation}\label{eqn:linearFaraday}
\delta\bb{B} = \frac{c}{\omega} \,\bb{k} \btimes \delta\bb{E} ,
\end{equation}
and so (\ref{eqn:linearVlasov2}) becomes
\begin{equation}\label{eqn:linearVlasov3}
\delta f_s = \frac{\imag}{\omega} \frac{q_s}{m_s}\int_0^\infty {\rm d}\tau\, \exp[ - \imag \phi(\tau) ] \!\left. \pD{\bb{v}}{F_s} \right|_{\bs{r}',\bs{v}'} \!\bcdot \!\left( -\imag \D{\tau}{\phi} \msb{I} + \imag \bb{k}\bb{v}' \right) \!\bcdot \delta\bb{E} ,
\end{equation}
where $\tau = t - t' \ge 0$ is the new time variable and we have introduced the phase function
\begin{equation}\label{eqn:phase}
\phi(\tau) \doteq \bb{k}\bcdot ( \bb{r} - \bb{r}' ) - \omega \tau .
\end{equation}
At this point, it is worth recalling that $\mc{Z}_s$ and $\mc{E}_s$, and therefore the equilibrium distribution $F_s(\mc{Z}_s,\mc{E}_s)$, are constant along the orbit $[\bb{r}'(t'), \bb{v}'(t')]$, i.e.~$\mc{E}'_s = \mc{E}_s$ and $\mc{Z}'_s = \mc{Z}_s$. We may use this fact alongside the chain rule to write
\begin{equation}
\left. \pD{\bb{v}}{F_s} \right|_{\bs{r}',\bs{v}'} = \left( \pD{\bb{v}}{\mc{E}_s} \pD{\mc{E}_s}{F_s} + \pD{\bb{v}}{\mc{Z}_s} \pD{\mc{Z}_s}{F_s} \right)_{\!\bs{r}',\bs{v}'} = m_s \bb{v}' \pD{\mc{E}_s}{F_s} - \frac{\ey}{\Omega_s} \pD{\mc{Z}_s}{F_s} .
\end{equation}
The phase-space derivatives of $F_s$ can then be pulled outside of the $\tau$-integral in (\ref{eqn:linearVlasov3}) to yield, after integration by parts and some regrouping,
\begin{equation}\label{eqn:deltaf}
\delta f_s = \frac{\imag}{\omega} \pD{\mc{Z}_s}{F_s} \frac{c}{B} \delta E_y + \frac{q_s}{T_s} \int_0^\infty {\rm d}\tau\, \exp[-\imag\phi(\tau)] \left( - T_s \pD{\mc{E}_s}{F_s} + \frac{k_y}{\omega} \frac{c T_s}{q_s B} \pD{\mc{Z}_s}{F_s} \right) \bb{v}' \bcdot \delta\bb{E} .
\end{equation}
This equation describes the velocity-space response of the distribution function to the fluctuating electric (and, thereby, magnetic) field. The differences between this expression and that pertaining to a homogeneous plasma lie in three places. The first term in (\ref{eqn:deltaf}) represents the advection of the distribution function by the perturbed particle drifts. It is ultimately the cause of the instabilities we discuss in Section \ref{sec:results}. The second addition is the final term in parentheses, which becomes comparable to the usual $-T_s \partial F_s / \partial \mc{E}_s$ term when $k_y v_{ds} \sim \omega$, i.e.~the velocity associated with the guiding-centre drifts must be comparable to the phase speed in the $y$ direction of the waves. Finally, the phase function in (\ref{eqn:deltaf}) takes into account the particle drifts, and so the fluctuating electric (and thereby magnetic) fields will ultimately be sampled at spatial locations a distance more than a Larmor orbit away perpendicular to the mean magnetic field.

To complete the derivation of the general linear theory, we multiply (\ref{eqn:deltaf}) by $q_s\bb{v}$, integrate over velocity space, and sum over species to find a linear relationship between the current density and the fluctuating electric field:
\begin{subequations}\label{eqn:current}
\begin{align}\label{eqn:currenta}
\bb{j} &= \frac{\imag}{\omega} \sum_s \frac{q^2_s}{T_s} \int{\rm d}^3\bb{v}\, \bb{v}\ey \frac{cT_s}{q_s B} \pD{\mc{Z}_s}{F_s} \bcdot \delta\bb{E} \nonumber\\*
\mbox{} &\quad+ \sum_s \frac{q^2_s}{T_s} \int_0^\infty {\rm d}\tau \int{\rm d}^3\bb{v}\, \bb{v} \bb{v}' \exp[-\imag\phi(\tau)] \left( - T_s \pD{\mc{E}_s}{F_s} + \frac{k_y}{\omega} \frac{cT_s}{q_s B} \pD{\mc{Z}_s}{F_s} \right) \!\bcdot \delta\bb{E} \\*
\label{eqn:currentb}
\mbox{} &\doteq \bb{\sigma} \bcdot \delta\bb{E} ,
\end{align}
\end{subequations}
the final equality defining the conductivity tensor $\bb{\sigma}$. Inserting the solution of Faraday's law (\ref{eqn:linearFaraday}) into Amp\`{e}re's law (\ref{eqn:ampere}), and eliminating the current density using (\ref{eqn:currentb}) yields
\begin{subequations}\label{eqn:DdotE}
\begin{equation}
\msb{D} \bcdot \delta\bb{E} = 0,
\end{equation}
where the dispersion tensor
\begin{equation}\label{eqn:dispersion}
\msb{D} \doteq \left( k^2 \msb{I} - \bb{k}\bb{k} - \imag \frac{4\upi\omega}{c^2} \bb{\sigma} \right) v^2_A
\end{equation}
\end{subequations}
and $v_A \doteq B / \sqrt{4\pi\varrho}$ is the Alfv\'{e}n speed. Setting the determinant of (\ref{eqn:dispersion}) to zero provides the sought-after dispersion relation. 

\subsection{Evaluation of the orbit and gyrophase integrals}\label{sec:orbitintegral}

To derive any practical use from (\ref{eqn:DdotE}), we must evaluate the integrals in (\ref{eqn:currenta}). There are two main obstacles to doing so. First, we must choose a basis in which to express the conductivity and dispersion tensors. For the linear Vlasov theory of a homogeneous plasma, the basis typically used is defined by the magnetic field and the wavevector -- the so-called Stix basis \citep{stix62}. For the problem investigated here, there is a preferred direction set by gravity and the equilibrium gradients of $F_s$, and we therefore opt instead to use the $(x,y,z)$ coordinate system. While this complicates the appearance of simple wave motion in the dispersion tensor, it does greatly simplify the terms responsible for buoyancy and particle drifts, and affords a straightforward comparison with previous work on the magnetohydrodynamic stability of thermally stratified atmospheres. We thus decompose our wavevector as
\begin{equation}\label{eqn:kvec}
\bb{k} = k_\parallel \ex + k_\perp \cos\psi \ey + k_\perp \sin\psi \ez
\end{equation}
for $\psi \doteq \tan^{-1}(k_z/k_y)$, so that the phase function (\ref{eqn:phase}) becomes
\begin{equation}\label{eqn:phase2}
\phi(\tau) = k_\parallel v_\parallel \tau - k_y v_{ds} \tau + \frac{k_\perp v_\perp}{\Omega_s} \bigl[ \sin(\vartheta + \Omega_s \tau - \psi) - \sin(\vartheta - \psi) \bigr] - \omega \tau .
\end{equation}
The second obstacle is that the equilibrium distribution function $F_s$ that resides in the integrand of (\ref{eqn:currenta}) is independent of gyrophase {\em at fixed guiding centre}, not at fixed position. As the conductivity tensor is determined by integrals over velocity $\bb{v}$ at fixed position $\bb{r}$, rather than integrals over $\mc{E}_s$ at fixed  $\mc{Z}_s$, we must make both sides of (\ref{eqn:currenta}) commensurate. To do so, we leverage the scale separation between the Larmor radii and the gradient lengthscales of the plasma by Taylor expanding $F_s(\mc{Z}_s,\mc{E}_s)$ about the particle position $z$ and the new velocity-space variable
\begin{align}\label{eqn:varepsilon}
\varepsilon_s &\doteq \frac{1}{2} m_s \bigl( v^2_\parallel + v^2_\perp \bigr) + q_s \Phi_s ,
\end{align}
which is simply the particle energy $\mc{E}_s$ written out to zeroth order to $\rho_s / H$ (cf.~(\ref{eqn:ZEv})). With $F_s$ then expressed as a function of our new phase-space variables $(z,\varepsilon_s)$, the integration over gyrophase angle $\vartheta$ becomes straightforward.

Both of these tasks -- writing (\ref{eqn:current}) in the $(x,y,z)$ coordinate system to facilitate the orbit integral over $\tau$, and expanding the equilibrium distribution function about the phase-space coordinates $(z,\varepsilon_s)$ to ease the integration over gyrophase angle $\vartheta$ -- are detailed in Appendix \ref{app:derivation}. Here, we spare the reader the details and simply state the result:
\begin{align}\label{eqn:sigma2}
\bb{\sigma} &= - \frac{\imag}{\omega} \sum_s \frac{q^2_s}{T_s} \frac{c}{B} \D{z}{\Phi_s} \int{\rm d}^3\bb{v}\, \ey\ey\frac{cT_s}{q_s B}\pD{z}{F_s(z,\varepsilon_s)}  \nonumber\\*
\mbox{} &+ \frac{\imag}{\omega} \sum_s \frac{q^2_s}{T_s} \sum_{n=-\infty}^\infty \int{\rm d}^3\bb{v}\, \frac{\omega \bb{u}_{n,s} \bb{u}^\ast_{n,s} }{\omega + k_y v_{ds} - k_\parallel v_\parallel - n\Omega_s} \left( - T_s \pD{\varepsilon_s}{} + \frac{k_y}{\omega} \frac{cT_s}{q_s B} \pD{z}{} \right) F_s(z,\varepsilon_s) \nonumber\\*
\mbox{} &- \frac{\imag}{\omega} \sum_s \frac{q^2_s}{T_s} \sum_{n=-\infty}^\infty \int{\rm d}^3\bb{v}\, \frac{v_\perp}{\Omega_s} \frac{\omega \bb{w}_{n,s} \bb{u}^\ast_{n,s} }{\omega + k_y v_{ds} - k_\parallel v_\parallel - n\Omega_s} \nonumber\\*
\mbox{} &\quad\qquad\qquad\qquad\times \left( - T_s \pD{\varepsilon_s}{} + \frac{k_y}{\omega} \frac{cT_s}{q_s B} \pD{z}{} \right) \left( \pD{z}{} + q_s \D{z}{\Phi_s} \pD{\varepsilon_s}{} \right) F_s(z,\varepsilon_s),
\end{align}
where ${\rm d}^3\bb{v} = 2\upi v_\perp {\rm d}v_\perp {\rm d}v_\parallel$ and $\ast$ denotes the complex conjugate. The velocities $\bb{u}_{n,s}$ and $\bb{w}_{n,s}$ are composed of combinations of $v_\parallel$ and $v_\perp$ with the $n$th-order Bessel function $J_n(a_s)$ and  derivatives with respect to its argument $a_s \doteq k_\perp v_\perp / \Omega_s$, and are given by (\ref{eqn:uns}) and (\ref{eqn:wns}), respectively. Note that the partial $z$-derivatives of $F_s$ in (\ref{eqn:sigma2}) are taken at fixed $\varepsilon_s$, {\em not} at fixed $v_\parallel$ and $v_\perp$.\footnote{It is of at least historical interest to note that, notwithstanding a few typos (or perhaps errors, as they seem to have propagated through the calculation) and his omission of the self-consistent equilibrium electric field, A.~B.~Mikhailovskii essentially obtained equation (\ref{eqn:sigma2}) 49 years ago in an apparently overlooked appendix to his monograph on {\it Oscillations of an Inhomogeneous Plasma} \citep{mikhailovskii67}. In it, he derived the dielectric tensor for an equilibrium Vlasov plasma in a constant gravitational field (see his equation (II.3)). Were he to have solved for the dispersion relation and analysed it, he might have discovered the MTI more than three decades before it entered the astrophysics literature.}

Up to this point, the only assumptions that have been made about the equilibrium distribution function are that it is isotropic in velocity space at fixed guiding-centre position, that it is independent of the $y$ coordinate, and that its gradient lengthscales are much larger than the Larmor radii of all species. In order to carry out the remaining integrals in (\ref{eqn:sigma2}), we must choose a particular form for $F_s(\mc{Z}_s,\mc{E}_s)$. In the next Section, we use (\ref{eqn:sigma2}) to compute the dispersion tensor for an isotropic Maxwell-Boltzmann distribution function.

\subsection{Dispersion tensor and perturbed distribution for a Maxwell-Boltzmann equilibrium}\label{sec:maxwellian}

A natural choice for our isotropic equilibrium distribution function is the quasi-Maxwell-Boltzmann distribution
\begin{equation}\label{eqn:quasimaxwell}
F_{M,s}(\mc{Z}_s,\mc{E}_s) \doteq \frac{\mc{N}_s(\mc{Z}_s)}{[2\upi T_s(\mc{Z}_s) / m_s]^{3/2}} \exp\biggl[ - \frac{\mc{E}_s}{T_s(\mc{Z}_s)} \biggr] .
\end{equation}
where $\mc{N}_s$ is the number density of guiding centres. Besides being physically motivated by the systems of interest likely being near local thermodynamic equilibrium, (\ref{eqn:quasimaxwell}) also has the convenient mathematical property that the integrals in (\ref{eqn:sigma2}) can be readily performed. To wit, the integrals over $v_\parallel$ may be written in terms of
\begin{equation}\label{eqn:plasmadisp}
Z_p(\zeta) \doteq \frac{1}{\sqrt{\pi}} \int^\infty_{-\infty} {\rm d}x\,x^p\frac{{\rm e}^{-x^2}}{x-\zeta} \quad ( p \ge 0) ,
\end{equation}
which for $p=0$ is the standard plasma dispersion function \citep{fc61}, while those over $v_\perp$ can be calculated with the aid of the \citet{watson66} relation
\begin{equation}\label{eqn:watson}
\int_0^\infty {\rm d}x\, x J_n(px) J_n(qx) \, {\rm e}^{-a^2 x^2} = \frac{1}{2a^2} \Gamma_n\Bigl(\frac{pq}{2a^2}\Bigr) \,{\rm e}^{-(p-q)^2/4a^2},
\end{equation}
where $\Gamma_n(\lambda) \doteq I_n(\lambda) \exp(-\lambda)$ and $I_n$ is the $n$th-order modified Bessel function. (Several particularly useful integrals derived from (\ref{eqn:plasmadisp}) and (\ref{eqn:watson}) are given in Appendix \ref{app:integrals}.)

As explained in Section \ref{sec:orbitintegral} and detailed in Appendix \ref{app:derivation}, we may use the scale separation between the Larmor radii and the gradient lengthscales of the plasma to write
\begin{align}\label{eqn:maxwellexpanded}
F_{M,s}(\mc{Z}_s,\mc{E}_s) = \left[ 1 - \frac{v_\perp\cos\vartheta}{\Omega_s} \left( \pD{z}{} + q_s \D{z}{\Phi_s} \pD{\varepsilon_s}{} \right) \right] F_{M,s}(z,\varepsilon_s) + \mc{O}\Bigl(\frac{\rho_s}{H}\Bigr)^2 ,
\end{align}
in which
\begin{subequations}\label{eqn:dFM}
\begin{align}
\label{eqn:dFdz}
\pD{z}{F_{M,s}(z,\varepsilon_s)} &= F_{M,s}(z,\varepsilon_s) \left( \D{z}{\ln\mc{N}_s} - \frac{3}{2} \D{z}{\ln T_s} + \frac{\varepsilon_s}{T_s} \D{z}{\ln T_s} \right) ,\\*
-T_s \pD{\varepsilon_s}{F_{M,s}(z,\varepsilon_s)} &= F_{M,s}(z,\varepsilon_s) .
\end{align}
\end{subequations}
Taking the zeroth moment of (\ref{eqn:maxwellexpanded}), we see that the number density of particles is related to that of the guiding centres via a Boltzmann factor:
\begin{equation}\label{eqn:ntoN}
n_s \doteq \int{\rm d}^3\bb{v}\, F_{M,s}(\mc{Z}_s,\mc{E}_s) \simeq \int{\rm d}^3\bb{v}\, F_{M,s}(z,\varepsilon_s) = \mc{N}_s(z) \exp\biggl[ - \frac{q_s\Phi_s}{T_s(z)}\biggr] ,
\end{equation}
so that the zeroth-order equilibrium distribution function $F_{M,s}(z,\varepsilon_s)$ is equivalent to the standard (gyrotropic) Maxwell-Boltzmann distribution
\begin{equation}\label{eqn:fMs}
f_{M,s}(z,v_\parallel,v_\perp) = \frac{n_s}{\upi v^2_{ths}} \exp\Biggl( - \frac{v^2_\perp}{v^2_{ths}} \Biggr) \frac{1}{\sqrt{\upi} v_{ths}} \exp\Biggl( - \frac{v^2_\parallel}{v^2_{ths}} \Biggr) ,
\end{equation}
where $v_{ths} \doteq ( 2 T_s / m_s )^{1/2}$ is the thermal speed of species $s$. This is useful. Taking the first moment of (\ref{eqn:maxwellexpanded}) and using (\ref{eqn:vvec}), (\ref{eqn:dFM}), and (\ref{eqn:ntoN}), one can also show that the mean flow of the equilibrium plasma
\begin{align}
\bb{u}_s &\doteq \frac{1}{n_s} \int{\rm d}^3\bb{v}\, \bb{v} F_{M,s}(\mc{Z}_s,\mc{E}_s) \nonumber\\*
\mbox{} &\simeq \frac{1}{n_s} \int{\rm d}^3\bb{v}\, \ey \left[ - v_{ds} - \frac{v^2_\perp \cos^2\vartheta}{\Omega_s}  \left( \pD{z}{} + q_s \D{z}{\Phi_s} \pD{\varepsilon_s}{} \right) \right] F_{M,s}(z,\varepsilon_s) \nonumber\\*
\mbox{} &= \frac{1}{\Omega_s} \!\left( \frac{q_s}{m_s} \bb{E}_0 + \bb{g} - \frac{1}{m_s n_s} \grad P_s \right) \!\btimes \eb_0 = 0.
\end{align}
In other words, the average equilibrium velocity is a superposition of the particle drifts, which are caused by the equilibrium electric field and gravity (see \S\ref{sec:drifts}), and the diamagnetic drift, which is driven by the equilibrium pressure gradient. The two balance in equilibrium, giving zero net flow.

Our equilibrium distribution function now specified, we substitute (\ref{eqn:dFM}) into (\ref{eqn:sigma2}), perform the integrals using (\ref{eqn:plasmadisp}) and (\ref{eqn:watson}), and plug the resulting conductivity tensor into (\ref{eqn:dispersion}) to find
\begin{equation}\label{eqn:Maxdispersion}
\msb{D} = \bigl( k^2\msb{I} - \bb{k}\bb{k} \bigr) v^2_ A - \ey\ey \sum_s  \frac{1}{\varrho} \D{z}{P_s} \D{z}{\ln T_s} - \sum_s  \frac{m_s n_s}{\varrho} \Omega^2_s \!\sum_{n=-\infty}^\infty \frac{2\omega}{k_\parallel v_{ths}} \bigl( \msb{U}_{n,s} + \msb{W}_{n,s} \bigr) .
\end{equation}
The species- and Bessel-order-dependent tensors $\msb{U}_{n,s}$ and $\msb{W}_{n,s}$, whose elements are given by (\ref{eqn:Utensor}) and (\ref{eqn:Wtensor}), respectively, are functions of the $n$th-order Landau pole $\zeta_{n,s} \doteq (\omega+k_y v_{ds} - n\Omega_s)/|k_\parallel| v_{ths}$ and the dimensionless square of the perpendicular wavenumber $\alpha_s \doteq (k_\perp \rho_s)^2/2$. The perturbed distribution function may be calculated from (\ref{eqn:deltaf}) using (\ref{eqn:dFM}). The general result is not particularly useful in the discussion that follows, and so we provide here only the leading-order (in $\rho_s/H$) terms:
\begin{align}\label{eqn:df}
\delta f_s &= \frac{\imag}{\omega} \pD{z}{F_s(z,\varepsilon_s)} \frac{c}{B} \delta E_y \nonumber\\*
\mbox{} &+ \frac{\imag}{\omega} \frac{q_s}{T_s} \sum_{m,\,n=-\infty}^\infty  {\rm e}^{\imag(m-n)(\vartheta-\psi)} \frac{\omega J_m(a_s) \bb{u}^\ast_{n,s}\bcdot \delta\bb{E}}{\omega + k_y v_{ds} - k_\parallel v_\parallel - n\Omega_s} \left( 1 + \frac{k_y}{\omega} \frac{cT_s}{q_s B} \pD{z}{} \right) F_s(z,\varepsilon_s) ,
\end{align}
with $\partial F_s / \partial z$ being given by (\ref{eqn:dFdz}).

\section{Results}\label{sec:results}

The dispersion relation that results from taking the determinant of (\ref{eqn:Maxdispersion}) is rather unwieldy, and it seems best not to write it down here in full. Fortunately, all of the interesting results we have found can be obtained by taking some relatively simple asymptotic limits, which are catalogued in table \ref{tab:ordering}. To make contact with previous MHD calculations of the linear stability of a thermally stratified atmosphere, we first consider the long-wavelength (``drift-kinetic'') limit, $k \rho_s \ll 1$ (\S\ref{sec:longwavelength}). Contact is made with the theory of kinetic MHD for a stratified plasma. For parallel propagating modes ($k_\perp = 0$), we obtain in Section \ref{sec:gyroviscous} the leading-order correction that incorporates finite-Lamor-radius (FLR) effects -- the so-called gyroviscosity. Proceeding to yet smaller scales, in Section \ref{sec:gyrokinetic} we allow for low-frequency fluctuations whose spatial scales perpendicular to the guide magnetic field are comparable to the Larmor radius of the particles. In this ``gyrokinetic'' limit, the physics of drift waves becomes important, as does the damping of modes at perpendicular Larmor scales. In Appendix \ref{app:gk}, we show that one may alternatively obtain the results of Section \ref{sec:gyrokinetic} directly from the nonlinear gyrokinetic theory for a stratified atmosphere. We follow this with a discussion of our results as they relate to the ion- and electron-temperature-gradient drift-wave instabilities well known to the magnetic-confinement-fusion community (\S\ref{sec:ITG}). 

In the results that follow, we have taken the thermal-pressure scale height to be species independent, {\em viz.}~${\rm d}\ln P_s / {\rm d}z = {\rm d}\ln P / {\rm d}z$. The same is true for the temperature scale height, ${\rm d}\ln T_s / {\rm d}z = {\rm d}\ln T/{\rm d}z$. While we have generally allowed the equilibrium temperatures of the various species to differ ({\em viz.}~$T_i \ne T_e$), in preparing the figures we have taken the plasma to be composed of electrons and a single species of singly charged ions ($q_i = e$, $m_i/m_e = 1836$) with identical equilibrium temperatures ($T_i / T_e = 1$), unless otherwise explicitly stated. Without loss of generality, we take $k_\parallel > 0$. Because we have astrophysical applications in mind, we primarily focus on high-$\beta$ plasmas, the discussion in Section \ref{sec:ITG} being a notable exception. 

\setlength{\tabcolsep}{8pt}
\begin{table}
\begin{center}
\def~{\hphantom{0}}
\begin{tabular}{cccccc} 
 Limit 		& $\omega/\Omega_s$ 	& $k_\parallel \rho_s$ 		& $k_\perp \rho_s$ 	& $\beta^{-1}_s$	& Section\\[3pt]
 drift-kinetic 	& $\mc{O}(\epsilon)$ 	& $\mc{O}(\epsilon)$ 		& $\mc{O}(\epsilon)$	& $\mc{O}(1)$ 		& \S\ref{sec:longwavelength}\\ 
gyroviscous 	& $\mc{O}(\epsilon)$ 	& $\mc{O}(\sqrt{\epsilon})$ 	& 0 				& $\mc{O}(\epsilon)$	& \S\ref{sec:gyroviscous} \\
gyrokinetic 	& $\mc{O}(\epsilon)$ 	& $\mc{O}(\epsilon)$ 		& $\mc{O}(1)$ 		& $\mc{O}(1)$		& \S\ref{sec:gyrokinetic} \\
\end{tabular}
  \caption{Ordering of parameters relative to $\epsilon \doteq \rho_s/H$ in the various limits taken in Section \ref{sec:results}. Subsidiary expansions in small $m_e/m_i$ and in small or large $\beta_i$ and $T_i/T_e$ can be taken after the $\epsilon$ expansion is done, as long as their values do not interfere with the primary expansion in $\epsilon$.}
  \label{tab:ordering}
\end{center}
\end{table}

\subsection{Drift-kinetic limit: $k \rho_s \ll 1$}\label{sec:longwavelength}

In the drift-kinetic limit, there is no practical difference between the particle position and the guiding-centre position. This simplifies things considerably. For one thing, the pressure tensor becomes diagonal in a coordinate system set by the magnetic-field direction (cf.~(\ref{eqn:gyrotropicP})), so that 
\begin{equation}
\delta\msb{P}_s = \delta P_{\parallel s} \ex\ex + \delta P_{\perp s} ( \ey\ey + \ez\ez ).
\end{equation}
In addition, the higher-order contribution to the conductivity tensor, $\msb{W}_{n,s}$, may be completely dropped from the analysis, and $\msb{U}_{n,s}$ takes on a relatively simple form in which contributions from $n=-1$, $0$, and $1$ are all that is needed. Let us proceed.

We start by applying the low-frequency, long-wavelength ordering
\begin{equation}\label{eqn:longwave_limit}
\frac{\omega}{\Omega_s} \sim k_\parallel \rho_s \sim k_\perp \rho_s \sim \frac{\rho_s}{H} \doteq \epsilon \ll 1,
\end{equation}
with $\beta_s \doteq 8\upi n_s T_s / B^2_0 \sim 1$, to (\ref{eqn:Maxdispersion}). (A subsidiary expansion in, e.g., large $\beta_s$ may be taken, so long as $\beta_s \nsim 1/\epsilon$.) Note that this ordering precludes drift waves, since $k_y v_{ds} \sim k_y v_{ths} (\rho_s / H ) \ll \omega$. Expanding the $Z_p(\zeta_{n,s})$ functions about $\zeta_s \doteq \omega / k_\parallel v_{ths}$ and using the lowest-order approximations $\Gamma_0(\alpha_s) \simeq - \Gamma'_0(\alpha_s) \simeq 1$ and $\Gamma_1(\alpha_s) \simeq \alpha_s/2$, the elements of the tensor $\msb{U}_{n,s}$ (cf.~(\ref{eqn:Utensor})) simplify considerably. They are
\begin{subequations}\label{eqn:longwavetensor}
\begin{align}
\msb{U}^{(xx)}_{n,s} &\simeq Z_2(\zeta_s) \,\delta_{n,0}, \\
\msb{U}^{(xy)}_{n,s} &\simeq \imag \frac{k_z \rho_s}{2} \left( 1 - \frac{\imag}{k_z} \D{z}{\ln P_s} \right) Z_1(\zeta_s) \,\delta_{n,0} ,\\
\msb{U}^{(xz)}_{n,s} &\simeq -\imag \frac{k_y \rho_s}{2} Z_1(\zeta_s) \,\delta_{n,0} , \\ 
\msb{U}^{(yy)}_{n,s} &\simeq \frac{k^2_\parallel \rho^2_s}{4}  \, \zeta_s \bigl( \delta_{n,1} + \delta_{n,-1} \bigr)  + \frac{k^2_z\rho^2_s}{4} \left[ 2 + \left(\frac{1}{k_z} \D{z}{\ln P_s}\right)^2 \right] Z_0(\zeta_s)  \,\delta_{n,0} , \\
\label{eqn:longwavelength_Uyz}
\msb{U}^{(yz)}_{n,s} &\simeq \imag \frac{k_\parallel \rho_s}{4} \bigl( \delta_{n,1} + \delta_{n,-1} \bigr) - \frac{k_y k_z \rho^2_s}{4} \left( 2 + \frac{\imag}{k_z} \D{z}{\ln P_s} \right) Z_0(\zeta_s) \,\delta_{n,0} ,\\
\msb{U}^{(zz)}_{n,s} &\simeq \frac{k^2_\parallel \rho^2_s}{4} \, \zeta_s \bigl( \delta_{n,1} + \delta_{n,-1} \bigr) + \frac{k^2_y \rho^2_s}{2}  Z_0(\zeta_s) \,\delta_{n,0} ,
\end{align}
\end{subequations}
where $\delta_{n,m}$ denotes the Kronecker delta. The un-written components of $\msb{U}_{n,s}$ (namely, $yx$, $zx$, and $zy$) are the same as their transpose counterparts but with $\imag \rightarrow -\imag$. The final term in (\ref{eqn:longwavelength_Uyz}), while formally one order in $\epsilon$ smaller than the first term in that expression, must be retained, as the latter eventually vanishes to leading order under quasi-neutrality.

Inserting (\ref{eqn:longwavetensor}) into (\ref{eqn:Maxdispersion}) and setting its determinant to zero yields the general dispersion relation in the drift-kinetic limit, which we analyze for a variety of wavevector geometries in the next four subsections (\S\S\ref{sec:longwavelength_kprl}--\ref{sec:longwavelength_kprp}). Before doing so, it will serve useful to have readily available the long-wavelength limit of the perturbed distribution function (cf.~(\ref{eqn:df})):
\begin{equation}\label{eqn:longwavelength_df}
\delta f_s = \frac{\imag}{\omega} \frac{c}{B} \pD{z}{F_s} \delta E_y - \left( \frac{\imag}{\omega} \frac{q_s}{T_s} \frac{\omega \bb{u}^\ast_{0,s}\bcdot\delta\bb{E}}{\omega - k_\parallel v_\parallel} + \frac{2\bb{v}_\perp\bcdot\delta\bb{u}_\perp}{v^2_{ths}} \right) T_s \pD{\varepsilon_s}{F_s} 
\end{equation}
where
\begin{equation}\label{eqn:longwavelength_uprp}
\delta\bb{u}_\perp \doteq \frac{c}{B} \delta\bb{E} \btimes \ex
\end{equation}
is the lowest-order mean perpendicular velocity -- the (species-independent!) perturbed $\bb{E}\btimes\bb{B}$ flow. The appearance of the combination $(2\bb{v}_\perp\bcdot\delta\bb{u}_\perp )(\partial F_s / \partial \varepsilon_s)$ in (\ref{eqn:longwavelength_df}) suggests that, in the long-wavelength limit, it would be advantageous to redefine our energy variable to account for the kinetic energy of this flow,
\begin{equation}\label{eqn:einw}
\varepsilon_s = \frac{1}{2} m_s v^2 + q_s \Phi_s \longrightarrow \frac{1}{2} m_s | \bb{v} - \delta\bb{u}_\perp |^2 + q_s \Phi_s \doteq \frac{1}{2} m_s w^2_\perp + q_s \Phi_s,
\end{equation}
such that the energies of all particles are measured {\em in the frame of the} $\bb{E}\btimes\bb{B}$ {\em velocity}. In other words, Alfv\'{e}nic fluctuations do not change the form of the distribution function, but rather define the moving frame in which any changes to it are to be measured. Physically, this is because particles in a magnetised plasma adjust on a cyclotron timescale to take on the local $\bb{E}\btimes\bb{B}$ velocity; in a sense, $\rho_i$ performs the role of the mean free path. This principle is what underlies Kulsrud's formulation of kinetic MHD, in which the perpendicular particle velocities are measured relative to the $\bb{E}\btimes\bb{B}$ drift, the latter being governed by a set of MHD-like fluid equations rather than by a kinetic equation \citep{kulsrud64,kulsrud83}. In this frame, the perturbed distribution is gyrotropic at fixed position -- a particularly useful result.

Interpreting the other terms in (\ref{eqn:longwavelength_df}) is eased by shifting from $F_s(v_\parallel,\varepsilon_s)$ to $f_s(v_\parallel,w_\perp)$ and by using Faraday's law (\ref{eqn:linearFaraday}) to write the perpendicular components of $\delta\bb{E}$ in terms of $\delta B_z$ and $\delta B_\parallel$:
\begin{align}\label{eqn:longwavelength_df2}
\delta f_s(v_\parallel,w_\perp) &= \frac{\imag}{k_\parallel} \pD{z}{f_s} \frac{\delta B_z}{B} + \frac{w^2_\perp}{v^2_{ths}} \frac{\delta B_\parallel}{B} f_s \nonumber\\*
\mbox{} &\quad - \frac{\imag}{k_\parallel} \left( \frac{q_s \delta E_\parallel}{T_s}   + \D{z}{\ln P_s} \frac{\delta B_z}{B} - \imag k_\parallel \frac{w^2_\perp}{v^2_{ths}} \frac{\delta B_\parallel}{B} \right) \frac{v_\parallel}{v_\parallel - \omega / k_\parallel }  f_s .
\end{align}
Each of the terms on the right-hand side of (\ref{eqn:longwavelength_df2}) has a straightforward physical interpretation. The first term arises from particles sampling the equilibrium gradients as they stream along the perturbed magnetic field, or, equivalently, from the distribution function being advected in a Lagrangian sense upwards along a vertically perturbed magnetic-field line. The second term merely serves to enforce adiabatic invariance of $\mu_s$, and would vanish if we were to have written our distribution function in terms of $\mu_s$ rather than $w_\perp$. The remaining three terms -- those on the second line of (\ref{eqn:longwavelength_df2}) -- are due to the acceleration of particles along the magnetic field by the parallel electric force $q_s \delta E_\parallel$, the divergence of the parallel pressure $-(1/n_s) \grad\bcdot(\eb P_{\parallel s}) \rightarrow - (\delta B_z/B) (1/n_s) {\rm d}P_s/{\rm d}z$, and the mirror force $-\mu_s \eb\bcdot\grad B \rightarrow -\mu_s \imag k_\parallel \delta B_\parallel$. Each of these forces may lead to collisionless damping if a significant fraction of the particles in the distribution function is Landau resonant, $\omega \sim k_\parallel v_\parallel$, thereby transferring the free energy stored in the lower-order moments of (\ref{eqn:longwavelength_df2}) (e.g.,~density, momentum, pressure) to the higher-order ``kinetic'' moments (i.e.~fine-scale structure in velocity space). Lower-order moments of particular interest in this section are the perturbed density of species $s$,
\begin{equation}\label{eqn:longwavelength_dn}
\frac{\delta n_s}{n_s} =  \frac{\imag}{k_\parallel} \left[ \D{z}{\ln n_s} - \D{z}{\ln P_s} Z_1(\zeta_s) \right] \frac{\delta B_z}{B} + \bigl[ 1 - Z_1(\zeta_s) \bigr] \frac{\delta B_\parallel}{B} - \frac{\imag}{k_\parallel} \frac{q_s \delta E_\parallel}{T_s} Z_1(\zeta_s) ,
\end{equation}
the perturbed perpendicular pressure of species $s$,
\begin{subequations}\label{eqn:longwavelength_Pprp}
\begin{align}
\frac{\delta P_{\perp s}}{P_s} &= \frac{\imag}{k_\parallel} \D{z}{\ln P_s} \bigl[ 1 - Z_1(\zeta_s) \bigr] \frac{\delta B_z}{B} + 2 \bigl[ 1 - Z_1(\zeta_s) \bigr] \frac{\delta B_\parallel}{B} - \frac{\imag}{k_\parallel} \frac{q_s \delta E_\parallel}{T_s} Z_1(\zeta_s) \\*
\label{eqn:longwavelength_dpprp}
\mbox{} &= \frac{\delta n_s}{n_s} + \frac{\imag}{k_\parallel} \D{z}{\ln T_s} \frac{\delta B_z}{B} + \bigl[ 1 - Z_1(\zeta_s) \bigr] \frac{\delta B_\parallel}{B} ,
\end{align}
\end{subequations}
and the perturbed parallel pressure of species $s$,
\begin{subequations}\label{eqn:longwavelength_Pprl}
\begin{align}
\frac{\delta P_{\parallel s}}{P_s} &=  \frac{\imag}{k_\parallel} \D{z}{\ln P_s} \bigl[ 1 - 2 Z_3(\zeta_s) \bigr] \frac{\delta B_z}{B} + \bigl[ 1 - 2Z_3(\zeta_s)\bigr] \frac{\delta B_\parallel}{B_0} - \frac{\imag}{k_\parallel} \frac{q_s \delta E_\parallel}{T_s} 2Z_3(\zeta_s) \\*
\label{eqn:longwavelength_dpprl}
\mbox{} &= \frac{\delta n_s}{n_s} + \frac{\imag}{k_\parallel} \frac{\delta B_z}{B} \left[ \D{z}{\ln P_s} - \D{z}{\ln n_s} \frac{2Z_3(\zeta_s)}{Z_1(\zeta_s)} \right] + \frac{2Z_3(\zeta_s) - Z_1(\zeta_s)}{Z_1(\zeta_s)} \left( \frac{\delta n_s}{n_s} - \frac{\delta B_\parallel}{B} \right) .
\end{align}
\end{subequations}
To obtain (\ref{eqn:longwavelength_dpprp}) and (\ref{eqn:longwavelength_dpprl}), we have used (\ref{eqn:longwavelength_dn}) to eliminate $\delta E_\parallel$ in favour of $\delta n_s$; this makes the calculation of the perturbed temperatures particularly easy.

The final preparatory exercise before proceeding with the analysis of the drift-kinetic dispersion relation is to compute the parallel electric field found in (\ref{eqn:longwavelength_df}). This is most easily accomplished by multiplying (\ref{eqn:longwavelength_dn}) by $q_s n_s$, summing the result over species, and using quasi-neutrality (\ref{eqn:quasineutrality}) to eliminate all but the Landau terms. The result is
\begin{equation}\label{eqn:eparallel}
\delta E_\parallel = { {\displaystyle \sum\nolimits_s q_s n_s Z_1(\zeta_s) } \over {\displaystyle \sum\nolimits_s \frac{q^2_s n_s}{T_s} Z_1(\zeta_s) } }  \left( \imag k_\parallel \frac{\delta B_\parallel}{B} - \D{z}{\ln P_s} \frac{\delta B_z}{B} \right) .
\end{equation} 
This equation amounts to a statement of parallel pressure balance, taking into account the forcing-out of large-pitch-angle particles (those with $\cos^{-1}(v_\parallel / v ) \sim \upi / 2$) from regions of increased magnetic-field strength and the squeezing of particles upwards along vertically displaced magnetic-field lines by the pressure gradient. 

The rapid establishment of quasi-neutrality by the parallel electric field (\ref{eqn:eparallel}) means, upon comparison with (\ref{eqn:longwavelength_uprp}), that $\delta E_\parallel / \delta E_\perp \sim k \rho_s \ll 1$. Faraday's law (\ref{eqn:linearFaraday}) then expresses the freezing of the magnetic flux into the mean perpendicular flow of the perturbed plasma,
\begin{equation}\label{eqn:fluxfreezing}
\delta\bb{B} = \frac{c}{\omega} \bb{k}\btimes \delta\bb{E}_\perp = - \frac{\bb{k}}{\omega}\btimes \bigl( \delta\bb{u}_\perp\btimes B\ex \bigr) ,
\end{equation}
a natural outcome at long wavelengths. Introducing the Lagrangian perpendicular displacement via $\delta\bb{u}_\perp \doteq {\rm d}\bb{\xi}_\perp/{\rm d}t \rightarrow -\imag\omega\bb{\xi}_\perp$, (\ref{eqn:fluxfreezing}) may be written
\begin{equation}\label{eqn:induction}
\frac{\delta\bb{B}}{B} = \imag k_\parallel \bb{\xi}_\perp - \imag \bb{k}_\perp\bcdot\bb{\xi}_\perp \ex .
\end{equation}
This particular form of the induction equation affords a better physical understanding of the second moments (\ref{eqn:longwavelength_dpprp}) and (\ref{eqn:longwavelength_dpprl}), as follows. Using (\ref{eqn:induction}) to replace $\delta B_z/B$ by $\imag k_\parallel \xi_z$ in (\ref{eqn:longwavelength_dpprp}) and (\ref{eqn:longwavelength_dpprl}), we find that the Lagrangian changes in the perpendicular and parallel temperatures as a fluid element of species $s$ is displaced are
\begin{subequations}\label{eqn:longwavelength_DT}
\begin{align}
\frac{\Delta T_{\perp s}}{T_s} &\doteq \frac{\delta T_{\perp s}}{T_s} + \xi_z \D{z}{\ln T_s} = \bigl[ 1 - Z_1(\zeta_s) \bigr] \frac{\delta B_\parallel}{B} , \\*
\frac{\Delta T_{\parallel s}}{T_s} &\doteq \frac{\delta T_{\parallel s}}{T_s} + \xi_z \D{z}{\ln T_s} = \frac{2Z_3(\zeta_s) - Z_1(\zeta_s)}{Z_1(\zeta_s)} \left( \frac{\Delta n_s}{n_s} - \frac{\delta B_\parallel}{B} \right) .
\end{align}
\end{subequations}
For $\zeta_s \gg 1$ (i.e., the fluctuation propagates/grows faster than the time it takes a particle of species $s$ to traverse its parallel wavelength), $Z_1(\zeta_s) \simeq -1/2\zeta^2_s$ and $Z_3(\zeta_s) \simeq -3/4\zeta^2_s$, and (\ref{eqn:longwavelength_DT}) returns the linearised version of the double-adiabatic theory, $T_\perp / B \simeq {\rm const}$ and $T_\parallel B^2/n^2 \simeq {\rm const}$, in which heat flows are assumed negligible \citep{cgl56}. In the opposite limit $\zeta_s \ll 1$, $Z_1(\zeta_s) \simeq 1 + \imag \sqrt{\pi}\zeta_s$ and $Z_3(\zeta_s) \simeq 1/2$, and so we have $\Delta T_{\perp s}/T_s \simeq -\imag\sqrt{\pi} \zeta_s \delta B_\parallel / B$ and $\Delta T_{\parallel s}/T_s \simeq -\imag\sqrt{\pi} \zeta_s ( \Delta n_s / n_s - \delta B_\parallel / B )$, corresponding to nearly isothermal fluctuations. Each of these limits will aid in the analysis of the drift-kinetic dispersion relation, which we now, finally, commence.

\subsubsection{Drift-kinetic limit: Parallel propagation $(\bb{k}\parallel\eb)$}\label{sec:longwavelength_kprl}

The simplest wavevector geometry to analyze is $k_\perp = 0$, in which case (\ref{eqn:Maxdispersion}) and (\ref{eqn:longwavetensor}) combine to give $\msb{D}^{(xx)} \sim \mc{O}(1)$, $\msb{D}^{(xy)} \sim \mc{O}(\epsilon)$, $\msb{D}^{(yy)} \sim  \msb{D}^{(zz)} \sim \mc{O}(\epsilon^2)$, and $\msb{D}^{(xz)} = \msb{D}^{(zx)} = 0$. It is straightforward to show further that the contributions from $\msb{U}^{(yz)}_{n,s}$ and $\msb{U}^{(zy)}_{n,s}$ to the dispersion tensor vanish to leading order by quasi-neutrality (\ref{eqn:quasineutrality}), a fact that will become important again in Section \ref{sec:gyroviscous} when we consider gyroviscous contributions to the wave dispersion. The long-wavelength dispersion relation then reads
\begin{equation}\label{eqn:longwavelength_kprl_disprel}
\msb{D}^{(zz)} \left[ \msb{D}^{(xx)} \msb{D}^{(yy)}  - \msb{D}^{(xy)} \msb{D}^{(yx)} \right] = 0 .
\end{equation}
The first solution, $\msb{D}^{(zz)} = 0$, corresponds to Alfv\'{e}n waves with frequencies $\omega = \pm k_\parallel v_A$, which are polarized with $\delta\bb{B}$ along the $y$-axis. They are unaffected by buoyancy, and are completely decoupled from the compressive fluctuations. The latter satisfy
\begin{equation}\label{eqn:collisionlessMTI_kprl}
\omega^2 = k^2_\parallel v^2_A - \sum_s \frac{1}{\varrho} \D{z}{P_s} \left[ \D{z}{\ln T_s} +  \zeta_s Z_0(\zeta_s) \D{z}{\ln P_s} \, \Upsilon_s  \right] , 
\end{equation}
where
\begin{equation}\label{eqn:eta}
\Upsilon_s \doteq 1 - { {\displaystyle \sum\nolimits_{s'} \frac{q_s q_{s'} n_{s'}}{T_s}  Z_1(\zeta_{s'})} \over {\displaystyle \sum\nolimits_{s'} \frac{q^2_{s'} n_{s'}}{T_{s'}} Z_1(\zeta_{s'})} } .
\end{equation}
The terms on the right-hand side of (\ref{eqn:collisionlessMTI_kprl}) correspond, respectively, to the effects of magnetic tension, buoyancy, and collisionless damping by way of Landau-resonant particles undergoing acceleration by a combination of parallel electric fields and parallel gravity (both of which are proportional to the pressure gradient, and thus have been combined by introducing $\Upsilon_s$). It is instructive to compare (\ref{eqn:collisionlessMTI_kprl}) with its more familiar Braginskii-MHD (i.e.~collisional, magnetised) counterpart \citep[cf.][]{balbus00,kunz11}, written in a rather auspicious guise:
\begin{equation}\label{eqn:collisionalMTI}
\omega^2 = k^2_\parallel v^2_A - \sum_s \frac{1}{\varrho} \D{z}{P_s} \left[ \D{z}{\ln T_s} + \frac{\imag\omega}{k^2_\parallel \kappa - (5/2) \imag\omega} \D{z}{\ln P_s} \right] ,
\end{equation}
where $\kappa$ is the (electron-dominated) thermal diffusivity of a collisional plasma. At large enough wavenumbers such that their final terms (those proportional to ${\rm d}\ln P_s/{\rm d}z$) are negligible, both solutions return what now bears the moniker ``magnetothermal instability'' \citep{balbus00,balbus01}:
\begin{equation}
\omega^2 \simeq k^2_\parallel v^2_A + g \D{z}{\ln T} ,
\end{equation}
which may be unstable in an atmosphere with ${\rm d}\ln T_s / {\rm d}z < 0$, provided the magnetic field is sufficiently subthermal (i.e. $\sqrt{\beta} \gg k_\parallel H \gg 1$).

That both the collisional and collisionless calculations return nearly the same result in the long-wavelength limit is encouraging -- the MTI in its simplest form passes unscathed through to the kinetic case. But while, in both cases, the MTI is driven by the rapid transport of heat along perturbed magnetic-field lines, there is quite different physics at play in what governs that transport. For the collisional case, in which particles interact with one another on the collisional mean free path $\lambda_{mfp}$, the MTI grows maximally once $k^2_\parallel \lambda_{mfp} H \gtrsim 1$, that is, for wavelengths intermediate between the mean free path and the thermal-pressure scale height. The lower bound of $\lambda_{mfp}$ is readily understood: fluid elements separated by a fluctuation wavelength must be able to equilibrate their temperatures via conduction, short-circuiting the usual adiabatic response and allowing, say, hot fluid elements to remain hot as they are convectively lifted upwards into cooler surroundings. The upper bound of $H$ results from a displaced fluid element needing to maintain pressure balance with its surroundings by radiating sound waves faster than the fluid element rises (i.e.~$k_\parallel v_{thi} \gg \omega \sim v_{thi} / H$).\footnote{This limit, the {\em Boussinesq approximation} (for which the perturbed pressure $\delta p / p \sim \mc{O}(\zeta^{-2})$), was used to obtain the collisional dispersion relation (\ref{eqn:collisionalMTI}). Such a limit cannot in general be taken in the collisionless case, because the pressure response is different parallel and perpendicular to the magnetic field (see (\ref{eqn:longwavelength_dpprp}) and (\ref{eqn:longwavelength_dpprl})).} In the collisionless case, conduction is not set by particle-particle interactions, and sounds waves are collisionlessly damped. The application of the collisional MTI to a collisionless plasma is not obvious {\it a priori}.

\begin{figure}
\centering
\includegraphics[width =0.6\textwidth]{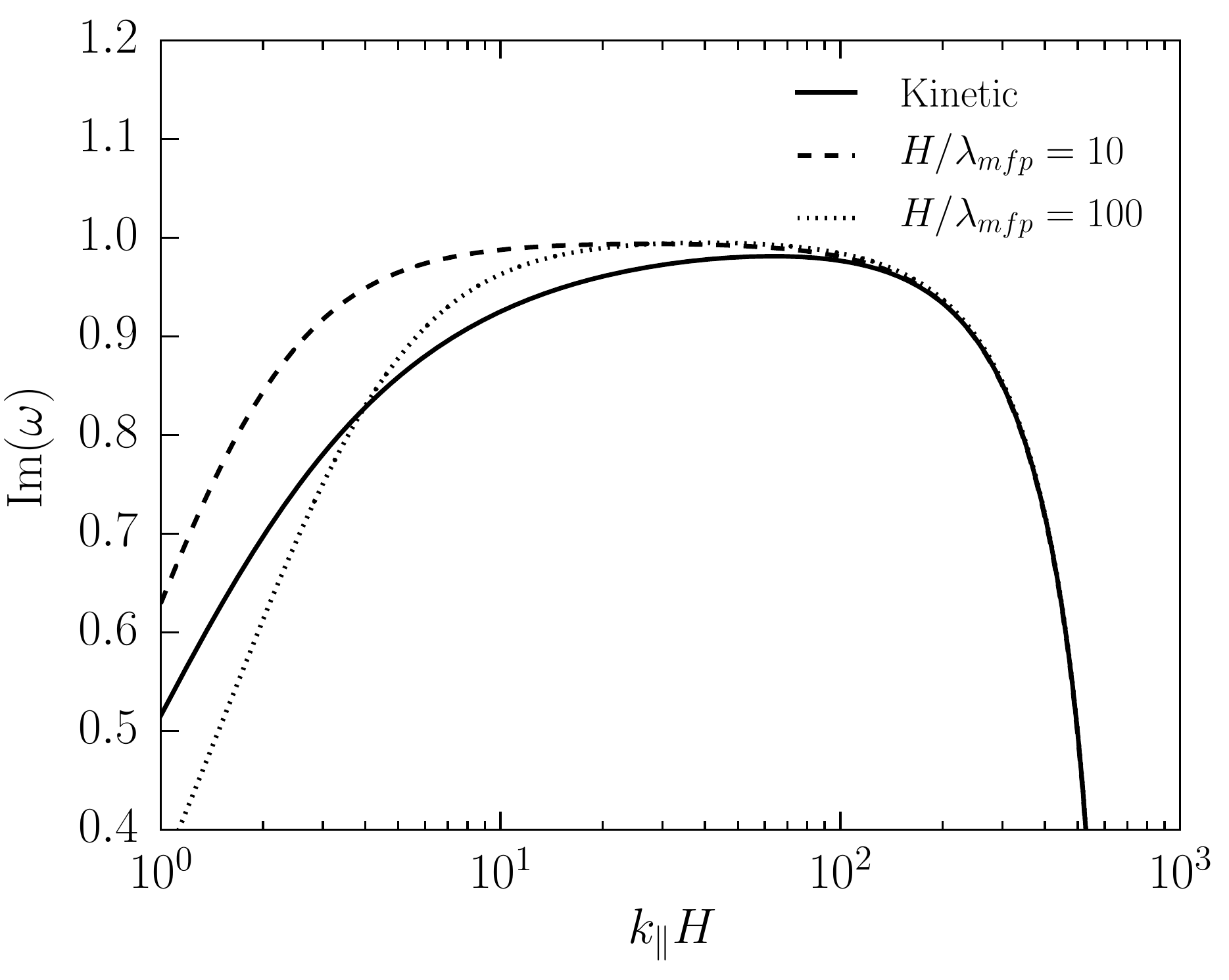}
\caption{Growth rates (normalised by $\sqrt{-g \,{\rm d}\ln T/{\rm d}z}$) as calculated from the collisionless drift-kinetic dispersion relation (\ref{eqn:collisionlessMTI_kprl}; solid line) and from the collisional Braginskii-MHD dispersion relation (dotted and dashed lines), both with $k_\perp = 0$, $\beta_i = 10^6$, and ${\rm d}\ln T / {\rm d}\ln P = 1/3$. See \S\ref{sec:longwavelength_kprl} for details.}
\label{fig:longwave_comp}
\end{figure}

In fact, the collisionless case is somewhat simpler than the collisional case, in that the same condition that allows displaced fluid elements to maintain pressure balance with their surroundings (that is, $\zeta_s \ll 1$) also ensures that such displacements proceed approximately isothermally. Indeed, (\ref{eqn:longwavelength_Pprp})--(\ref{eqn:longwavelength_DT}) with $k_\perp = 0$ and $\zeta_s \ll 1$ imply
\begin{equation}\label{eqn:DTcollisionless}
\frac{\delta P_{\perp}}{P} \sim \mc{O}(\zeta_i) , ~\frac{\delta P_{\parallel}}{P} \sim \mc{O}(\zeta^2_i) ,~ \frac{\Delta T_{\perp}}{T} = 0,~ \frac{\Delta T_{\parallel}}{T} \sim \mc{O}(\zeta_i) \quad \textrm{(collisionless, magnetised).}
\end{equation}
Simply put, conduction and buoyancy in a collisionless plasma are inextricably linked. This is {\em not} true in a collisional, magnetised plasma, for which \citep[cf.][]{kunz11}
\begin{equation}\label{eqn:DTcollisional}
\frac{\delta P}{P} \sim \mc{O}(\zeta^2_i) ,~ \frac{\Delta T}{T} \simeq \xi_z \D{z}{\ln P}  \frac{-\imag \omega}{k^2_\parallel \kappa - (5/2) \imag \omega} \quad \textrm{(collisional, magnetised).}
\end{equation}
In this case, pressure balance can be achieved regardless of whether conduction is rapid enough to ensure nearly isothermal displacements (just take $k_\parallel \lambda_{mfp} \rightarrow 0$). By contrast, in a collisionless plasma, particles redistribute heat along field lines not by communicating thermodynamic information via collisions with other particles, but by free-streaming along perturbed magnetic fields. This naturally occurs on roughly the same timescale on which a sound wave propagates and establishes pressure balance. If the perturbed field lines are not isothermal, it is for the same reason that a fluid element cannot maintain pressure equilibrium with its surroundings, that is, an appreciable number of particles are Landau-resonant with the fluctuation.

Figure \ref{fig:longwave_comp} displays the growth rates as calculated from (solid line) the collisionless drift-kinetic dispersion relation (\ref{eqn:collisionlessMTI_kprl}) and (dotted and dashed lines) the collisional Braginskii-MHD dispersion relation \citep[e.g.~equation 24 of][]{kunz11} with $k_\perp = 0$. All solutions asymptote to a value close to the maximum growth rate $\sqrt{-g\,{\rm d}\ln T/{\rm d}z}$ before being cutoff by magnetic tension at $k_\parallel H \approx \sqrt{\beta\, {\rm d}\ln T / {\rm d}\ln P}$. Their approach to the maximum growth rate, however, is quite different. These trends may be retrieved by balancing the final (stabilizing) terms in (\ref{eqn:collisionlessMTI_kprl}) and (\ref{eqn:collisionalMTI}) with the destabilizing temperature gradient, which gives $\omega \propto k_\parallel$ for the collisionless case and $\omega \propto k^2_\parallel$ for the collisional case. As explained above, this is related to the scale on which conduction can efficiently isothermalize displaced magnetic-field lines (cf.~(\ref{eqn:DTcollisionless}) and (\ref{eqn:DTcollisional})).

\subsubsection{Drift-kinetic limit: Oblique propagation with $k_\perp = k_z$ $(\psi = \upi/2)$}\label{sec:longwavelength_kz}

Next, we relax our assumption that $k_\perp = 0$ and allow for wavevectors with a vertical component: $\bb{k} = k_\parallel \ex + k_z \ez$. The first thing to notice is that we no longer have $\delta B_\parallel = 0$, but rather $\delta B_\parallel/B = -\imag k_z \xi_z$ (see (\ref{eqn:induction})), and so the Lagrangian change of the perpendicular temperature  (\ref{eqn:longwavelength_DT}) as a fluid element is displaced upwards or downwards no longer identically vanishes. Instead, {\em changes in temperature go hand-in-hand with changes in magnetic-field strength}. Because the latter are collisionlessly damped \citep{barnes66}, this makes the physics of the MTI much richer than in the collisional case, for which $\Delta T \simeq 0$ in the fast-conduction limit regardless of wavevector orientation. This is particularly true at large $k_z$, as we now show.

For $k_\perp = k_z$, we have $\msb{D}^{(xx)} \sim \mc{O}(1)$, $\msb{D}^{(xy)} \sim \mc{O}(\epsilon)$, and $\msb{D}^{(xz)} \sim \msb{D}^{(yy)} \sim \msb{D}^{(zz)} \sim \mc{O}(\epsilon^2)$. (As in \S\ref{sec:longwavelength_kprl}, the contributions from $\msb{U}^{(yz)}_{n,s}$ and $\msb{U}^{(zy)}_{n,s}$ vanish by quasi-neutrality.) The leading-order dispersion relation is thus of the same form as (\ref{eqn:longwavelength_kprl_disprel}), with the Alfv\'{e}nic fluctuations $\omega = \pm k_\parallel v_A$ completely decoupled from the compressive ones. The dispersion relation for the latter may be manipulated into a form similar to the $k_\perp = 0$ result (\ref{eqn:collisionlessMTI_kprl}),
\begin{align}\label{eqn:collisionlessMTI_kz}
\omega^2 = k^2 v^2_A - \sum_s \frac{1}{\varrho} \D{z}{P_s} \left[ \D{z}{\ln T_s} + \zeta_s Z_0(\zeta_s) \D{z}{\ln P_s} \widetilde{\Upsilon}_s \right] ,
\end{align}
the only substantive difference between the two being that $\Upsilon_s$ (see equation (\ref{eqn:eta})) is replaced by
\begin{equation}\label{eqn:eta-tilde}
\widetilde{\Upsilon}_s \doteq \Upsilon_s + k^2_z H^2 (1+\Upsilon_s).
\end{equation}
The additional term in $\widetilde{\Upsilon}_s$ that is proportional to $k^2_z H^2$ arises from the Barnes damping of magnetic-field-strength fluctuations, an effect we anticipate to stabilize the drift-kinetic MTI once that term becomes comparable to the destabilizing temperature gradient -- that is, once $\zeta_i k^2_z H^2 \sim 1$, or $k_z H \sim \sqrt{k_\parallel H}$. Just before that, where $\zeta_i \ll (k_z/k_\parallel)^2$, the two terms in brackets in (\ref{eqn:collisionlessMTI_kz}) must balance and the growth rate ought to decrease as $k^{-2}_z$. 

\begin{figure}
\centering
\includegraphics[width =0.48\textwidth]{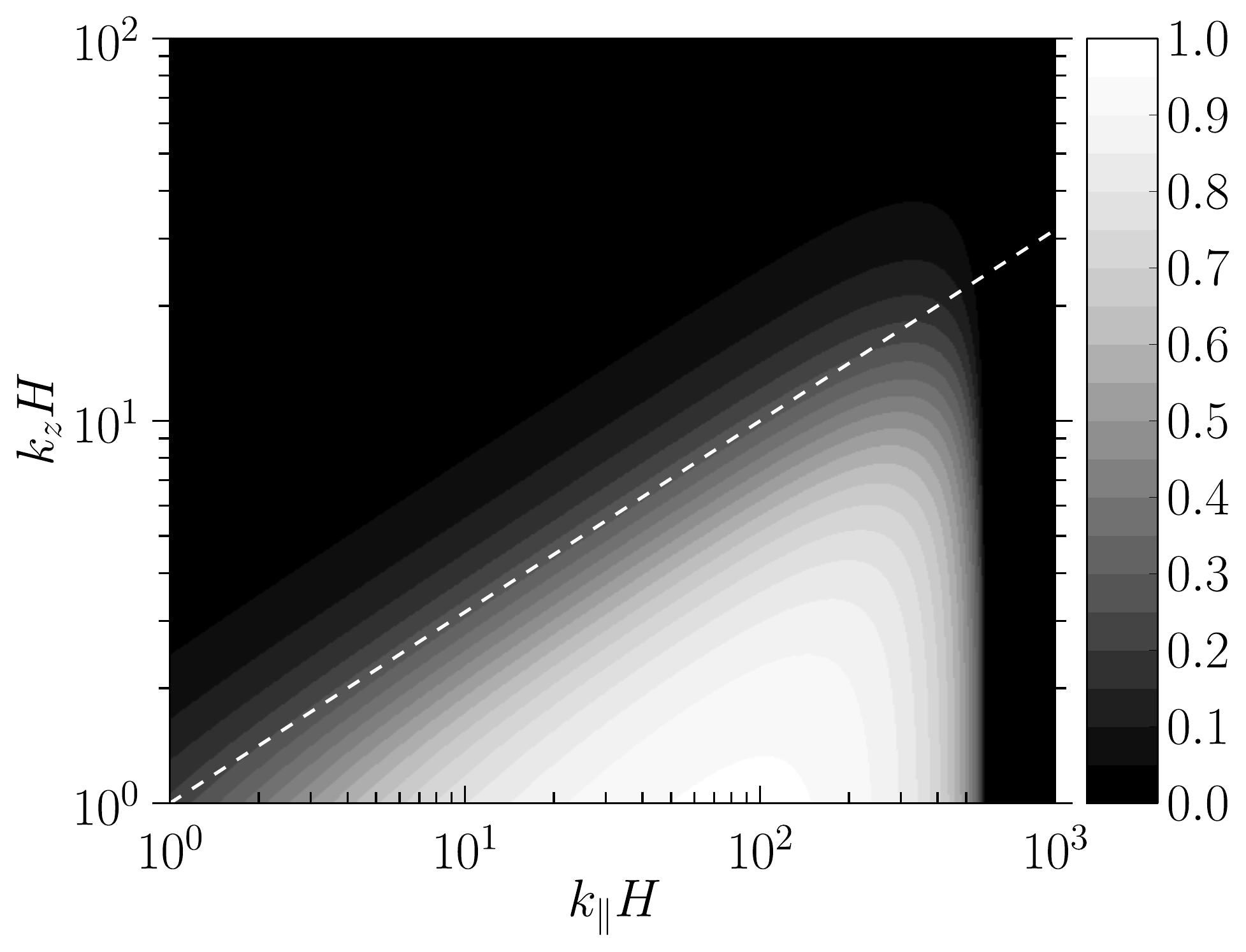}
\quad
\includegraphics[width =0.48\textwidth]{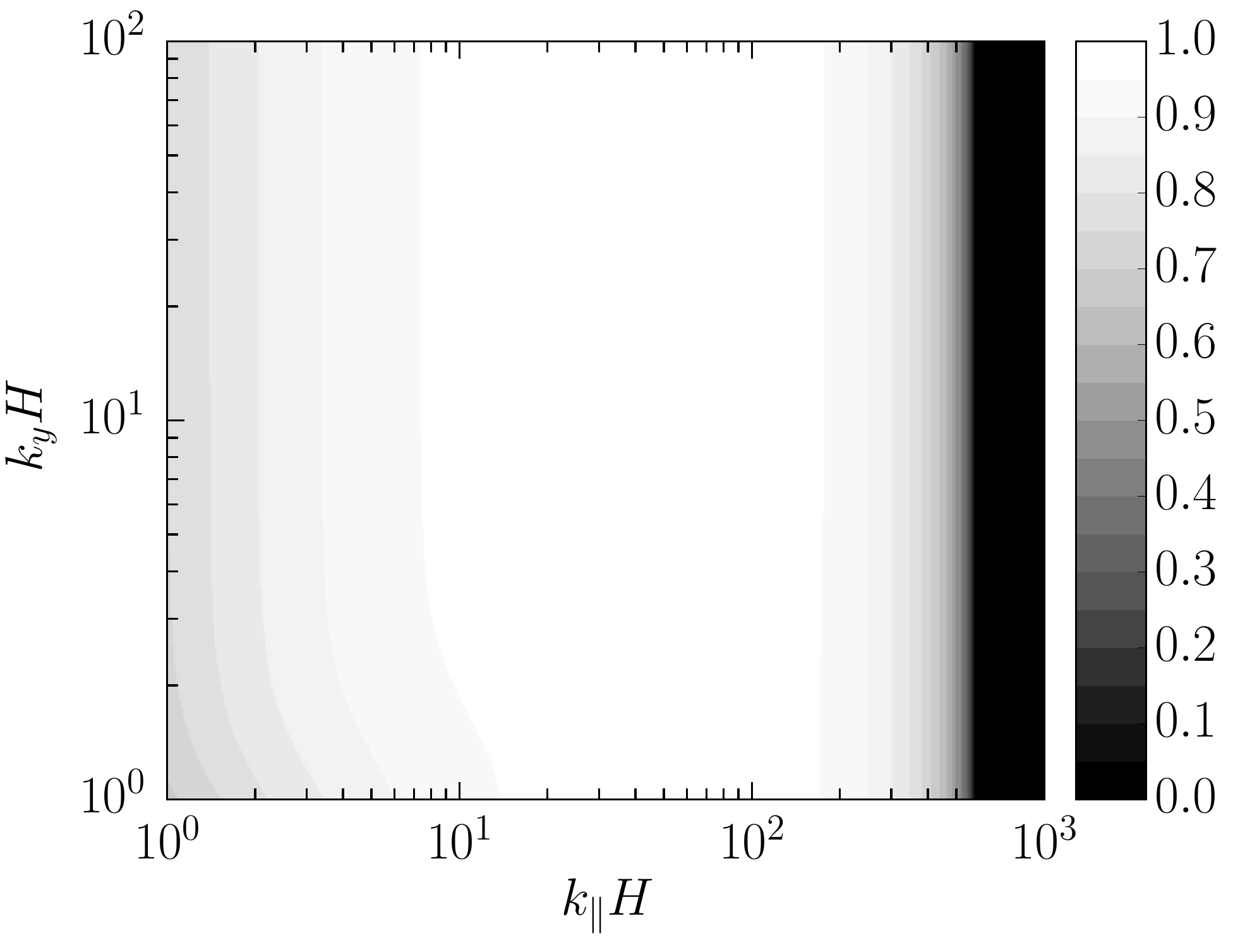}
\caption{Kinetic MTI growth rate (normalised by $\sqrt{-g \,{\rm d}\ln T/{\rm d}z}$) as calculated from the collisionless drift-kinetic dispersion relation with $\beta_i = 10^6$, ${\rm d}\ln T / {\rm d}\ln P = 1/3$, and either ({\em left}) $k_y = 0$ (see (\ref{eqn:collisionlessMTI_kz})) or ({\em right}) $k_z = 0$ (see (\ref{eqn:collisionlessMTI_ky})). The white dashed line on the left panel is $k_z H \sim \sqrt{k_\parallel H}$, which traces the boundary where Barnes damping stabilizes the drift-kinetic MTI. See \S\ref{sec:longwavelength_kz} and \S\ref{sec:longwavelength_ky} for details.}
\label{fig:longwave}
\end{figure}

This is indeed what we find in Figure \ref{fig:longwave}a, which shows the instability growth rate as a function of $k_\parallel$ and $k_z$. As for the $k_\perp = 0$ modes, the growth rate increases with $k_\parallel$ until it attains its maximum value $\approx$$\sqrt{-g \, {\rm d}\ln T/{\rm d}z}$ at $k_\parallel H \sim 100$. Beyond $k_\parallel H \approx \sqrt{\beta\, {\rm d}\ln T / {\rm d}\ln P}$, magnetic tension stabilizes the mode and it propagates as a pseudo-Alfv\'{e}n wave undergoing weak Landau damping. As anticipated, increasing $k_z$ reduces the growth rate $\propto$$k^{-2}_z$, strongly suppressing the drift-kinetic MTI once $k_z H \sim \sqrt{k_\parallel H}$ (demarcated by the white dashed line in the figure). Note that, in the collisional case, the limiting $k_z$ is set by the Braginskii viscosity and is instead given by $k_z H \sim \sqrt{H/\lambda_{mfp}}$, independent of parallel wavenumber \citep[see \S 4.2.1 and fig.~5b of][]{kunz11}. The difference is because, in the collisional case, wave damping is caused by the parallel diffusion of momentum (occurring at a rate $\sim$$k^2_\parallel \lambda_{mfp} v_{ths}$) rather than by collisionless wave-particle interactions ($\sim$$k_\parallel v_{ths}$).

\subsubsection{Drift-kinetic limit: Oblique propagation with $k_\perp = k_y$ $(\psi = 0)$}\label{sec:longwavelength_ky}

The story gets a bit more complicated when the wavevector has a component perpendicular to both the mean magnetic field and the equilibrium gradients, $\bb{k} = k_\parallel \ex + k_y \ey$. First, the contributions from $\msb{U}^{(yz)}_{n,s}$ and $\msb{U}^{(zy)}_{n,s}$ to the dispersion tensor no longer vanish. This is crucial, as these matrix elements are responsible for coupling Alfv\'{e}n waves and slow modes, a feature heretofore lacking (and one that will reappear in \S\ref{sec:gyroviscous}). Secondly, with $\msb{D}^{(xx)} \sim \mc{O}(1)$, $\msb{D}^{(xy)} \sim \msb{D}^{(xz)} \sim \mc{O}(\epsilon)$, and $\msb{D}^{(yy)} \sim \msb{D}^{(yz)} \sim \msb{D}^{(zz)} \sim \mc{O}(\epsilon^2)$, all the elements of the dispersion tensor ultimately contribute to the dispersion relation, and the algebra gets considerably more thorny. Surprisingly, a fortuitous rearrangement of terms leads to a rather compact dispersion relation:
\begin{align}\label{eqn:collisionlessMTI_ky}
\Biggl\{ \omega^2 &- k^2_\parallel v^2_A + \sum_s \frac{1}{\varrho} \D{z}{P_s} \left[ \D{z}{\ln T_s} + \zeta_s Z_0(\zeta_s) \D{z}{\ln P_s} \Upsilon_s \right] \Biggr\} \\*
\mbox{} &\times \left[ \omega^2 - k^2 v^2_A + k^2_y \sum_s \zeta_s Z_0(\zeta_s) \frac{P_s}{\varrho} \bigl( 1 + \Upsilon_s \bigr) \right] = k^2_y \left[ \sum_s \zeta_s Z_0(\zeta_s) \frac{1}{\varrho} \D{z}{P_s} \Upsilon_s \right]^2 \nonumber .
\end{align}
The first term on the left-hand side -- the one in braces -- should look familiar from Section \ref{sec:longwavelength_kprl} (cf.~(\ref{eqn:collisionlessMTI_kprl})). It captures the effects of magnetic tension, buoyancy, and collisionless damping on fluctuations polarized with $\delta\bb{B}$ in the $z$ direction. The second term on the left-hand side of (\ref{eqn:collisionlessMTI_ky}) should also look familiar; it includes the contribution from Barnes damping of $\delta B_\parallel \ne 0$ fluctuations ($\propto k^2_y ( 1 + \Upsilon_s )$) that featured prominently in Section \ref{sec:longwavelength_kz}. The right-hand side is new, and deserves some discussion.

Recall from Sections \ref{sec:longwavelength_kprl} and \ref{sec:longwavelength_kz} that, for $k_\perp = 0$ or $k_\perp = k_z$, Alfv\'{e}n waves polarized with $\delta\bb{B}$ in the $y$ direction are completely decoupled from all other fluctuations. This is because they change neither the density nor the temperature (see (\ref{eqn:longwavelength_dn})--(\ref{eqn:longwavelength_Pprl})), and therefore are not subject to buoyancy forces. This is clearly not the case when $k_y \ne 0$. These fluctuations are now accompanied by compressions and rarefactions in the magnetic-field lines in order to preserve the divergence-free constraint, $\delta B_\parallel = -(k_y/k_\parallel) \delta B_y$. Accompanying these magnetic-field-strength fluctuations are changes in temperature and density (see (\ref{eqn:longwavelength_DT})), which influence the behaviour of vertical displacements subject to gravity. As a result, all components of the magnetic field are coupled.

In particular, consider the limit $k^2_y H^2 \gg k_\parallel H$. The first bracket on the second line of (\ref{eqn:collisionlessMTI_ky}) -- the one associated with $y$-polarized magnetic-field fluctuations -- is then dominated by the $\propto$$k^2_y$ term. This corresponds to strong Barnes damping of $\delta B_\parallel$ fluctuations. In the previous section (\S\ref{sec:longwavelength_kz}), this damping was detrimental to the growth of the drift-kinetic MTI, since, there, $\delta B_z = -(k_z/k_\parallel) \delta B_\parallel$. However, in the present arrangement, the system has a mechanism to minimize the collisionless damping: simply counter the mirror force $-\mu_s \grad_\parallel \delta B_\parallel$ with the parallel gravitational force $m_s\bb{g}\bcdot\delta\eb$. As a Landau-resonant particle accelerates downwards along a vertically perturbed magnetic-field line into a region of lower potential energy, arrange the magnetic field so that the particle is also entering a region of enhanced magnetic-field strength. The mathematical manifestation of this arrangement is the cancellation of $k^2_y$ from the Barnes-damping term in (\ref{eqn:collisionlessMTI_ky}) with the $k^2_y$ on that equation's right-hand side, the result being that
\begin{equation}\label{eqn:collisionlessMTI_largeky}
\omega^2 \simeq k^2_\parallel v^2_A - \sum_s \frac{1}{\varrho} \D{z}{P_s} \left[ \D{z}{\ln T_s} + \zeta_s Z_0(\zeta_s) \D{z}{\ln P_s} \widehat{\Upsilon}_s \right] ,
\end{equation}
where
\begin{equation}\label{eqn:eta-hat}
\widehat{\Upsilon}_s \doteq \frac{\sum_{s'} \zeta_{s'} Z_0(\zeta_{s'}) P_{s'}  \Upsilon_s}{\sum_{s'} \zeta_{s'} Z_0(\zeta_{s'}) P_{s'} \bigl( 1 + \Upsilon_{s'} \bigr)} \simeq \frac{\Upsilon_s}{2} ~{\rm for}~\zeta_s \ll 1.
\end{equation}
Comparing (\ref{eqn:collisionlessMTI_largeky}) with (\ref{eqn:collisionlessMTI_kprl}), we see that the collisionless damping (captured by the final term in those equations) is reduced from the $k_\perp = 0$ case by a factor of $\simeq$$2$ at large $k_\parallel H$ -- that is, the growth rate is actually {\it larger} for $k_y \ne 0$, independent of $k_y$ (so long as $k^2_y H^2 \gg k_\parallel H \gg 1$).

\begin{figure}
\centering
\includegraphics[width=\textwidth]{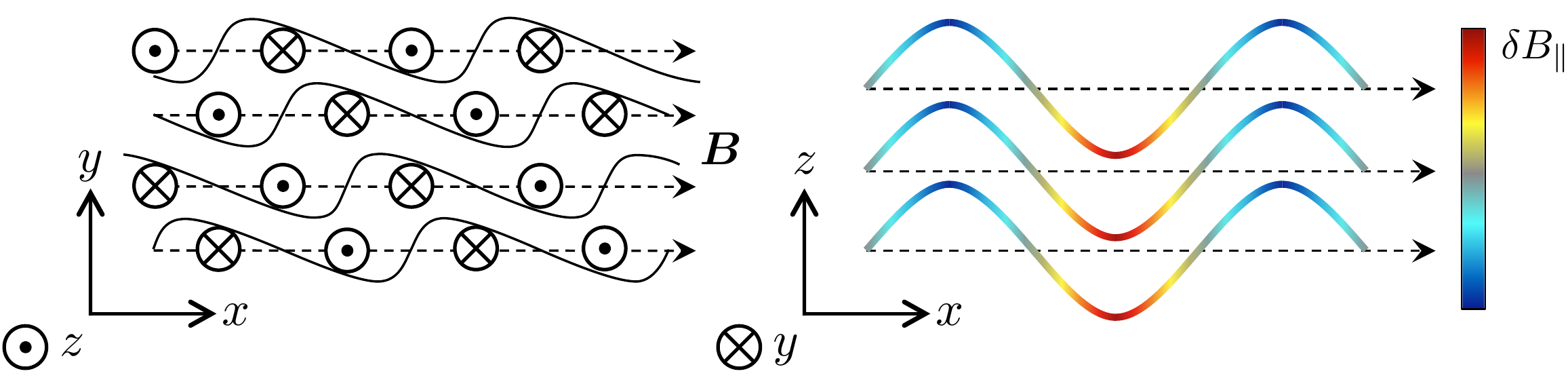}
\caption{Geometry of perturbed magnetic-field lines for an MTI-unstable mode with $k_\parallel H = k_y H \gg 1$ in the drift-kinetic limit. Field lines are shown as viewed ({\it left}) from above in the $x$-$y$ plane and ({\it right}) from the side in the $x$-$z$ plane. The horizontal field fluctuations satisfy $\delta B_\parallel = - \delta B_y$, with vertical displacements of field lines anti-correlated with the strength of the perturbed field (indicated by the coloring of field lines on the right). The mirror force $-\mu\nabla_\parallel \delta B_\parallel$ thus acts oppositely to the parallel gravitational force $m_s \bb{g}\bcdot\delta\eb$. See \S\ref{sec:longwavelength_ky} for details.}
\label{fig:kydiagram}
\end{figure}

This scenario is diagrammed in Figure \ref{fig:kydiagram}, which shows the geometry of perturbed magnetic-field lines for an MTI-unstable mode with $k_\parallel H = k_y H \gg 1$. The horizontal field fluctuations satisfy $\delta B_\parallel = -\delta B_y$ in order to preserve the solenoidality constraint (shown in the left panel), with vertical displacements of field lines anti-correlated with the strength of the perturbed field (indicated by the coloring of field lines on the right). As a result of this arrangement, the mirror force acts oppositely to the parallel gravitational force, reducing the parallel acceleration experienced by Landau-resonant particles. Perhaps the simplest way of understanding this advantageous field-line topology is to examine the energy of a $\mu$-conserving particle in the frame of the perturbed $\bb{E}\btimes\bb{B}$ drift, $(1/2) m_s v^2_\parallel + \mu_s B + q_s \Phi_s$. The change in energy of this particle as it interacts with the wave is reduced by ensuring that increasing (decreasing) magnetic-field strengths occur in regions of decreasing (increasing) potential.

At the risk of belaboring this point, let us use (\ref{eqn:longwavelength_dn}) and (\ref{eqn:induction}) to write the evolution equation for the vertical displacement of a fluid element (neglecting magnetic tension): 
\begin{align}\label{eqn:xiky}
\DD{t}{\xi_z} &= - g \sum_s \frac{m_s n_s}{\varrho} \frac{\delta n_s}{n_s} \nonumber\\*
\mbox{} &= - g \D{z}{\ln T} \xi_z - g \sum_s \frac{m_s n_s}{\varrho} \bigl[ 1 - \Upsilon_s Z_1(\zeta_s) \bigr]  \left(   \frac{q_s}{T_s} \D{z}{\Phi_s} \xi_z + \frac{\delta B_\parallel}{B} \right) .
\end{align}
Growth is maximized by reducing the amplitude of the final term, which is achieved by having magnetic-field-strength fluctuations $\delta B_\parallel$ anti-correlated with upward displacements $\xi_z > 0$ into regions of larger potential (${\rm d}\Phi_s/{\rm d}z > 0$). This is precisely what is shown in Figure \ref{fig:kydiagram}.

\subsubsection{Drift-kinetic limit: Arbitrary wavevector orientation}\label{sec:longwavelength_kprp}

Having now considered all simple wavevector geometries, we finally examine the general drift-kinetic dispersion relation, which, after some straightforward but tedious algebra, may be written in the following convenient form:
\begin{align}\label{eqn:collisionlessMTI}
\Biggl\{ \omega^2 &- \bigl( k^2_\parallel + k^2_z \bigr) v^2_A + \sum_s \frac{1}{\varrho} \D{z}{P_s} \left[ \D{z}{\ln T_s} + \zeta_s Z_0(\zeta_s) \D{z}{\ln P_s} \widetilde{\Upsilon}_s \right] \Biggr\} \nonumber\\*
\mbox{} &\times \left[ \omega^2 - \bigl( k^2_\parallel + k^2_y \bigr) v^2_A + k^2_y \sum_s \zeta_s Z_0(\zeta_s) \frac{P_s}{\varrho} \bigl( 1 + \Upsilon_s \bigr) \right] \nonumber\\*
\mbox{} &= k^2_y \left[ \sum_s \zeta_s Z_0(\zeta_s) \frac{1}{\varrho} \D{z}{P_s} \Upsilon_s \right]^2 + k^2_y k^2_z \left[ \sum_s \zeta_s Z_0(\zeta_s) \frac{P_s}{\varrho} \bigl( 1 + \Upsilon_s \bigr) \right]^2 ,
\end{align}
with $\widetilde{\Upsilon}_s$ given by (\ref{eqn:eta-tilde}). Setting the first term in braces to zero returns the dispersion relation analyzed in Section \ref{sec:longwavelength_kz}; it includes the effects of collisionless damping due to the parallel electric field, the magnetic mirror force, and the parallel gravitational force. The second term in brackets (the one on the second line of (\ref{eqn:collisionlessMTI})) is identical to the second term in (\ref{eqn:collisionlessMTI_ky}), and is related to fluctuations polarized with $\delta\bb{B}$ in the horizontal plane; it includes the contribution from Barnes damping of $\delta B_\parallel \ne 0$ fluctuations. The first term on the right-hand side, responsible for coupling these two branches, was discussed thoroughly in the previous section. The final term on the right-hand side is new, requiring both $k_y \ne 0$ and $k_z \ne 0$.

Motivated by the results in the previous section, we take the limit $k^2_y H^2 \gg k_\parallel H$ (while keeping $k_z \lesssim k_y$). Equation (\ref{eqn:collisionlessMTI}) becomes
\begin{equation}\label{eqn:collisionlessMTI_largekykz}
\omega^2 \simeq ( k^2_\parallel + k^2_z ) v^2_A - \sum_s \frac{1}{\varrho} \D{z}{P_s} \left[ \D{z}{\ln T_s} + \zeta_s Z_0(\zeta_s) \D{z}{\ln P_s} \widehat{\Upsilon}_s \right] ,
\end{equation}
which, aside from the tension term, is exactly the same as the $k_z = 0$ solution (\ref{eqn:collisionlessMTI_largeky})! The stabilization of the drift-kinetic MTI at moderate $k_z$ due to Barnes damping seen in Section \ref{sec:longwavelength_kz} has been removed by the same physics as was the focus of the previous section -- the mirror force, by which Landau-resonant particles bleed energy from the magnetic-field-strength fluctuations, is offset by the divergence of the parallel pressure.

\begin{figure}
\centering
\includegraphics[width =0.48\textwidth]{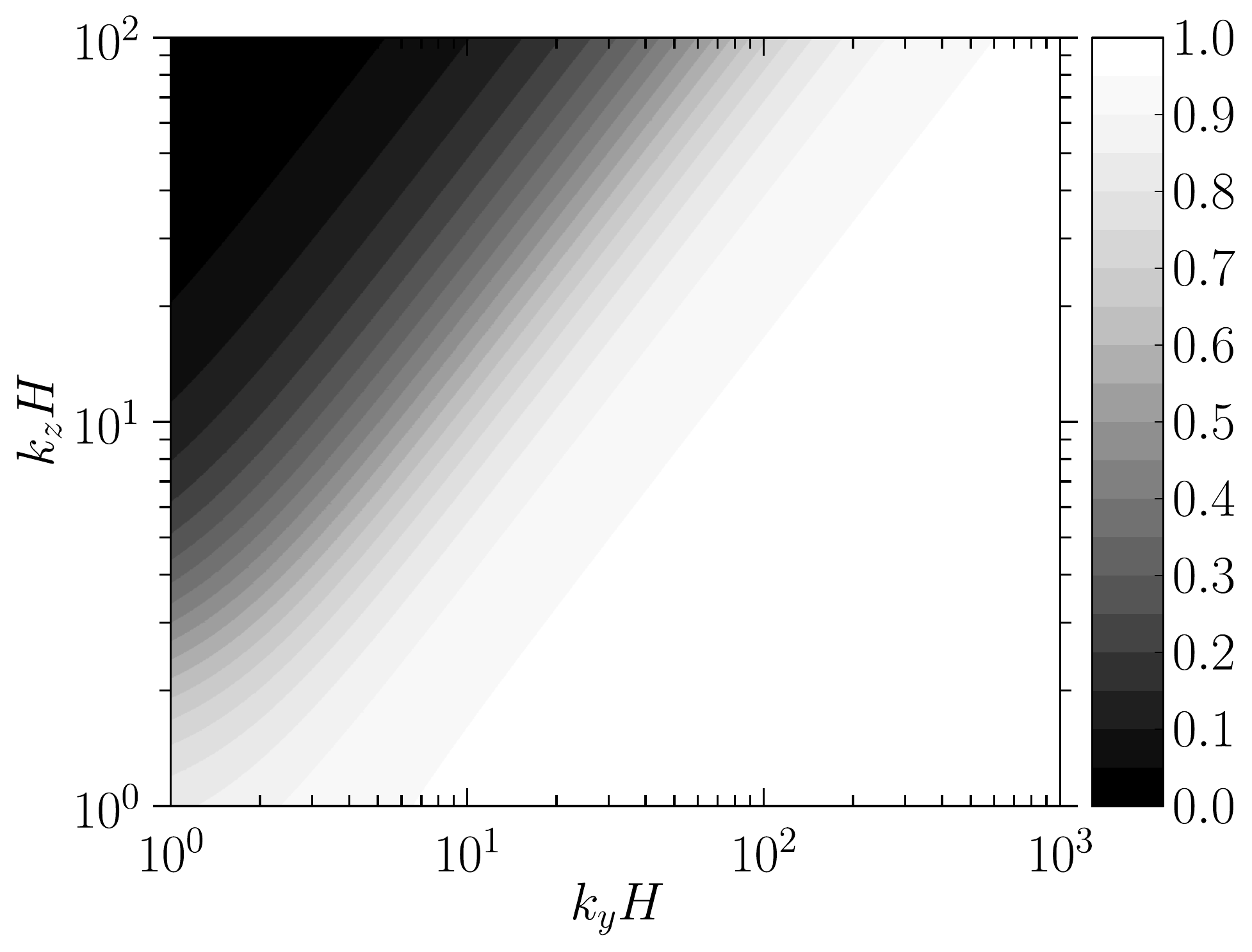}
\caption{Kinetic MTI (normalised by $\sqrt{-g \,{\rm d}\ln T/{\rm d}z}$) as a function of $k_yH$ and $k_zH$, as calculated from the collisionless drift-kinetic dispersion relation (\ref{eqn:collisionlessMTI}) with $\beta_i = 10^6$, ${\rm d}\ln T/{\rm d}\ln P = 1/3$, and $k_\parallel H=61.4$ (corresponding to the fastest-growing $k_\perp=0$ mode). See \S\ref{sec:longwavelength_kprp} for details.}
\label{fig:longwave_general}
\end{figure}

Such behaviour can be seen in Figure \ref{fig:longwave_general}, which displays the instability growth rate calculated from (\ref{eqn:collisionlessMTI}) as a function of $k_y H$ and $k_z H$ for $k_\parallel H = 61.4$ (corresponding to the fastest-growing $k_\perp = 0$ mode for $\beta_i = 10^6$). At large $k_y$, the growth rate is maximal and approximately constant with $k_z$, at least until $k_y \sim k_z$. At that point the ordering leading to (\ref{eqn:collisionlessMTI_largekykz}) breaks down, and for $k^2_y / k^2_z \ll 1$ we have from (\ref{eqn:collisionlessMTI})
\begin{equation}
\omega^2 \simeq ( k^2_\parallel + k^2_y) v^2_A - \frac{k^2_y}{k^2_z} \sum_s \frac{1}{\varrho} \D{z}{P_s} \D{z}{\ln T_s} 
\end{equation}
to leading order in $k^2_y/k^2_z$, i.e.~the growth decreases as $\propto$$k^{-1}_z$. The physical reason for this decrease is that, when $k_z \gg k_y$, the mirror force can no longer be  compensated adequately by the divergence of the parallel pressure, and Barnes damping wins out. The mathematical manifestation of this is an additional term in the evolution equation for the vertical displacement of a fluid element due to the perturbed perpendicular pressure (cf.~\ref{eqn:xiky}); neglecting magnetic tension and pressure,
\begin{align}\label{eqn:xikykz}
\DD{t}{\xi_z} &= - g \sum_s \frac{m_s n_s}{\varrho} \frac{\delta n_s}{n_s}  - \imag k_z \sum_s \frac{P_s}{\varrho} \frac{\delta P_{\perp s}}{P_s} \nonumber\\*
\mbox{} &= - g \D{z}{\ln T} \xi_z - \sum_s \frac{P_s}{\varrho} \left( \imag k_z +  \frac{m_s g}{T_s} \right) \bigl[ 1 - \Upsilon_s Z_1(\zeta_s) \bigr]  \left(   \frac{q_s}{T_s} \D{z}{\Phi_s} \xi_z + \frac{\delta B_\parallel}{B} \right) \nonumber\\*
\mbox{} &\quad  -\imag k_z \frac{\delta B_\parallel}{B} \sum_s \frac{P_s}{\varrho} \bigl[ 1 - Z_1(\zeta_s) \bigr] .
\end{align}
Minimizing the second term in the final equality of (\ref{eqn:xikykz}) by having upward displacements go hand-in-hand with weak magnetic-field strengths -- an arrangement exploited by the $k^2_y H^2 \gg k_\parallel H$ modes in Section \ref{sec:longwavelength_ky} -- causes the final term in (\ref{eqn:xikykz}) to strongly damp the fluctuations when $k^2_z H^2 \gg k_\parallel H$.

\subsection{Gyroviscous limit: $k^2_\parallel \rho_s H \sim 1$}\label{sec:gyroviscous}

Having examined the long-wavelength limit, we now progress to smaller scales, namely, those intermediate between the atmospheric scale height and the species' Larmor radii. To keep things simple at first, we focus only on fluctuations whose wavevectors are aligned with the mean magnetic field, $\bb{k} = k_\parallel \ex$. A natural question to ask is what would happen to the growth rate of the MTI if the magnetic tension were insufficient to provide a parallel-wavenumber cutoff, thus allowing for growth at relatively small scales. Indeed, many numerical simulations of the (collisional) MTI presented in the published literature had adopted rather large initial plasma beta parameters $\beta \sim 10^8$--$10^{12}$ \citep{ps07,psl08,mccourt11,parrish12}, one of the motivations being to test whether the MTI in its nonlinear phase might act as an efficient dynamo capable of explaining the presence of the $\sim$$\mu{\rm G}$ magnetic fields currently observed in nearby galaxy clusters \citep[e.g.][]{ct02}. At such high $\beta$, the ion Larmor radius can become large enough for the characteristic scale at which FLR effects are important -- roughly the geometric mean of the ion Larmor radius and the scale height, {\it viz.}
\begin{equation}
k^{-1}_{\rm FLR} \sim (\rho_i H)^{1/2} \approx 10 \left( \frac{\beta_i}{10^{12}} \right)^{1/4} \left( \frac{10^{-3}~{\rm cm}^{-3}}{n} \right)^{1/4} \left( \frac{H}{500~{\rm kpc}} \right)^{1/2} ~{\rm pc} 
\end{equation}
-- to be comparable to the size of some fluctuations, i.e.~$k^2_\parallel \rho_i H \sim 1$. Because FLR effects introduce wave dispersion and thus are typically a stabilizing influence, one might inquire at what magnetic-field strength does such stabilization become more important than that provided by magnetic tension. The answer, as it turns out, is when $\beta_i \gtrsim H/\rho_i$ or, for parameters characteristic of galaxy cluster outskirts,
\begin{equation}
B \lesssim 0.2 \left(\frac{n}{10^{-3}~{\rm cm^{-3}}}\right)^{1/3} \left( \frac{T}{10~{\rm keV}}\right)^{1/2} \left( \frac{H}{500~{\rm kpc}} \right)^{-1/3} ~{\rm nG} .
\end{equation}
Such a field is within the range of observationally allowed primordial magnetic-field strengths \citep[e.g.][]{neronov10,planck15}. 

\begin{figure}
\centering
\begin{minipage}[b]{0.48\textwidth}
\includegraphics[width =\textwidth]{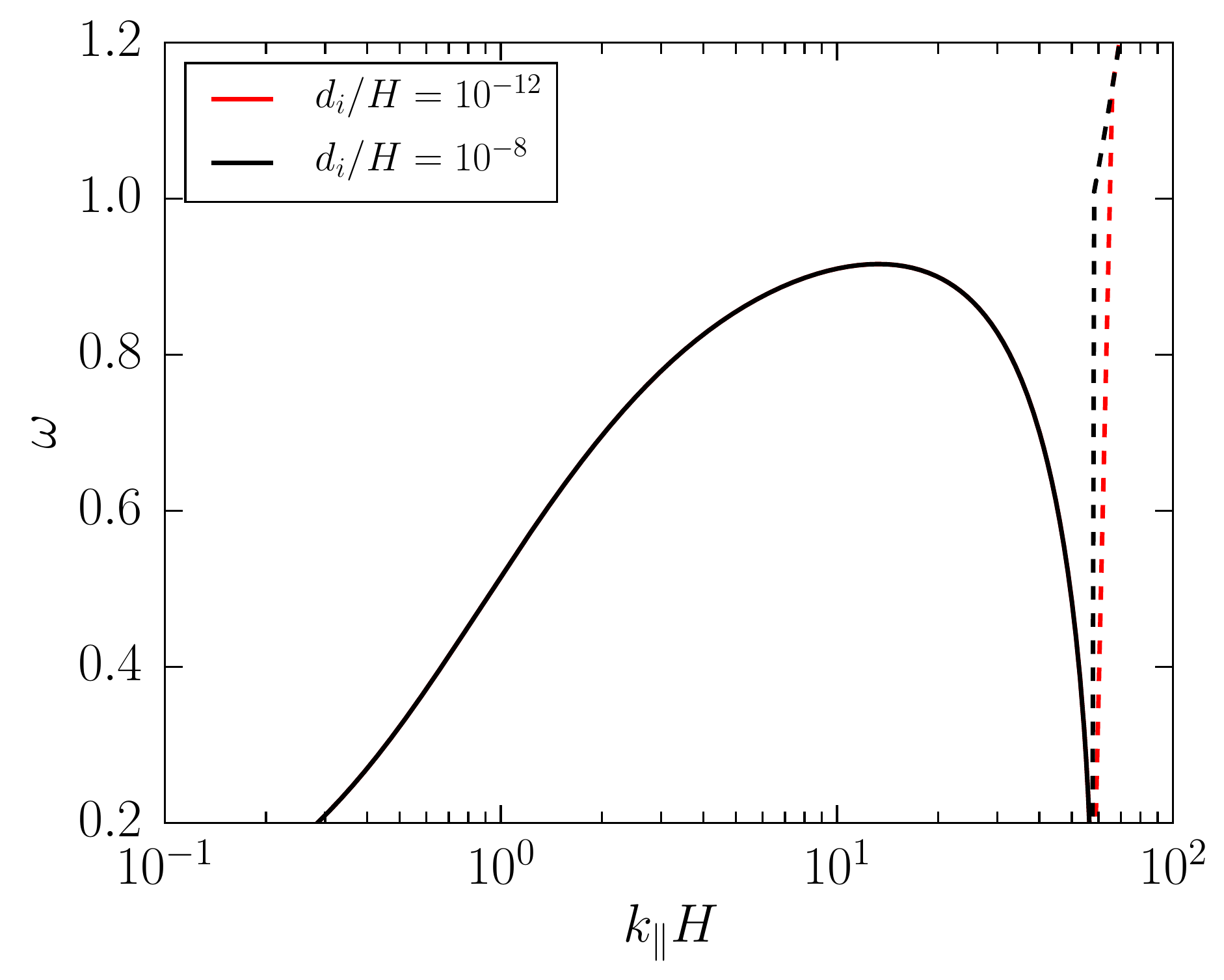}
\end{minipage}
\quad
\begin{minipage}[b]{0.48\textwidth}
\includegraphics[width =\textwidth]{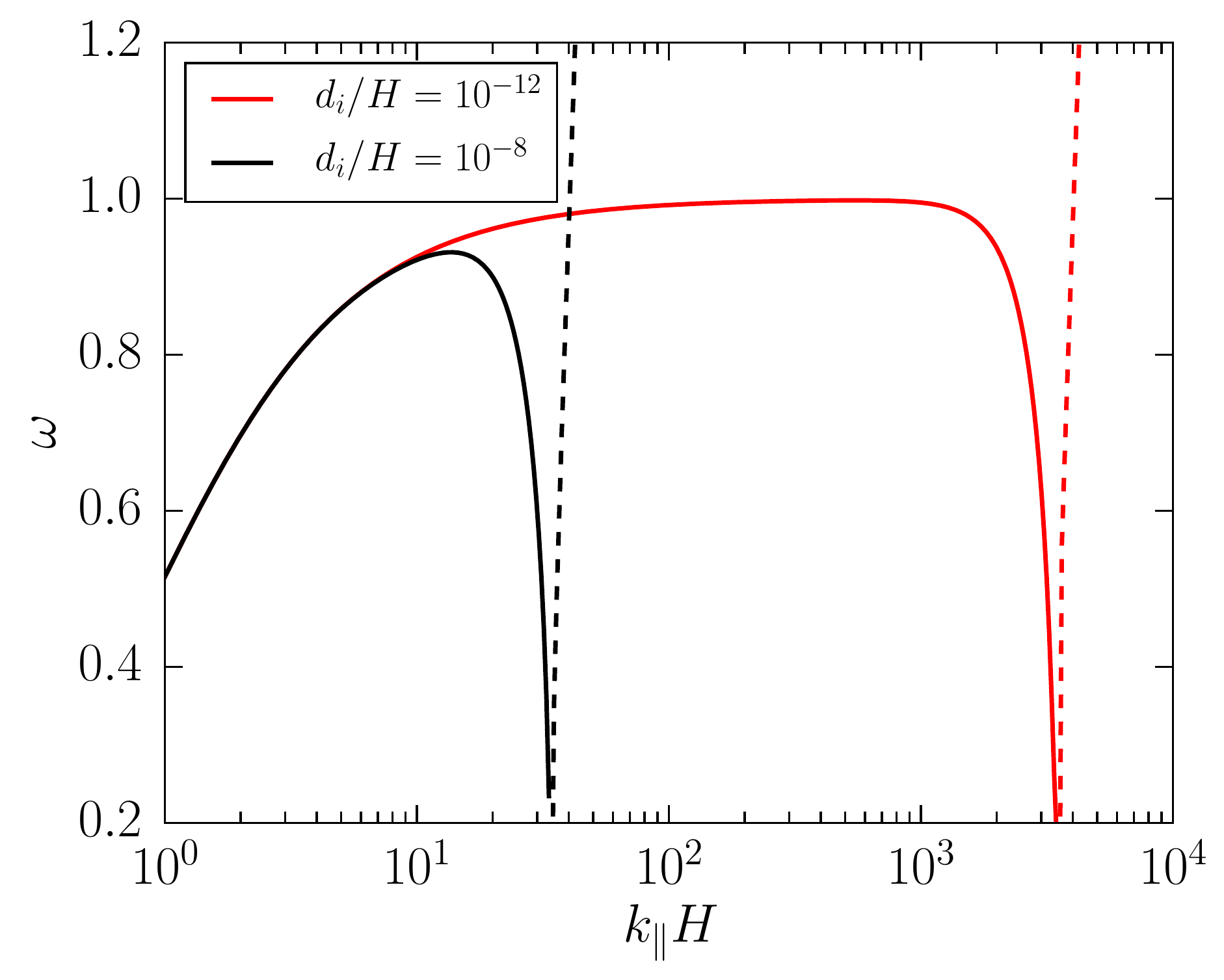}
\end{minipage}
\caption{Real (dashed) and imaginary (solid) parts of the wave frequency (normalised by $\sqrt{-g \,{\rm d}\ln T/{\rm d}z}$) for ({\it left}) $\beta_i=\beta_e=10^4$ and ({\it right}) $\beta_i =10^{10}$ for two ratios of ion skin depth $d_i$ and thermal-pressure scale height $H$ (note that the ion Larmor radius $\rho_i = \sqrt{\beta_i} d_i$). Growing modes are stabilised at large $k_\parallel H$ by either magnetic tension ({\it left}) or gyroviscosity ({\it right}). See \S\ref{sec:gyroviscous} for details.}
\label{fig:kprl}
\end{figure}

This short preamble concluded, we begin our reduction of the general dispersion relation (\ref{eqn:Maxdispersion}) in this ``gyroviscous'' limit by adopting the ordering
\begin{equation}\label{eqn:gv_ordering}
\frac{\omega}{\Omega_s} \sim \frac{1}{(k_\parallel H )^2} \sim \frac{1}{\beta_s} \sim \frac{\rho_s}{H} \doteq \epsilon \ll 1 ,
\end{equation}
under which the components of the dispersion tensor satisfy $\msb{D}^{(xx)} \sim \mc{O}(\epsilon)$, $\msb{D}^{(xy)} \sim \msb{D}^{(yy)} \sim \msb{D}^{(yz)} \sim \msb{D}^{(zz)} \sim \mc{O}(\epsilon^2)$. The slight complication to obtaining these estimates is that the $Z_0(\zeta_{n,s})$ function for $n = \pm 1$ must be expanded in its large arguments to third order in $k_\parallel \rho_s$, so that, e.g.
\begin{equation}\label{eqn:gyroUyz}
\msb{U}^{(yz)}_{n,s} \simeq \imag \frac{k_\parallel \rho_s}{4} \left( 1 + \frac{k^2_\parallel\rho^2_s}{2}\right) \bigl( \delta_{n,1} + \delta_{n,-1} \bigr) .
\end{equation}
(Recall from \S\ref{sec:longwavelength_kprl} that the first term in (\ref{eqn:gyroUyz}) ultimately vanishes by quasi-neutrality. A similar cancellation occurs for the leading-order term in $\msb{U}^{(xy)}$.) Taking these subtleties into account, the dispersion relation to leading order in $\epsilon$ is
\begin{equation}\label{eqn:gy_dsp}
\msb{D}^{(xx)} \left[ \msb{D}^{(yy)} \msb{D}^{(zz)} - \msb{D}^{(yz)} \msb{D}^{(zy)} \right] = 0 .
\end{equation}
Since $\msb{D}^{(xx)}$ is in general non-zero, the term in brackets must vanish, yielding
\begin{align}\label{eqn:gyroviscous_kprl_disprel}
\bigl(\omega^2 - k^2_\parallel v^2_A \bigr)  \left(\omega^2 - k^2_\parallel v^2_A + \sum_s \frac{1}{\varrho} \D{z}{P_s} \D{z}{\ln T_s} \right) = \omega^2 \left( \sum_s \frac{m_s n_s}{\varrho} \frac{k^2_\parallel v^2_{ths}}{2\Omega_s} \right)^2 ,
\end{align}
where we have used the fact that the ordering (\ref{eqn:gv_ordering}) precludes Landau resonances (i.e.~$\zeta_s \sim \mc{O}(\sqrt{\epsilon})$). The first term in parentheses on the left-hand size of (\ref{eqn:gyroviscous_kprl_disprel}) corresponds to Alfv\'{e}n waves ($\omega = \pm k_\parallel v_A$) whose $\delta\bb{B}$ is polarized in the $y$ direction. The second term in parentheses is readily identifiable as the pseudo-Alfv\'{e}n-wave branch, which, in the absence of stabilizing magnetic tension, is unstable to the MTI for ${\rm d}\ln T/{\rm d}z < 0$. Its magnetic-field fluctuations are polarized in the $z$ direction, thus enabling rapid field-aligned conduction for vertical displacements. As in Sections \ref{sec:longwavelength_ky} and \ref{sec:longwavelength_kprp}, these two branches are coupled by a non-zero right-hand side, but here the cause is ``gyroviscosity'' \citep{kaufman60,braginskii65,ramos05} -- an FLR-induced dissipationless cross-field transport of momentum that arises from Larmor-scale spatial variations in the guiding-centre $\bb{E}\btimes\bb{B}$ drifts. It is dispersive and stabilizing, an effect that can be seen clearly in Figure \ref{fig:kprl}b, which shows the wave frequency versus parallel wavenumber for a variety of $\beta_i$ and thermal-pressure scale heights. For $\beta_i \gtrsim H / \rho_i$, the gyroviscosity is the dominant stabilizing effect (over the magnetic tension), providing a wavenumber cutoff
\begin{equation}\label{eqn:kmaxGV}
( k_\parallel H )_{\rm max} \simeq \left( \frac{4H^2}{d^2_i} \frac{1}{\beta_i} \D{\ln P}{\ln T} \right)^{1/4} ,
\end{equation}
where $d_i \doteq ( m_i c^2 / 4\upi e^2 n_i )^{1/2} = \rho_i / \sqrt{\beta_i}$ is the ion skin depth. For $n_i = 10^{-3}~{\rm cm}^{-3}$, $H = 500~{\rm kpc}$, and ${\rm d}\ln T / {\rm d}\ln P = 1/3$, (\ref{eqn:kmaxGV}) gives $ (k_\parallel H)_{\rm max} \approx 5\times 10^4 \, ( 10^{12} / \beta_i )^{1/4}$.

The physics driving the gyroviscous stabilization of the MTI may be elucidated by examining the leading-order perturbed distribution function (\ref{eqn:df}) in the frame of the perturbed $\bb{E}\btimes\bb{B}$ flow under the gyroviscous ordering (\ref{eqn:gv_ordering}):
\begin{equation}\label{eqn:gv_df}
\delta f_s(v_\parallel, w_\perp) = \frac{\imag}{k_\parallel} \left[ \left( \pD{z}{\ln f_s} - \D{z}{\ln P_s} \right) \frac{\delta B_z}{B} + 2k_\parallel  \frac{\omega}{\Omega_s} \frac{v_\parallel\bb{v}_\perp}{v^2_{ths}} \bcdot \biggl(\frac{\delta\bb{B}}{B} \times\ex \biggr) \right] f_s .
\end{equation}
The final term in this expression makes the perturbed distribution function $\delta f_s(v_\parallel, w_\perp)$ gyrophase-dependent (a feature absent in the drift-kinetic limit), which endows the pressure tensor with off-diagonal elements:
\begin{equation}\label{eqn:gvP}
\delta\msb{P}= \sum_s \delta\msb{P}_s = \sum_s P_s  \,\frac{\imag\omega}{\Omega_s} \left[ \bigl( \ex\ey + \ey\ex \bigr) \frac{\delta B_z}{B} - \bigl( \ex\ez + \ez\ex \bigr) \frac{\delta B_y}{B} \right] .
\end{equation}
Substituting this expression into the perturbed momentum equation and taking the limit $m_e / m_i \rightarrow 0$, we find that the Lagrangian displacement perpendicular to the mean magnetic field $\bb{\xi}_\perp$ evolves according to
\begin{equation}\label{eqn:gvmom}
\left( \DD{t}{} + k^2_\parallel v^2_A  \right) \bb{\xi}_\perp = - g \D{z}{\ln T} \bb{\xi}_z - \frac{k^2_\parallel v^2_{thi}}{2\Omega_i} \D{t}{\bb{\xi}_\perp} \btimes \ex .
\end{equation}
The first term on the right-hand side -- the buoyancy term -- drives vertical displacements, which would grow exponentially (provided ${\rm d}\ln T / {\rm d}z < 0$) were it not for the magnetic tension and the final (gyroviscous) term on the right-hand side. The latter rotates a vertical displacement into the horizontal plane (i.e.~vertical momentum is transported in the cross-field $y$ direction). If $k^2_\parallel \rho_i H \gtrsim 1$, this rotation occurs on a faster timescale than buoyancy can grow the displacement and the MTI becomes stabilised.

A final note is perhaps in order regarding the gyroviscosity and the contributions to it that sometimes arise in the presence of temperature gradients, the so-called \citet{mt71,mt84} terms \citep[see also][]{cs04}. For an isotropic equilibrium distribution function, the gyroviscous contribution to the pressure tensor of species $s$ is
\begin{align}\label{eqn:gyroviscosity}
\msb{G}_s = \frac{1}{4\Omega_s} &\Bigl\{ \eb \btimes \bigl[ ( P_{\perp s} \grad\bb{u}_s + \grad \bb{Q}_{\perp s}) + ( P_{\perp s}\grad\bb{u}_s + \grad\bb{Q}_{\perp s} )^{\rm T} \bigr] \bcdot \bigl( \msb{I} + 3\eb\eb \bigr) \nonumber\\*
\mbox{} & - \bigl( \msb{I} + 3\eb\eb \bigr) \bcdot \bigl[ ( P_{\perp s}\grad\bb{u}_s + \grad \bb{Q}_{\perp s} ) + ( P_{\perp s} \grad\bb{u}_s + \grad\bb{Q}_{\perp s} )^{\rm T} \bigr] \btimes \eb \Bigr\} ,
\end{align}
where the superscript ${\rm T}$ denotes the matrix transpose and $\bb{Q}_{\perp s} = Q_{\perp s} \eb$ is the parallel flux of perpendicular heat ({\it viz.}~$Q_{\perp s} = m_s \int {\rm d}^3\bb{v} \, v_\parallel (v^2_\perp / 2 ) f_s$). (We refer the reader to equation 6 and appendix A of \citealt{schekochihin10} for a simple derivation of (\ref{eqn:gyroviscosity})). The $\grad\bb{u}_s$ terms, first obtained (in the collisional limit) by \citet{braginskii65}, simplify considerably for our problem. With $\bb{k} = k_\parallel \eb$, they are $\imag k_\parallel (P_s/\Omega_s) ( \ex\btimes\delta\bb{u}_\perp\ex - \ex\ex\btimes\delta\bb{u}_\perp )$, which, using flux freezing (\ref{eqn:fluxfreezing}), returns precisely (\ref{eqn:gvP}). The gyroviscous stress associated with these terms has a non-zero divergence, and therefore exerts a force on the plasma (the final term in (\ref{eqn:gvmom})). The $\grad\bb{Q}_{\perp s}$ terms, on the other hand, vanish identically for $\bb{k} = k_\parallel\eb$, and so heat-flux contributions to the gyroviscosity do not affect the modes investigated in this section. For arbitrary wavevector orientation, however, these terms become
\begin{equation}
\imag k_\perp \frac{\delta Q_{\perp s} }{\Omega_s} \Bigl[ \cos\psi \bigl( \ex\ez + \ez\ex \bigr) - \sin\psi \bigl(\ex\ey+\ey\ex \bigr) \Bigr] ,
\end{equation}
where $\delta Q_{\perp s} = m_s \int{\rm d}^3\bb{v} v_\parallel ( v^2_\perp / 2) \delta f_s$. Its divergence is $k_\parallel \bb{k}_\perp(\delta Q_{\perp s}/\Omega_s) \btimes\ex$, and so the perturbed heat flux exerts a cross-field force on the plasma that is larger at smaller scales. One can show by directly computing
\begin{align}
\delta Q_{\perp s} = \imag v_{ths} q_s& \int{\rm d}^3\bb{v}\, \frac{v_\parallel}{v_{ths}} \frac{v^2_\perp}{v^2_{ths}} \frac{J_n(a_s) \bb{u}^\ast_{n,s}\bcdot\delta\bb{E}}{\omega + k_y v_{ds} - k_\parallel v_\parallel - n \Omega_s} \nonumber\\*
\mbox{} &\times \left[ 1 + \frac{\omega_{Ts}}{\omega} \left( \frac{5}{2} - \frac{v^2_\parallel + v^2_\perp}{v^2_{ths}} \right) \right] f_s(z,v_\parallel,v_\perp) 
\end{align}
and taking the relevant limit $\zeta_s \ll 1$, that this force is only important at perpendicular scales of order the ion Larmor scale. This is precisely our next destination.

\subsection{Gyrokinetic limit: $k_\parallel \rho_s \ll k_\perp \rho_s \sim 1$}\label{sec:gyrokinetic}

We have found that proceeding further to yet smaller parallel scales (e.g., $k_\parallel \rho_i \sim 1$) is not a particularly fruitful venture, and so we now turn our attention to fluctuations with small (i.e.~kinetic-scale) perpendicular wavelengths. To do so in an analytically tractable way, we adopt the gyrokinetic ordering \citep{rf68,th68,al80,ctb81,fc82}:
\begin{equation}\label{eqn:gkordering}
\frac{\omega}{\Omega_s} \sim \frac{k_\parallel}{k_\perp} \sim k_\parallel \rho_s \sim \frac{\rho_s}{H} \doteq \epsilon \ll 1,
\end{equation}
which selects low-frequency fluctuations that are highly oblique with respect to the magnetic-field direction. Note that, in this ordering, the perpendicular lengthscale of the fluctuations $k^{-1}_\perp$ satisfies $k_\perp \rho_s \sim k_\parallel H \sim 1$, and so the fluctuations are permitted to have (perpendicular) scales on the order of the Larmor radius. As a result, $k_y v_{ds} \sim k_\parallel v_{ths} \sim \omega$, i.e.~drift waves (and the various instabilities typically associated with them) are allowed. In common with the long-wavelength limit taken in Section \ref{sec:longwavelength}, collisionless damping via the Landau resonance is captured (but cyclotron resonances are not).

The procedure for deriving the dispersion relation in the gyrokinetic limit is similar to that employed in the long-wavelength and gyroviscous limits: we apply the ordering (\ref{eqn:gkordering}) to the dispersion tensor (\ref{eqn:Maxdispersion}), retain its leading-order terms, and set its determinant to zero. Unfortunately, the resulting equation is rather formidable, requiring myriad algebraic gymnastics to massage into a parseable form. Instead, we have found that a better route -- one suggested by previous work on linear and nonlinear gyrokinetic theory -- is to use (\ref{eqn:potentials}) and (\ref{eqn:linearFaraday}) to shift our independent variables from the field components $(\delta E_\parallel, \delta E_y, \delta E_z)$ to the potentials $(\varphi, A_\parallel, \delta B_\parallel)$ and then variously combine the three rows of the transformed dispersion tensor to obtain as our new rows what amounts to the quasi-neutrality constraint, the parallel component of Amp\`{e}re's law, and the perpendicular component of Amp\`{e}re's law (which, in gyrokinetics, is equivalent to a statement of perpendicular pressure balance). The determinant, and thus the dispersion relation, is of course identical following this profitable change of basis. 

The technical details of these manipulations are not given here; instead, we provide in Appendix \ref{app:gk} an alternate derivation of the linear gyrokinetic theory starting from a derivation of the nonlinear gyrokinetic theory of a thermally stratified atmosphere. Apart from some tedious bookkeeping and adroit consolidation of terms, the essential ingredients of the calculation are an expansion of the $Z_p(\zeta_{n,s})$ functions about $\zeta_{0,s} = (\omega + k_y v_{ds})/k_\parallel v_{ths}$ and several uses of the identity $\sum_{n=1}^\infty 2\Gamma_n(\alpha_s) = 1 - \Gamma_0(\alpha_s)$. After much effort, (\ref{eqn:DdotE}) becomes
\begin{equation}\label{eqn:gkDdotE}
\left[
\begin{array}{ccc}
\mc{A} & \mc{A} - \mc{B} & \mc{C} \\ [1.8em]
\mc{A} - \mc{B} & \mc{A} - \mc{B} + \mc{F} - {\displaystyle \frac{\alpha_i}{\overline{\omega}^2_i} } & \mc{C} + \mc{E} \\ [1.8em]
\mc{C} & \mc{C} + \mc{E} & \mc{D} - {\displaystyle \frac{2}{\beta_i} }
\end{array}
\right] \negthickspace \left[
\begin{array}{c}
\varphi \\ [1.5em]
- {\displaystyle \frac{\omega A_\parallel}{k_\parallel c}} \\ [1.5em]
{\displaystyle \frac{T_i}{q_i} \frac{\delta B_\parallel}{B} }
\end{array}
\right]
= 0,
\end{equation}
where $\overline{\omega}_i \doteq \omega / k_\parallel v_{Ai}$ and we have employed the shorthand notation (cf.~\S2.6 of \citet{howes06})
\begin{subequations}\label{eqn:howescoeffs}
\begin{align}\label{eqn:howesA}
\mc{A} &\doteq \sum_s \frac{q^2_s n_s / T_s}{q^2_i n_i / T_i} \nonumber\\*
\mbox{} &+ \sum_s \frac{q^2_s n_s / T_s}{q^2_i n_i / T_i} \frac{\omega}{k_\parallel v_{ths}} Z_0(\zeta_{0,s}) \Gamma_0(\alpha_s) \left\{ 1 + \frac{\omega_{Ts}}{\omega} \left[ \frac{3}{2} - \frac{Z_2(\zeta_{0,s})}{Z_0(\zeta_{0,s})} - \alpha_s \frac{\Gamma'_0(\alpha_s)}{\Gamma_0(\alpha_s)} \right] \right\}  ,\\*
\mc{B} &\doteq \sum_s \frac{q^2_s n_s / T_s}{q^2_i n_i / T_i}\left\{1 -  \Gamma_0(\alpha_s) \left[ 1 + \frac{\omega_{Ts}}{\omega} \left( 1 - \alpha_s \frac{\Gamma'_0(\alpha_s)}{\Gamma_0(\alpha_s)}\right) \right] \right\} \nonumber\\*
\mbox{}&- \sum_s \frac{q^2_s n_s / T_s}{q^2_i n_i / T_i} \frac{\omega_{Ps}}{k_\parallel v_{ths}} Z_0(\zeta_{0,s}) \Gamma_0(\alpha_s)  \left\{ 1 + \frac{\omega_{Ts}}{\omega} \left[ \frac{3}{2} - \frac{Z_2(\zeta_{0,s})}{Z_0(\zeta_{0,s})} - \alpha_s \frac{\Gamma'_0(\alpha_s)}{\Gamma_0(\alpha_s)} \right] \right\} ,\\*
\mc{C} &\doteq - \sum_s \frac{q_s n_s}{q_i n_i} \frac{\omega}{k_\parallel v_{ths}} Z_0(\zeta_{0,s}) \Gamma'_0(\alpha_s) \left\{ 1 + \frac{\omega_{Ts}}{\omega} \left[ \frac{1}{2} - \frac{Z_2(\zeta_{0,s})}{Z_0(\zeta_{0,s})} - \alpha_s \frac{\Gamma''_0(\alpha_s)}{\Gamma'_0(\alpha_s)} \right] \right\} ,\\*
\mc{D} &\doteq - \sum_s \frac{n_s T_s}{n_i T_i} \frac{2\omega}{k_\parallel v_{ths}} Z_0(\zeta_{0,s}) \Gamma'_0(\alpha_s) \left\{ 1 + \frac{\omega_{Ts}}{\omega} \left[ -\frac{1}{2} - \frac{Z_2(\zeta_{0,s})}{Z_0(\zeta_{0,s})} - \alpha_s \frac{\Gamma''_0(\alpha_s)}{\Gamma'_0(\alpha_s)} \right] \right\} ,\\*
\mc{E} &\doteq - \sum_s \frac{q_s n_s}{q_i n_i} \Gamma'_0(\alpha_s) \left[ 1 - \frac{\omega_{Ts}}{\omega} \alpha_s \frac{\Gamma''_0(\alpha_s)}{\Gamma'_0(\alpha_s)} \right] \nonumber\\*
\mbox{} &- \sum_s \frac{q_s n_s}{q_i n_i} \frac{\omega_{Ps}}{k_\parallel v_{ths}} Z_0(\zeta_{0,s}) \Gamma'_0(\alpha_s) \left\{ 1 + \frac{\omega_{Ts}}{\omega} \left[ \frac{1}{2} - \frac{Z_2(\zeta_{0,s})}{Z_0(\zeta_{0,s})} - \alpha_s \frac{\Gamma''_0(\alpha_s)}{\Gamma'_0(\alpha_s)} \right] \right\} , \\*
\mc{F} &\doteq \sum_s \frac{q^2_s n_s / T_s}{q^2_i n_i / T_i} \frac{\omega_{Ps}}{\omega} Z_1(\zeta_{0,s}) \Gamma_0(\alpha_s) \left\{ 1 + \frac{\omega_{Ts}}{\omega} \left[ \frac{3}{2} - \frac{Z_3(\zeta_{0,s})}{Z_1(\zeta_{0,s})} - \alpha_s \frac{\Gamma'_0(\alpha_s)}{\Gamma_0(\alpha_s)} \right] \right\} ,
\end{align}
\end{subequations}
where
\begin{equation}\label{eqn:omegaT}
\omega_{Ts} \doteq - k_y \frac{cT_s}{q_s B} \D{z}{\ln T_s} \quad {\rm and} \quad \omega_{Ps} \doteq k_y v_{ds} = -k_y \frac{cT_s}{q_s B} \D{z}{\ln P_s}  \tag{\theequation {\it a,b}}
\end{equation}
are the temperature-gradient and diamagnetic drift frequencies of species $s$.\footnote{The sign convention employed here is such that $\omega_{Ts}$ and $\omega_{Ps}$ are both positive for the interesting case of ${\rm d}\ln T_s/{\rm d}z < 0$ and ${\rm d}\ln P_s/{\rm d}z < 0$. Note that this is different than that used in most of the literature on temperature-gradient-driven instabilities in tokamaks.} 

Setting the determinant of the $3\times3$ matrix in (\ref{eqn:gkDdotE}) to zero gives the dispersion relation in the gyrokinetic limit:
\begin{equation}\label{eqn:gkdsp}
\left( \frac{\alpha_i\mc{A}}{\overline{\omega}^2_i}  - \mc{A} \mc{F} - \mc{A} \mc{B} + \mc{B}^2 \right) \left( \frac{2\mc{A}}{\beta_i} - \mc{A}\mc{D} + \mc{C}^2 \right) = ( \mc{A}\mc{E} + \mc{B}\mc{C} )^2 .
\end{equation}
The left-hand side of (\ref{eqn:gkdsp}) contains two factors, the first corresponding to the Alfv\'en-wave branch and the second to the slow-wave branch. The right-hand side represents coupling of the two branches that occurs either at finite Larmor radius (see \S 2.6 of \citet{howes06} for a thorough discussion of this effect in a homogeneous plasma) or in pressure-stratified plasmas when $k_y \ne 0$. The latter has already been discussed as part of the long-wavelength limit in \S\ref{sec:longwavelength_ky}. We now examine the short-wavelength regime. As for the previous sections, having the leading-order expression for the perturbed distribution function will serve useful:
\begin{equation}\label{eqn:gkdf}
\delta f_s = - \frac{q_s\varphi}{T_s} f_s + \frac{\omega}{\omega+\omega_{Ps}-k_\parallel v_\parallel} \frac{q_s\langle\chi\rangle_{\gas}}{T_s} \left[ 1 + \frac{\omega_{Ts}}{\omega} \left( \frac{5}{2} - \frac{v^2_\parallel+v^2_\perp}{v^2_{ths}} \right) \right] f_s ,
\end{equation}
where $\chi \doteq \varphi - v_\parallel A_\parallel/c - \bb{v}_\perp \bcdot \bb{A}_\perp/c$ is the gyrokinetic potential (see (\ref{eqn:gkpotential})); its ring average $\langle \chi \rangle_{\gas}$ is given by (\ref{eqn:gkpotentialk}). Equation (\ref{eqn:gkdf}) may be obtained by applying the gyrokinetic ordering to (\ref{eqn:df}) or, alternatively, by consulting Appendix \ref{app:gk}.

\begin{figure}
\centering
\begin{minipage}[b]{0.48\textwidth}
\includegraphics[width =\textwidth]{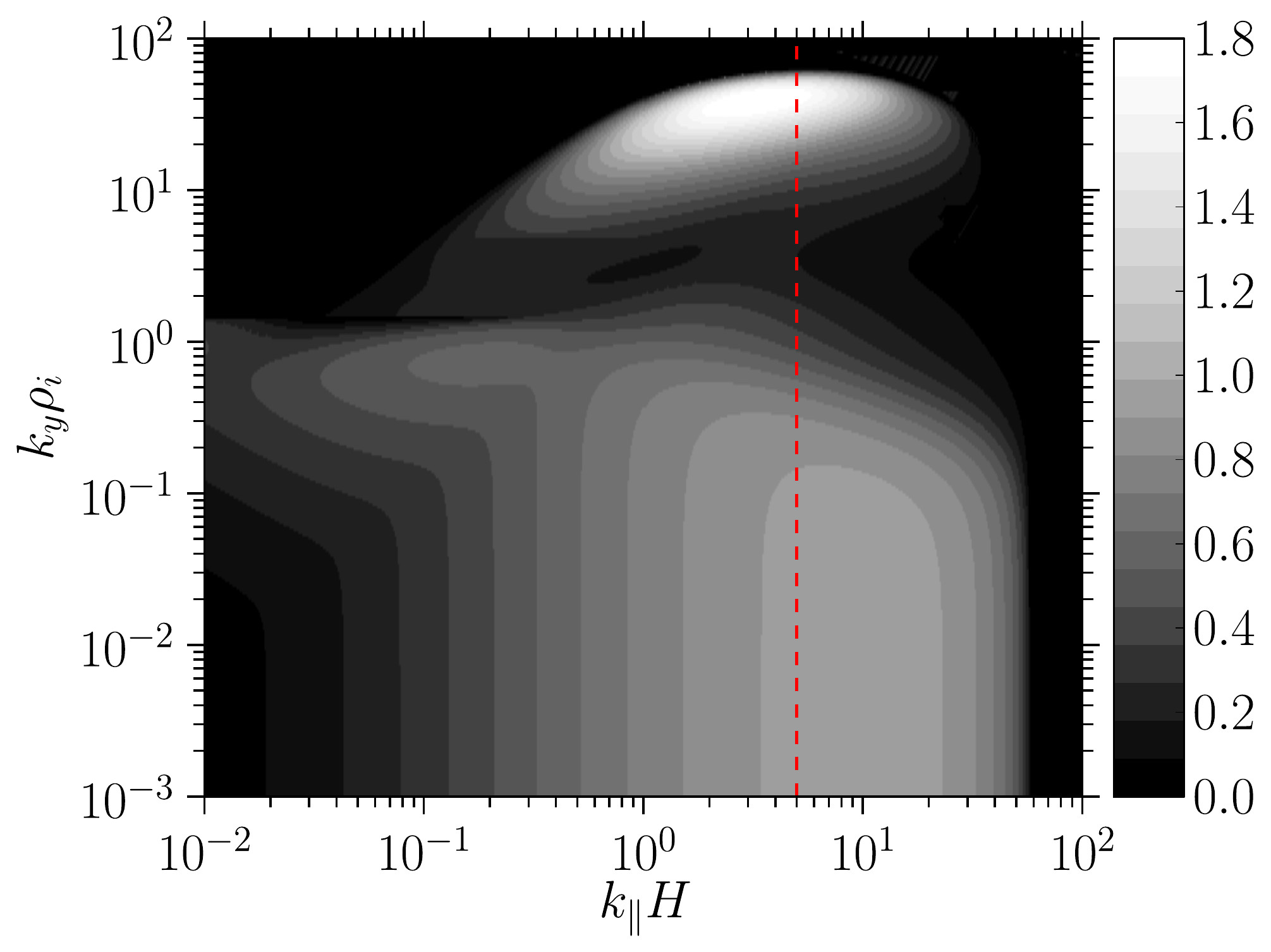}
\end{minipage}
\quad
\begin{minipage}[b]{0.46\textwidth}
\includegraphics[width =\textwidth]{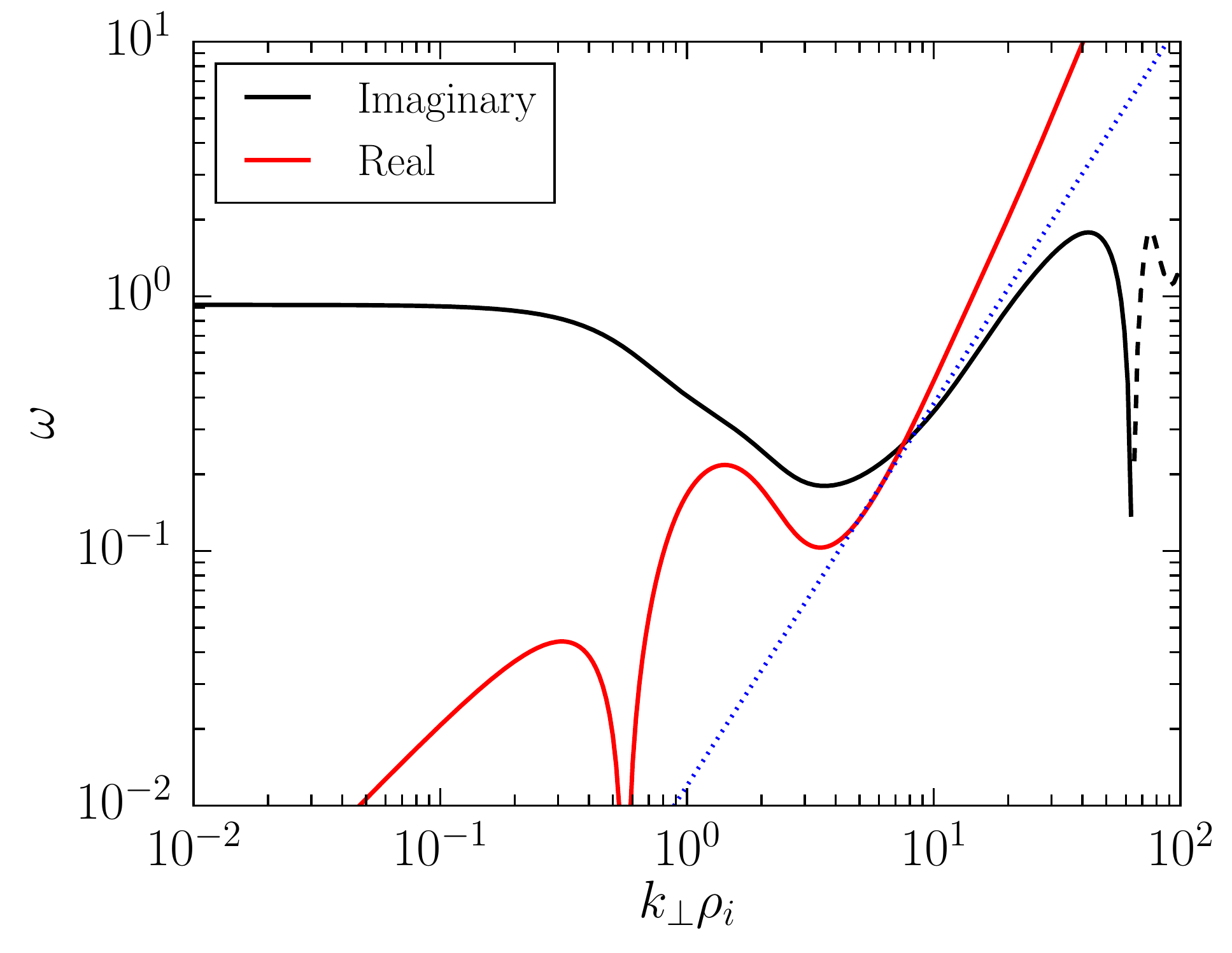}
\end{minipage}
\caption{({\it left}) Instability growth rate (normalised by $\sqrt{-g \,{\rm d}\ln T/{\rm d}z}$) in the $k_\parallel H$-$k_y\rho_i$ plane, as calculated from the gyrokinetic dispersion relation (\ref{eqn:gkdsp}) with ${\rm d}\ln T / {\rm d}\ln P = 1/3$, $\beta_i =10^4$, $k_z = 0$ and $m_e/m_i = 1/1836$. ({\it right}) Real (red) and imaginary (black) parts of $\omega / \sqrt{-g\,{\rm d}\ln T/{\rm d}z}$ for $k_\parallel H=5$, which is indicated by the red dotted line in the left panel. At long wavelengths, the plasma is unstable to the Alfv\'{e}nic MTI. As the ion Larmor radius is approached ($k_y \rho_i \sim 1$), the Alfv\'{e}nic MTI becomes coupled to the collisionlessly damped slow mode and thus its growth rate decreases. At sub-ion-Larmor scales, the ion response is nearly Boltzmann, and kinetic Alfv\'{e}n waves are destabilised by the electron temperature gradient. This is the eMTI. The dotted blue line is the approximate analytic solution (\ref{eqn:highbetagk}). At electron Larmor scales, ${\rm Im}(\omega) < 0$ (denoted by the dashed black line) and the eMTI is damped. See \S\ref{sec:gyrokinetic} for details.}
\label{fig:gk}
\end{figure}

In Figure \ref{fig:gk}, we plot solutions of (\ref{eqn:gkdsp}) for a hydrogenic plasma with $\beta_i = 10^4$, $\psi = 0$ (i.e.~$k_z = 0$), and ${\rm d}\ln T/{\rm d}\ln P = 1/3$. At $k_y \rho_i \ll 1$, the growth rate remains roughly constant with $k_y \rho_i$, an effect discussed in \S\ref{sec:longwavelength_ky}. As $k_y \rho_i$ gets larger, the growth rate decreases and $\omega$ acquires a real part comparable to its imaginary part. The decrease in growth rate may be obtained from (\ref{eqn:gkdsp}) by expanding in $\zeta_i, \alpha_i \ll 1$ and $\zeta_e, \alpha_e \rightarrow 0$; in this limit, $\mc{B} \simeq \alpha_i$, $\mc{E} \simeq -(3/2) \alpha_i$, and $\mc{F} \simeq (\omega_{Pi}\omega_{Ti} / \omega^2) (1 + T_e/T_i)$, and so (\ref{eqn:gkdsp}) becomes
\begin{equation}\label{eqn:gklongwavelength}
\omega^2 = \left( k^2_\parallel v^2_A - g \D{z}{\ln T}\frac{k^2_y}{k^2_\perp} \right) \left( 1 + \frac{9}{8} \frac{k^2_\perp \rho^2_i}{\Lambda} \right)^{-1} ,
\end{equation}
where we have employed the shorthand $\Lambda \doteq 2/\beta_i - \mc{D} + \mc{C}^2 / \mc{A}$ to represent the slow-mode piece of (\ref{eqn:gkdsp}). The first term in parentheses is readily identified as the Alfv\'{e}nic MTI (cf.~(\ref{eqn:prelim3})). The second term in (\ref{eqn:gklongwavelength}) captures the FLR stabilization, with the growth rate eventually falling off as $\sim$$(k_\perp \rho_i)^{-1}$. As $k_\perp \rho_i$ approaches unity, the Alfv\'{e}nic MTI becomes more and more coupled to the collisionlessly damped slow-mode branch, and the dispersion relation becomes approximately $(\mc{D} - 2/\beta_i ) \mc{F} \simeq \mc{E}^2$ -- a weakly Barnes-damped slow mode (i.e.~$\omega \simeq -\imag k_\parallel v_A / \sqrt{\pi\beta_i}$) coupled by FLR effects ($\mc{E}^2$) to a weakly unstable Alfv\'{e}nic-MTI mode ($\mc{F}$). This is a result of the ions gyroaveraging over Larmor-scale fluctuations and thus reducing the connection between the electromagnetic fluctuations and the ion distribution function. The destabilised Alfv\'{e}n wave, for example, loses its character altogether, as the ions drift across the magnetic-field lines and the mode becomes dispersive.

On the other side of  $k_\perp \rho_i \sim 1$, the growth rate begins to increase, eventually attaining values comparable to (and even exceeding!) those achieved at long wavelengths. This trend continues unabated until kinetic effects intervene at the electron Larmor scale, where the wave is ultimately damped. To uncover the physics driving the sub-ion-Larmor-scale growth, we expand (\ref{eqn:gkdsp}) in the region $k_\perp \rho_e \ll 1 \ll k_\perp \rho_i$, in which $\Gamma_0(\alpha_i)$ and its derivatives asymptote to zero and $\Gamma_0(\alpha_e) \simeq \Gamma_1(\alpha_e) \simeq 1$, whence $\mc{A} \simeq 1 + T_i/T_e$, $\mc{B} \simeq 1 - (T_i/T_e)(\omega_{Te}/\omega)$, $\mc{C} \simeq 0$, $\mc{E} \simeq -1$, and $\mc{F} \simeq (T_i/T_e)(\omega_{Pe}/\omega)[1+(\omega_{Te}/\omega)]$. As we are generally concerned with $\beta_i \gg 1$, the coefficient $\mc{D}$ warrants some care. Usually, in such sub-ion-Larmor expansions, $\beta_i$ is taken to be order unity relative to the ion-to-electron mass ratio, and $\mc{D}$ can be safely taken to be $\simeq$$0$. Specifically, the ion contribution to $\mc{D}$ scales with $\Gamma'_0(\alpha_i) \sim -1/\sqrt{8\pi\alpha^3_i}$, which is small, and the electron contribution scales with $(m_e/m_i)^{1/2} $, which is small as well. Indeed, this expansion is what, in a homogeneous plasma, leads to the usual dispersion relation describing kinetic Alfv\'{e}n waves \citep[e.g.][]{kingsep90}:
\begin{equation}\label{eqn:homoKAW}
\omega = \pm \frac{k_\parallel v_A k_\perp \rho_i}{\sqrt{\beta_i + 2 / ( 1 + T_e/T_i )}} .
\end{equation}
But, with our interest in high-$\beta$ plasmas, $\beta_i$ can be $\sim$$m_i / m_e$ or perhaps even more, and the electron contribution to $\mc{D}$ may be $\gtrsim$$2/\beta_i$ (its main competition in (\ref{eqn:gkdsp})). For that reason, the mass-ratio expansion involved in taking $k_\perp \rho_e \ll 1$ cannot be performed independently of taking $\beta_i \gg 1$.

These complications taken into account, we retain both $1/\beta_i$ and $\mc{D}$ in (\ref{eqn:gkdsp}) while adopting the asymptotic values of $\mc{A}$, $\mc{B}$, $\mc{C}$, $\mc{E}$, and $\mc{F}$ given in the preceding paragraph. Equation (\ref{eqn:gkdsp}) becomes
\begin{equation}\label{eqn:GKasymp}
\left( \frac{\alpha_i\mc{A}}{\overline{\omega}^2_i}  - \mc{A} \mc{F} - \mc{A} \mc{B} + \mc{B}^2 \right) \left( \frac{2}{\beta_i} - \mc{D} \right) = \mc{A} ,
\end{equation}
or
\begin{align}\label{eqn:inhomoKAW}
&\omega^2 \left( \frac{2\beta_i}{2-\beta_i \mc{D}} + \frac{2}{1+T_e/T_i} \right) - k^2_\parallel v^2_A k^2_\perp \rho^2_i \nonumber \\*
\mbox{} &\qquad= - \frac{2T_i}{T_e} \left[ \omega \left( \omega_{Pe} - \omega_{Te} \, \frac{1 - T_e/T_i}{1+T_e/T_i} \right) + \omega_{Te} \left(  \omega_{Pe} - \frac{\omega_{Te}}{1+ T_e/T_i}  \right)  \right] .
\end{align}
For $\alpha_i \gg (m_i/m_e)^{1/3}$ -- a limit well-satisfied in the regime of interest -- $\mc{D}$ is dominated by its electron contribution $\simeq$$2\imag\sqrt{\pi} (T_e/T_i) (\omega / k_\parallel v_{the}) [ 1 - (\omega_{Te}/2\omega)]$. The first term in parentheses on the left-hand side of (\ref{eqn:inhomoKAW}) then reduces to either $2/(1+T_e/T_i)$ if $\beta_i \ll 1$, $\beta_i$ if $1 \ll \beta_i \ll (m_i/m_e)^{1/2}$, or $2\imag (T_i/T_e) (k_\parallel v_{the}/\sqrt{\pi}) / ( 2\omega - \omega_{Te} )$ if $\beta_i \gg (m_i/m_e)^{1/2} \gg 1$. (Here, we have taken $k_y \rho_i \sim k_\parallel H$ with respect to the subsidiary expansion in $\beta_i$ and mass ratio.) In all cases, a necessary (but not sufficient) condition for instability is
\begin{equation}\label{eqn:eMTIcriterion}
k^2_\parallel v^2_A k^2_\perp \rho^2_i -  \omega_{Te} \frac{2T_i}{T_e} \left( \omega_{Pe} - \frac{\omega_{Te}}{1+T_e/T_i} \right) < 0 ,
\end{equation}
or, equivalently,
\begin{equation}
k^2_\parallel v^2_A - c^2_s \frac{k^2_y}{k^2_\perp} \D{z}{\ln T_e} \left( \D{z}{\ln P_e} - \D{z}{\ln T_e} \frac{1}{1+T_e/T_i} \right) < 0 ,
\end{equation}
where $c_s \doteq (T_e / m_i)^{1/2}$ is the ion-acoustic speed. With ${\rm d}\ln P_e/{\rm d}z < 0$, we require ${\rm d}\ln T_e / {\rm d} z < 0$ (as in the drift-kinetic and gyroviscous limits). This is a kinetic-Alfv\'{e}n drift wave, destabilised by rapid electron conduction along perturbed magnetic-field lines; we name this the {\em electron magnetothermal instability}, or eMTI, due to its reliance on the electron temperature gradient and the relatively passive role played by the ion species. Indeed, at these scales, the ion response is nearly Boltzmann, $\delta f_i \approx -(e\varphi/T_i) f_i$ (cf.~(\ref{eqn:gkdf})).

\begin{figure}
\centering
\begin{minipage}[b]{0.48\textwidth}
\includegraphics[width =\textwidth]{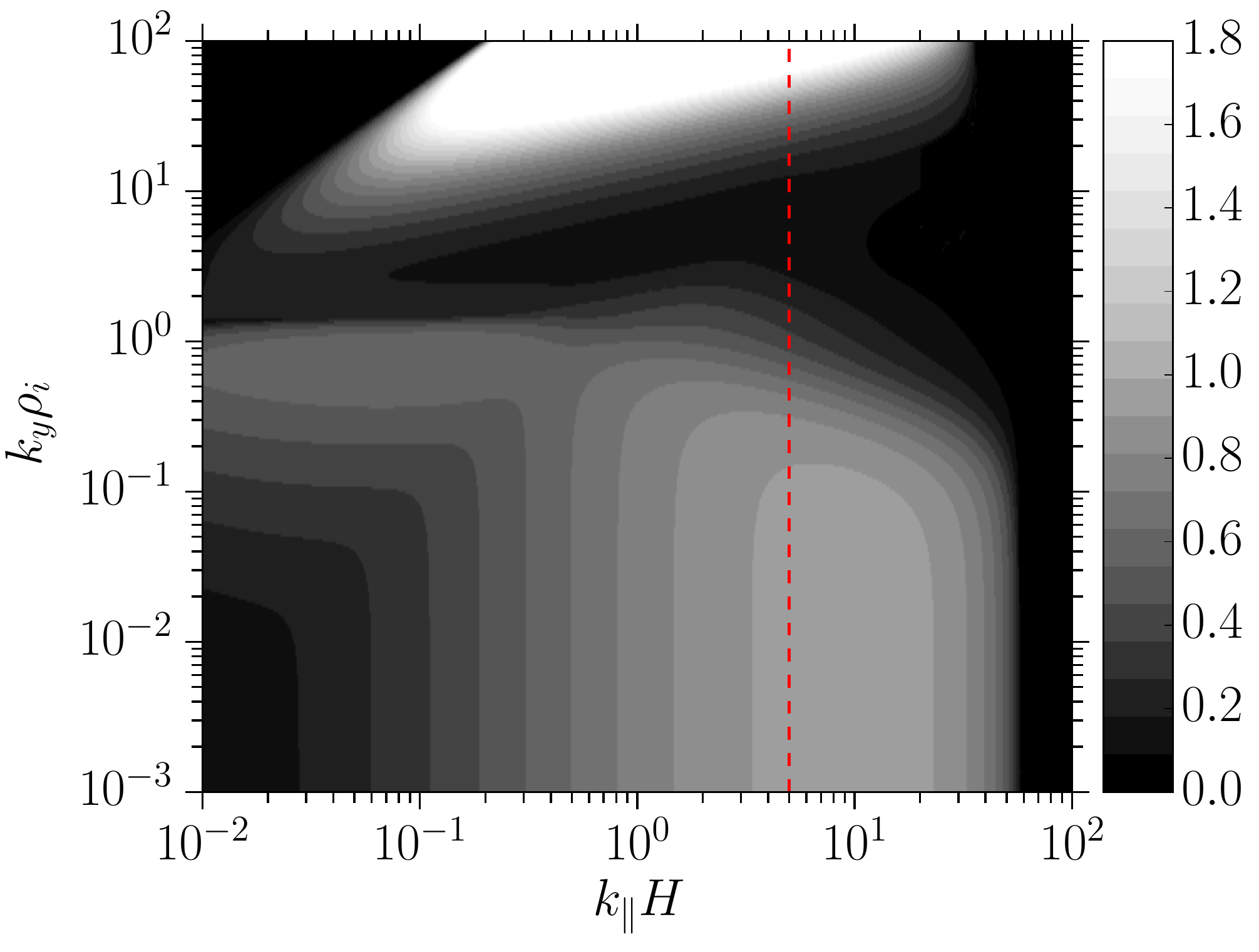}
\end{minipage}
\quad
\begin{minipage}[b]{0.46\textwidth}
\includegraphics[width =\textwidth]{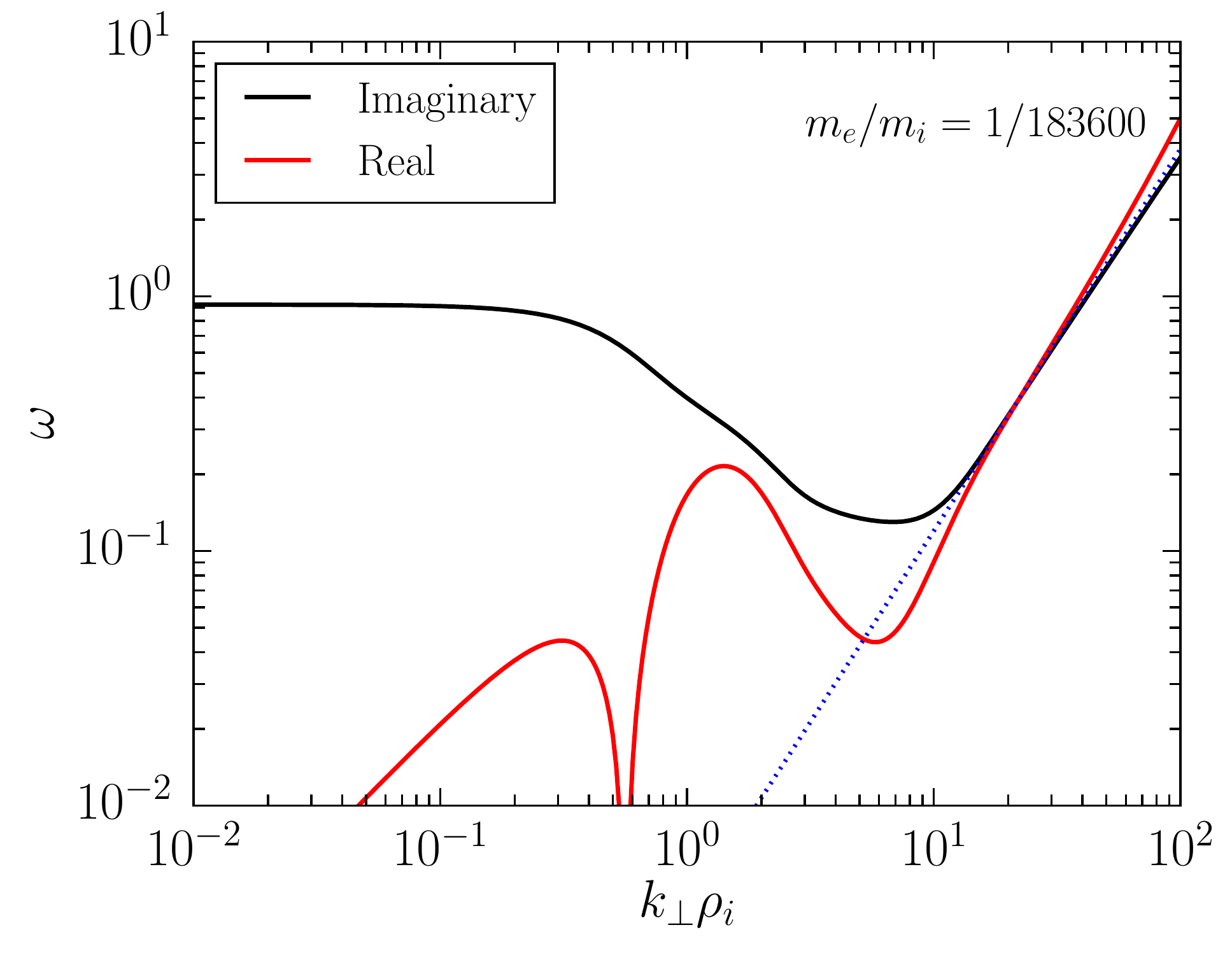}
\end{minipage}
\caption{Same as Figure \ref{fig:gk} but with an artificially suppressed electron-to-ion mass ratio ($m_e / m_i = 1/183600$), in order to demonstrate the accuracy of the approximate analytic solution (the dotted blue line, equation (\ref{eqn:highbetagk})) and the effect of electron Landau damping.}
\label{fig:gk1}
\end{figure}

\begin{figure}
\centering
\includegraphics[width=0.46\textwidth]{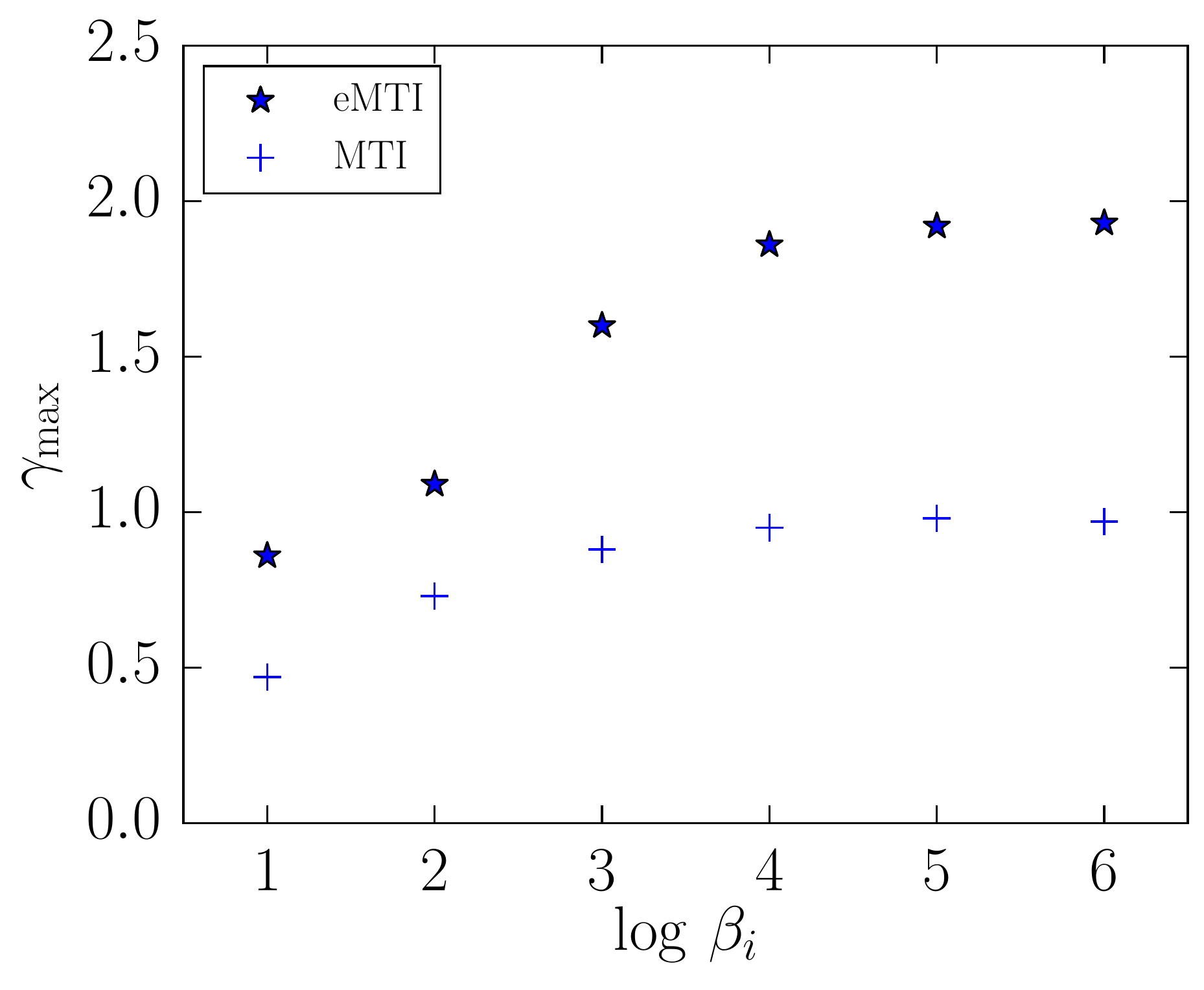}
\caption{Maximum growth rates (normalised by $\sqrt{-g \,{\rm d}\ln T/{\rm d}z}$) of sub-ion-Larmor-scale eMTI (stars) and long-wavelength kinetic MTI (crosses) as a function of $\beta_i$ across all $k_\parallel$ and $k_y$ ($k_z = 0$); ${\rm d}\ln T/{\rm d}\ln P = 1/3$ and $T_i / T_e = 1$ for all points. See the penultimate paragraph of \S\ref{sec:gyrokinetic} for details.}
\label{fig:maxgrowth}
\end{figure}

\begin{figure}
\centering
\includegraphics[width=0.46\textwidth]{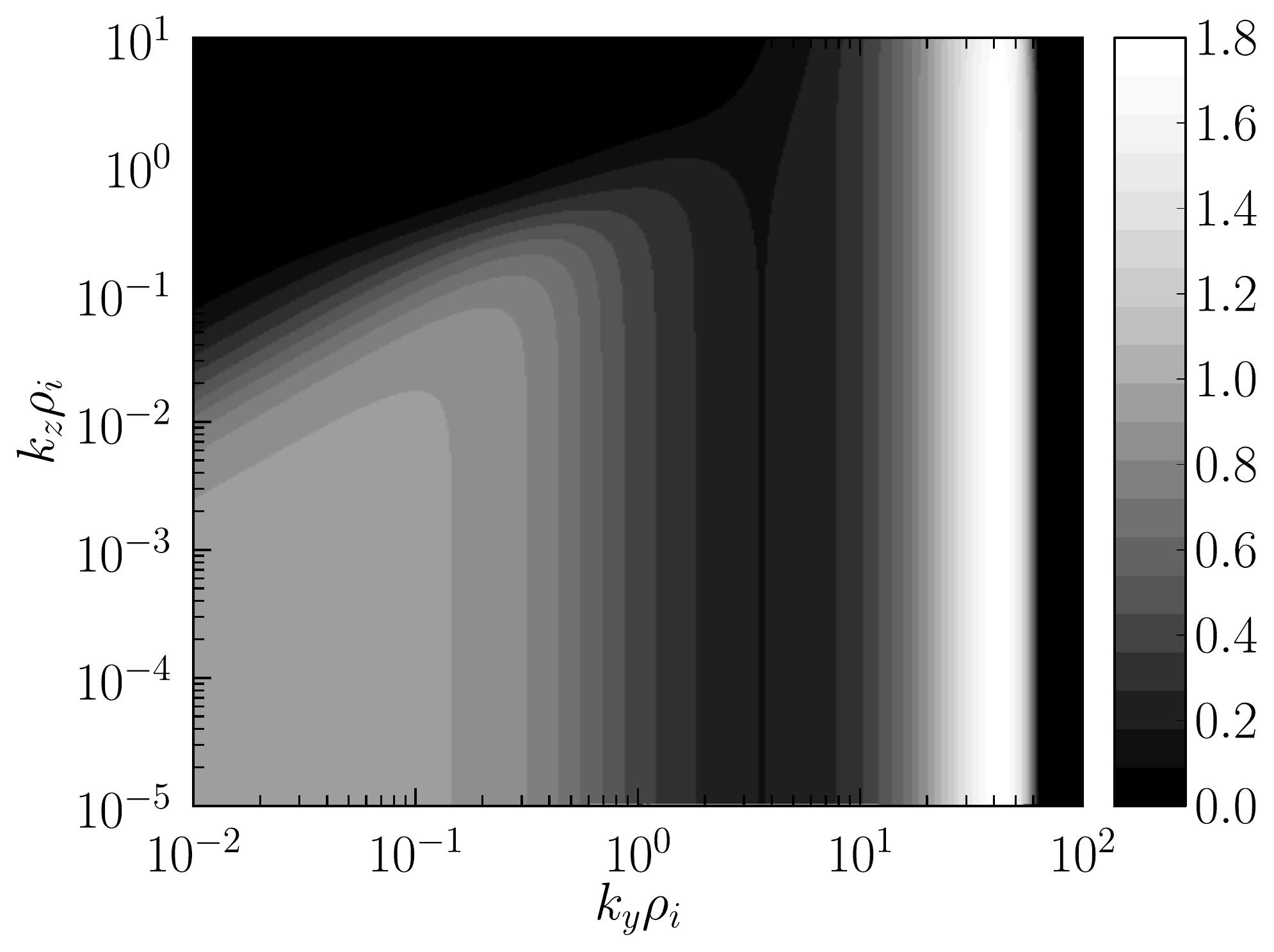}
\caption{Kinetic MTI and eMTI growth rates (normalised by $\sqrt{-g \,{\rm d}\ln T/{\rm d}z}$) as a function of $k_y\rho_i$ and $k_z\rho_i$, as calculated from the gyrokinetic dispersion relation (\ref{eqn:gkdsp}) with $\beta_i = 10^4$, ${\rm d}\ln T/{\rm d}\ln P = 1/3$, and $k_\parallel H = 5$. See the final paragraph of \S\ref{sec:gyrokinetic} for details.}
\label{fig:gk2d}
\end{figure}

We now obtain some approximate analytic solutions to (\ref{eqn:inhomoKAW}). In the high-$\beta_i$ limit featured in Figures \ref{fig:gk}--\ref{fig:gk2d}, equation (\ref{eqn:inhomoKAW}) becomes
\begin{equation}
\omega^2 \simeq \imag\sqrt{\pi} \frac{\omega_{Te}}{k_\parallel v_{the}} \biggl[ \omega \bigl( \omega_{Pe} - \omega_{Te} \bigr) - \omega_{Te} \left( \omega_{Pe} - \frac{\omega_{Te}}{1+T_e/T_i} \right) - k^2_\parallel v^2_A k^2_\perp \rho^2_{sound} \biggl( \frac{2\omega}{\omega_{Te}} - 1 \biggr) \biggr] ,
\end{equation}
where $\rho_{sound} \doteq c_s / \Omega_i$ is the ion sound radius. We may expand this solution in the parameter $(k_y \rho_e / k_\parallel H)^{1/2} \sim (m_e/m_i)^{1/4} \ll 1$ to find the leading-order expression
\begin{equation}\label{eqn:highbetagk}
\omega \simeq (1 + \imag ) \left[ \frac{\sqrt{\pi}}{2} \frac{\omega^2_{Te}}{k_\parallel v_{the}} \left( \frac{\omega_{Te}}{1+T_e/T_i} - \omega_{Pe} \right) \right]^{1/2} ;
\end{equation}
i.e.~the mode is {\em overstable}, with a growth rate $\sim$$k_y \rho_{sound} (c_s/H) \sqrt{ k_y \rho_e / k_\parallel H}$.\footnote{Equation (\ref{eqn:highbetagk}) is the inhomogeneous counterpart of equation (63) in \citealt{howes06}.} Despite all the approximations, the growth rate agrees rather well with the numerical solution (see Fig.~\ref{fig:gk}). The real part of (\ref{eqn:highbetagk}) differs from that seen in the figure, which is not too surprising -- $(m_e/m_i)^{1/4}$ is not that small of a number! -- but the accuracy of the solution improves as $m_e/m_i \rightarrow 0$. This is shown in Figure \ref{fig:gk1}, which gives the solution to (\ref{eqn:gkdsp}) for $m_e/m_i = 1/183600$. The blue dotted line, equation (\ref{eqn:highbetagk}), is a clearly an excellent fit to both the real and imaginary parts of $\omega$. On the other hand, the solution to (\ref{eqn:inhomoKAW}) for $1 \ll \beta_i \ll (m_i/m_e)^{1/2}$ is approximately
\begin{align}\label{eqn:GKomega}
\omega &\simeq - \frac{1}{\beta_e} \biggl( \omega_{Pe} - \omega_{Te} \, \frac{1 - T_e/T_i}{1+T_e/T_i} \biggr) \nonumber\\*
\mbox{} &\quad+ \imag \, \sqrt{\frac{2}{\beta_e}} \left[ \omega_{Te} \biggl( \omega_{Pe} - \frac{\omega_{Te}}{1+T_e/T_i} \biggr) - k^2_\parallel v^2_A k^2_\perp \rho^2_{sound}   \right]^{1/2} .
\end{align}
In this regime, the kinetic-Alfv\'{e}nic nature of the mode is readily apparent from the final term in the square root.

In all of these limits, the maximum growth rate of the eMTI is set by electron FLR effects (the mode is damped at $k_y \rho_e \sim 1$ -- see right panel of Fig.~\ref{fig:gk}) and is generally larger than the growth rate for the (long-wavelength) kinetic MTI. This is shown explicitly in Fig.~\ref{fig:maxgrowth}, which gives the respective maximum growth rates $\gamma_{\rm max}$ evaluated across the full range of $k_\parallel$ and $k_y$ ($k_z = 0$) for fixed temperature gradient and ion-to-electron temperature ratio. The difference is probably related to the relatively mundane role that the inertia-bearing ions play in the eMTI relative to the MTI (recall that the ion response is Boltzmann for the eMTI). It is the electrons which behave similarly in both cases: indeed,  (\ref{eqn:gkdf}) with $(k_y \rho_e)^2 \ll 1$ implies
\begin{equation}
\frac{\delta T_{\perp e}}{T_e} \simeq \frac{e}{T_e} \frac{v_{the} A_\parallel}{c} \frac{\omega_{Te}}{k_\parallel v_{the}} = \frac{\imag}{k_\parallel} \frac{\delta B_z}{B_0} \D{z}{\ln T_e} ,
\end{equation}
which states that the Lagrangian change in a fluid element's perpendicular electron temperature is nearly zero as the flux-frozen electrons carry the magnetic field upwards or downwards. While there appears to be an asymptotic eMTI maximum growth rate as $\beta_i \rightarrow \infty$, which is roughly twice as large as the maximum MTI growth rate, the complicated dependence of $k_\parallel H$ and $k_y \rho_e$ on $\beta_i$ at maximum eMTI growth precludes a straightforward analytic calculation for it.

We close this section by commenting on the transition between the drift-kinetic solution for $k_y, k_z \ne 0$ explored in Section \ref{sec:longwavelength_kprp} -- see Figure \ref{fig:longwave_general} in particular -- and the gyrokinetic solution. Recall from that section that the growth rate at long wavelengths is maximal and roughly constant for $k_y / k_z \lesssim 1$. In Figure \ref{fig:gk2d} we see that this trend continues to small wavelengths until $k_y \rho_i \sim k_z \rho_i \sim 1$, at which point ion FLR effects sharply reduce the MTI growth rate. Thereafter, the eMTI emerges with an even larger growth rate, which is independent of $k_z$ until $k_z \rho_e \sim 1$. This independence is not particularly surprising: for $k^2_\perp \rho^2_i \gg 1$ the gyrokinetic dispersion relation (\ref{eqn:gkdsp}) only depends upon $k_z$ through $\Gamma_0(\alpha_e)$ and its derivatives, the drift frequency depending only upon $k_y$. As with the long-wavelength solution (\ref{eqn:collisionlessMTI_largekykz}), the mirror force, by which Landau-resonant particles bleed energy from the magnetic-field-strength fluctuations, is offset by the divergence of the parallel pressure.

\subsection{Relationship between the magnetothermal and the ion- and electron-temperature-gradient instabilities}\label{sec:ITG}

At this point in the manuscript, those readers familiar with drift-wave instabilities in magnetic-confinement-fusion devices might be curious as to the relationship between the various incarnations of the magnetothermal instability elucidated above and the ion- and electron-temperature-gradient instabilities long known to plague fusion plasmas (the so-called ``ITG'' and ``ETG''). Indeed, both the MTI and ITG/ETG are convective instabilities driven by the free energy stored in the background temperature gradient. Their relationship is complicated somewhat by the usual restriction of ITG/ETG analyses to the electrostatic $\beta\rightarrow 0$ limit, in which the magnetic field is assumed rigid, and the geometrical complexity of tokamak magnetic fields, which introduces its own drift-wave instabilities apart from the ``slab'' ITG/ETG instabilities (e.g., toroidal ITG/ETG, trapped electron modes).\footnote{The slab and toroidal ITG and ETG have been generalised to include electromagnetic fields in the limit of small but finite $\beta$, so that $A_\parallel$ is retained while $\delta B_\parallel = 0$ \citep[e.g.][]{khd93,reynders94,sh01}. Finite-beta effects generally stabilize these instabilities due to drift-wave coupling to the shear Alfv\'{e}n wave.} Nevertheless, the transition between the ITG/ETG, on the one hand, and the MTI, on the other, as the plasma beta is increased, is an interesting question. In this section, we explore their connection.

Following conventional work on the slab ITG/ETG \citep{rs61,cks91}, we begin by taking the electrostatic limit, in which the electric field $\bb{E} = -\grad\varphi$. The relevant dispersion relation is obtained by substituting the perturbed distribution function (\ref{eqn:linearVlasov2}) with $\delta\bb{B} = 0$ into the quasi-neutrality constraint (\ref{eqn:quasineutrality}), adopting plane-wave solutions, performing the resulting integrals (after conveniently setting $F_s = F_{M,s}(\mc{Z}_s, \mc{E}_s)$ -- see (\ref{eqn:quasimaxwell})), and demanding nontrivial solutions. The result is that
\begin{align}\label{eqn:electrostatic}
0 &= \sum_s \frac{q^2_s n_s/T_s}{q^2_i n_i/T_i} \nonumber\\*
\mbox{} &+\sum_s \frac{q^2_s n_s/T_s}{q^2_i n_i/T_i} \sum_{n=-\infty}^\infty \frac{\omega}{k_\parallel v_{ths}} Z_0(\zeta_{n,s}) \Gamma_n(\alpha_s) \left\{ 1 + \frac{\omega_{Ts}}{\omega} \left[ \frac{3}{2} - \frac{Z_2(\zeta_{n,s})}{Z_0(\zeta_{n,s})} - \alpha_s \frac{\Gamma'_n(\alpha_s)}{\Gamma_n(\alpha_s)} \right] \right\} .
\end{align}
In the low-frequency limit, we need only retain the $n=0$ contributions to the summation above, and (\ref{eqn:electrostatic}) reduces to $\mc{A} = 0$ (cf.~(\ref{eqn:howesA})). We adopt this limit throughout the remainder of this section.

Explicit solutions of the dispersion relation $\mc{A} = 0$ may be calculated in the analytically tractable limit $k_\parallel v_{the} \gg \omega,\, \omega_{Pi} > k_\parallel v_{thi}$, whence   $Z_0(\zeta_{0,i})$ and $Z_2(\zeta_{0,i})$ can be expanded in their large argument. The result is
\begin{align}\label{eqn:generalITG}
&\bigl( \omega + \omega_{Pi} \bigr)^3 \left[ \frac{T_i}{T_e} + 1 - \Gamma_0(\alpha_i) \right] - \bigl( \omega + \omega_{Pi} - \omega_{Ti} \bigr) \frac{k^2_\parallel v^2_{thi}}{2} \Gamma_0(\alpha_i) \nonumber\\*
\mbox{} &\quad= \Gamma_0(\alpha_i) \Biggl[ \bigl( \omega + \omega_{Pi} \bigr)^2 + \frac{k^2_\parallel v^2_{thi}}{2} \Biggr] \biggl\{ -\omega_{Pi} + \omega_{Ti} \biggl[ 1 - \alpha_i \frac{\Gamma'_0(\alpha_i)}{\Gamma_0(\alpha_i)} \biggr] \biggr\} .
\end{align}
We now make some standard simplifying assumptions, which are known to be quantitatively imprecise but nevertheless afford a compact analytic solution whose qualitative physics may be readily deduced. Taking ${\rm d}\ln n_i / {\rm d}\ln T_i \ll (k_\parallel H / k_y \rho_i)^{2/3} \ll (T_i/T_e)^{1/3}$, we find that the entire right-hand side of (\ref{eqn:generalITG}) may be dropped. Expanding $\Gamma_0(\alpha_i) \simeq - \Gamma'_0(\alpha_i) \simeq 1$, we find that
\begin{equation}\label{eqn:ITG}
\omega \simeq - \omega_{Pi} + \bigl( -\omega_{Ti} k^2_\parallel c^2_s \bigr)^{1/3} ,
\end{equation}
which is unstable for $\omega_{Ti} > 0$ or, equivalently, $\bb{k}\bcdot(\eb\btimes\grad T_i) > 0$. This is the ITG instability, Doppler-shifted by the equilibrium diamagnetic drift. The physics of the instability is explained in detail by \citet{cks91}, to which we refer the reader. In brief, parallel flow along fields lines compresses the ions. An electrostatic potential is produced in order to enforce quasi-neutrality (the electron response is Boltzmann), which drives an $\bb{E}\btimes\bb{B}$ flow that advects cool ions into the compressed region. This lowers the pressure there, pulling in yet more ions along the field lines and reinforcing the original density enhancement. The result is a feedback loop that leads to instability. The key requirement here is for the mode frequency $\omega$ to be larger than the rate of pressure relaxation along field lines $k_\parallel c_s$, so that the Lagrangian change in an ion fluid element's pressure vanishes to leading order. Other requirements are that the density gradient not be too strong and $k_y \rho_i$ not be too large. 

\begin{figure}
\centering
\begin{minipage}[b]{0.48\textwidth}
\includegraphics[width =\textwidth]{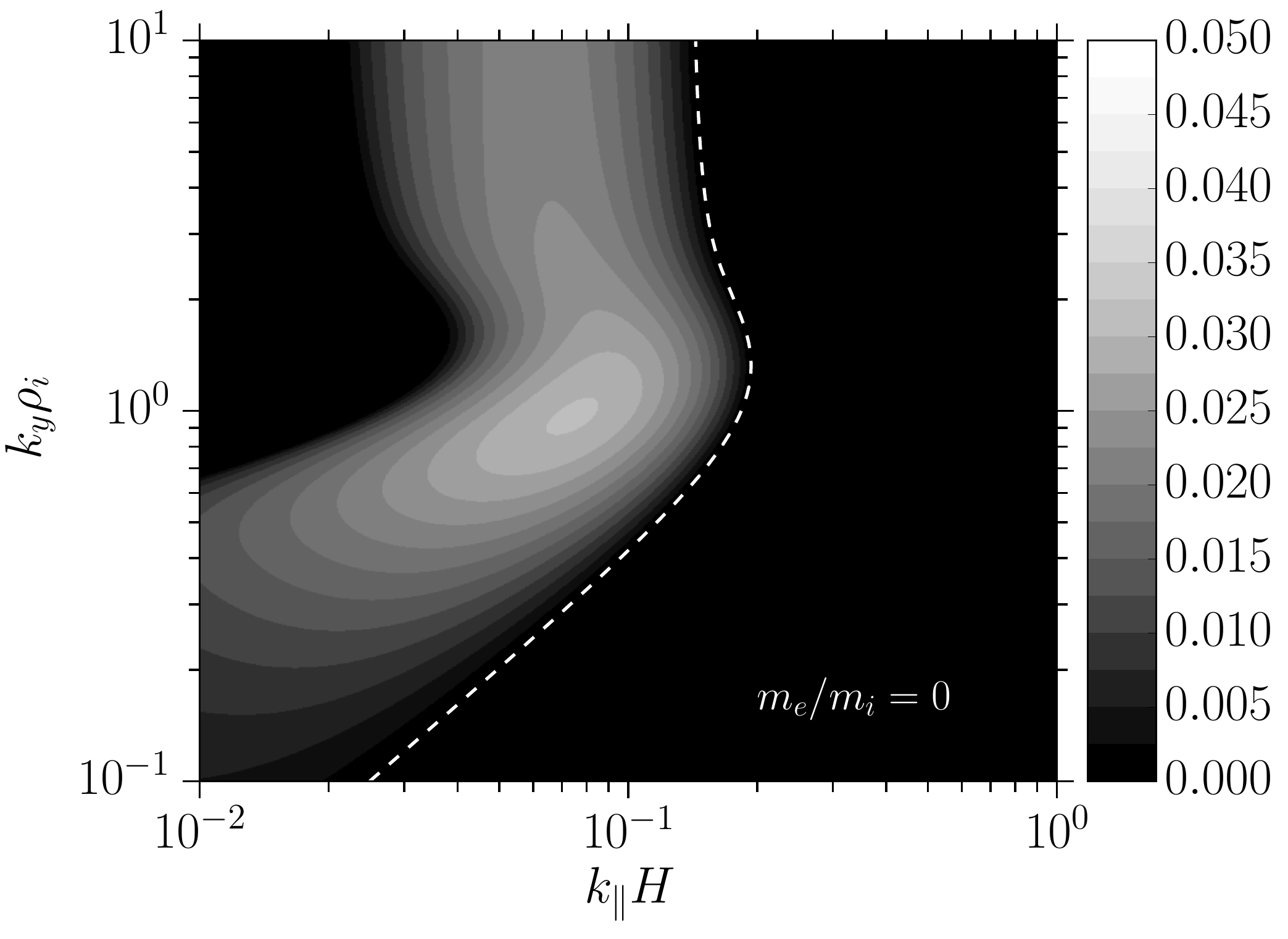}
\end{minipage}
\quad
\begin{minipage}[b]{0.48\textwidth}
\includegraphics[width =\textwidth]{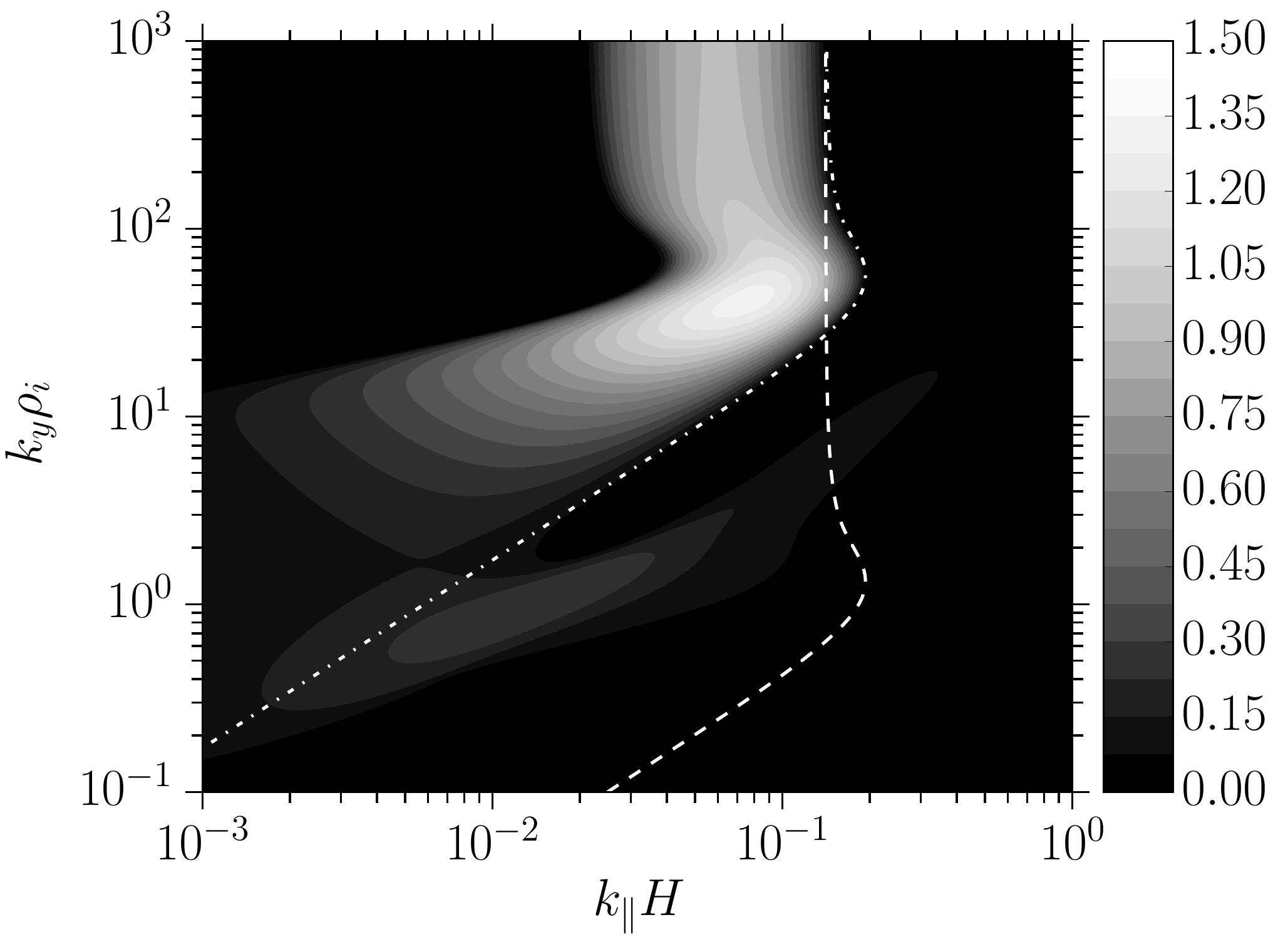}
\end{minipage}
\caption{Electrostatic ITG/ETG instability growth rate (normalised by $\sqrt{-g \,{\rm d}\ln T/{\rm d}z}$) for ${\rm d}\ln T_i / {\rm d}\ln P_i = 1$ using ({\it left}) $m_e=0$ and ({\it right}) $m_e/m_i=1/1836$; massless electrons are used in the left panel to eliminate the ETG and thus accentuate ITG growth. The white dashed (dot-dashed) lines trace the boundary of marginal stability to ITG (ETG). See \S\ref{sec:ITG} for details.}
\label{fig:ITG}
\end{figure}

These stabilizing effects can be seen in Figure \ref{fig:ITG}a, which shows the growth rate of the electrostatic ITG in the $k_\parallel H$-$k_y \rho_i$ plane obtained from numerically solving (\ref{eqn:electrostatic}) with $m_e/m_i = 0$ (so that ETG is suppressed -- see below). The dashed line shows the marginal stability threshold, which may be obtained analytically by setting ${\rm Im}(\omega) = 0$ in the dispersion relation $\mc{A} = 0$, demanding that ${\rm Im}(\mc{A}) = 0$ (which determines the real frequency of the mode at marginal stability), and substituting the result back into $\mc{A} = 0$:
\begin{equation}\label{eqn:marginalstability}
\frac{k_\parallel H}{k_y \rho_i} \D{\ln T_i}{\ln P_i} = \pm \sqrt{ \frac{1/2 - {\rm d}\ln n_i/{\rm d}\ln T_i - \alpha_i \Gamma'_0(\alpha_i) / \Gamma_0(\alpha_i) }{[ 2 ( 1 + T_i/T_e ) / \Gamma_0(\alpha_i) - 1 ]^2 -1 }} .
\end{equation}
(Here we have taken ${\rm Re}(\omega) / k_\parallel v_{the} \ll 1$.) Consider the numerator inside the square root on the right-hand side of (\ref{eqn:marginalstability}), which must be positive for a sensible solution. This shows the standard result that, at long wavelengths (i.e.~$\alpha_i \rightarrow 0$), the density gradient can be no steeper than ${\rm d}\ln n_i / {\rm d}\ln T_i = 1/2$, or ${\rm d}\ln T_i/{\rm d}\ln P_i > 2/3$ for instability. (In the fusion literature, this is expressed as $\eta_i \doteq {\rm d}\ln T_i / {\rm d}\ln n_i > 2$.) FLR effects change this slightly: using the asymptotic result $\Gamma_0(\alpha_i) \approx 1/\sqrt{2\pi\alpha_i}$ in (\ref{eqn:marginalstability}) gives a critical density gradient ${\rm d}\ln n_i/{\rm d}\ln T_i \approx 1$, or ${\rm d}\ln T_i/{\rm d}\ln P_i \gtrsim 1/2$. (Note that our fiducial MTI-unstable atmosphere with ${\rm d}\ln T_i/{\rm d}\ln P_i = 1/3$ is stable to ITG/ETG.)

Equilibrium gradients in the electron species also give rise to instability. A calculation analogous to the ITG may be performed at sub-ion-Larmor scales ($\alpha_i \gg 1$) by assuming $\omega / k_\parallel v_{the} > 1$ and setting $\Gamma_0(\alpha_i) = \Gamma'_0(\alpha_i) = 0$. The dispersion relation is identical to (\ref{eqn:generalITG}), but with the ion and electron subscripts reversed. Accordingly, the electron counterpart to (\ref{eqn:ITG}) is
\begin{equation}\label{eqn:ETG}
\omega \simeq - \omega_{Pe} + \left( -\omega_{Te} k^2_\parallel \frac{T_i}{m_e} \right)^{1/3} ,
\end{equation}
which is unstable for $\omega_{Te} > 0$ or, equivalently, $\bb{k}\bcdot(\eb\btimes\grad T_e) < 0$. This is the ETG instability. The numerical solution to (\ref{eqn:generalITG}) for $m_i/m_e = 1836$ is given in Figure \ref{fig:ITG}b, which shows the ETG peaking at $k_\parallel H \sim 0.1$ and $k_y \rho_e \sim 1$.

The common characteristic of both the ITG and ETG instabilities is that the unstable mode drifts across stationary magnetic-field lines. This is allowed because, at the low values of $\beta$ required for the electrostatic approximation to hold, the electron skin depth is at least as large as the ion Larmor radius, and so the ion (and electron) species can slip through the otherwise rigid field lines. By contrast, at large $\beta$, the magnetic field is unfrozen only by electron FLR effects, and so a super-Larmor-scale displacement in the plasma carries the magnetic-field lines with it. It is this displacement of field lines, and the accompanying rapid streaming of particles along it, that both stabilises the plasma to ITG/ETG and destabilises it to the MTI/eMTI. While the physics is markedly different, the end result is the same: the free energy stored in the temperature gradient drives convective instability both at long wavelengths and, more vigorously, at Larmor scales.

\section{Summary}\label{sec:summary}

Efforts to provide a proper accounting of kinetic effects in high-$\beta$ astrophysical plasmas are still in their infancy, particularly when it comes to the influence of plasma microphysics on the macroscale dynamics -- convection being but one example. In this paper we have studied the linear stability properties of a collisionless, magnetised, thermally stratified atmosphere. This is an extension of previous work on collisional, magnetised, stratified atmospheres \citep{balbus00,balbus01,quataert08,kunz11}, which had shown them to be linearly unstable when the temperature increases in the direction of gravity. This instability -- the magnetothermal instability (MTI) -- is facilitated by rapid, field-aligned conduction obviating adiabatic fluid displacements. In this fast-conduction limit, slow-mode perturbations exhibit almost no (Lagrangian) change in their temperature; for ${\rm d}\ln T / {\rm d}z < 0$, it is thus energetically favourable for vertically displaced fluid elements to continue rising or falling in the gravitational potential. In certain situations, the accompanying field-aligned viscous momentum transport can drive Alfv\'{e}nic fluctuations unstable as well.

This paper focuses instead on the case in which the heat and momentum transport that enables these instabilities is not due to collisional processes, but rather stems from the complex interplay between field-aligned particle streaming, collisionless wave damping, and finite-Larmor-radius effects. Because prior work on the MTI relied on the buoyant ion species being thermally equilibrated with the inertialess parallel-conducting electrons, the application of those results to a collisionless system is not by any means straightforward. To what extent they do carry over to the collisionless case is a particularly intriguing question in the context of the hot, diffuse plasma in the outskirts of galaxy clusters -- the intracluster medium -- in which the collisional mean free path can approach the thermal-pressure scale height. Radiatively inefficient black-hole accretion flows, such as that onto Sgr A$^\ast$ at the Galactic center, are also a natural venue for application of our results, although the situation there is further complicated by differential rotation and the consequent kinetic magnetorotational instability \citep{qdh02,shq03,islam14,hq14,qht15}.

Our results may be summarized as follows. First, at long wavelengths, the fluid MTI carries over relatively unscathed to the kinetic case, enjoying the same stability criterion (${\rm d}\ln T/ {\rm d}z > 0$) and maximum growth rate. The physics driving the requisite isothermalisation of perturbed field lines is, of course, quite different in each case, but that appears to be of little concern to buoyancy, which does its job regardless. That being said, it is rather important to note that conduction and buoyancy in a collisionless plasma are inextricably linked, a feature not present in a collisional fluid. In a collisionless plasma, approximately isothermal displacements are tandem with the ability of those displacements to maintain pressure balance with their surroundings. Both processes suffer from collisionless damping if an appreciable number of particles are Landau-resonant with the fluctuation.

Second, because of adiabatic invariance, changes in temperature go hand-in-hand with changes in magnetic-field strength. Since the latter are collisionlessly damped, this makes the physics of the MTI much richer than in the collisional case, for which $\Delta T \simeq 0$ in the fast-conduction limit (provided $\bb{B}_0 \perp \bb{g}$). One consequence is that fluctuations whose wavevectors have a component along gravity ($k_z\ne 0$) are strongly damped once $k_z H \gtrsim \sqrt{k_\parallel H}$. This damping can be mitigated by allowing $\bb{k}\bcdot(\bb{g}\btimes\bb{B}_0)\ne 0$, which introduces an additional degree of freedom for the magnetic-field perturbations. This freedom allows the magnetic field to seek out favourable arrangements, in which field-strength fluctuations are anti-correlated with upward (downward) displacements into regions of greater (lesser) potential. The result is that the mirror force acts oppositely to the parallel gravitational force, and so the collisionless damping of field-strength fluctuations is reduced (see Fig.~\ref{fig:kydiagram}). 

Third, for extremely sub-thermal magnetic fields characteristic of those found in primordial galaxy clusters and used in some published (fluid) simulations of the MTI, gyroviscosity -- not magnetic tension -- sets the (parallel) wavenumber cutoff. It does so by providing a dissipationless cross-field transport of momentum, which redirects buoyant motions unfavourably into the horizontal plane.

Fourth, and perhaps most importantly, we have shown that the plasma is overstable at sub-ion-Larmor scales ($k_\perp \rho_i \gg 1 \gg k_\perp \rho_e$) to a kinetic-Alfv\'{e}n drift wave instability -- the electron MTI, or eMTI -- which exhibits a growth rate even larger than does the long-wavelength MTI. It is driven by the electron temperature gradient and relies on (approximately) Boltzmann ions and isothermal-along-perturbed-field-lines electrons. To give this last result some context, we briefly reviewed the slab ITG and ETG -- two of the drift-wave instabilities long known to the magnetic-confinement-fusion community for driving the anomalous transport limiting tokamak performance. While our focus throughout the manuscript has been on applications to the high-$\beta$ intracluster medium, these drift-wave instabilities may also find applications in the hot and tenuous coronae of stars.

As a closing thought, it is tempting to speculate on the nonlinear evolution in a high-$\beta$ plasma of the instabilities investigated herein. One cannot go too far in extrapolating, of course, if only because the velocity-space anisotropy generated by adiabatic invariance as the instabilities grow will quickly drive secondary Larmor-scale instabilities (firehose, mirror), which have been shown to be efficient at isotropising the distribution function and thus providing an effective ``viscosity'' (\citealt{kss14}, \citealt{riquelme15}, \citealt{sn15}, \citealt{melville16}; in the context of the $\beta\sim 1$ solar wind, e.g., \citealt{ht08,ht15}). If the production of such Larmor-scale parasites also reduces the efficacy of electron conduction (as recently suggested by \citealt{komarov16} and \citealt{riquelme16}), one might imagine a scenario in which the MTI, reliant upon the efficient field-aligned transport of heat, grows only as fast as allowed by the parasites it triggers. However, even without the complications introduced by MTI-driven firehose and mirror instabilities, the linear system itself presents an interesting puzzle. We have a plasma in which the same temperature gradient drives growth simultaneously at large scales (drift-kinetic MTI) and small scales (eMTI), the latter with growth rates larger than the former. This implies that the large-scale MTI will likely be dependent upon the saturation of the eMTI, particularly if the latter greatly impacts the nature and efficacy of electron heat transport. Such speculation awaits well-constructed kinetic simulations.

\vspace{0.2in}

\noindent We thank Ian Abel for sharing with us his expertise on the gyrokinetic theory of stratified plasmas; Alex Schekochihin, Greg Hammett, and Steve Cowley for useful discussions on the ITG and ETG instabilities; and Henrik Latter and the anonymous referees for comments that lead to an improved presentation.

\appendix

\section{Details concerning the derivation of the conductivity tensor (\ref{eqn:sigma2})}\label{app:derivation}

In this Appendix, we flesh out some mathematical details from the derivation of the conductivity tensor that were mercifully expunged from Section \ref{sec:orbitintegral}. The first task of that section was to perform the $\tau$ integral in (\ref{eqn:currenta}), a task that is greatly aided by using the Jacobi-Anger expansion
\begin{align}\label{eqn:ustarns}
\bb{v}' {\rm e}^{-\imag a_s \!\sin(\vartheta+\Omega_s\tau-\psi) } &= \sum_{n=-\infty}^\infty {\rm e}^{-\imag n ( \vartheta + \Omega_s\tau - \psi )} \biggl[ \bigl( v_\parallel \ex - v_{ds} \ey \bigr) J_n \nonumber\\*
\mbox{} &\quad+ v_\perp \biggl( \cos\psi \frac{nJ_n}{a_s} - \imag\sin\psi J'_n \biggr) \ey + v_\perp \biggl( \sin\psi \frac{nJ_n}{a_s} + \imag\cos\psi J'_n \biggr) \ez \biggr] \nonumber\\*
\mbox{} &\doteq \sum_{n=-\infty}^\infty {\rm e}^{-\imag n ( \vartheta + \Omega_s\tau - \psi )} \,\bb{u}^\ast_{n,s} ,
\end{align}
where $a_s \doteq k_\perp v_\perp / \Omega_s$ is the argument of the $n$th-order Bessel function $J_n$, the prime denotes differentiation with respect to that argument, and $\ast$ denotes the complex conjugate. The integration is then straightforward, the result being that the conductivity tensor defined by (\ref{eqn:current}) simplifies to
\begin{align}\label{eqn:sigma}
\bb{\sigma} &= \frac{\imag}{\omega} \sum_s \frac{q^2_s}{T_s} \int{\rm d}^3\bb{v}\, \bb{v}\ey \frac{cT_s}{q_s B} \pD{\mc{Z}_s}{F_s}\nonumber\\*
\mbox{} &\quad+ \frac{\imag}{\omega} \sum_s \frac{q^2_s}{T_s} \sum_{n=-\infty}^\infty \int{\rm d}^3\bb{v}\, \bb{v} \bb{u}^\ast_{n,s} \,\frac{ \omega \,{\rm e}^{-\imag n(\vartheta-\psi)+\imag a_s\! \sin(\vartheta-\psi)}}{\omega + k_y v_{ds} - k_\parallel v_\parallel - n\Omega_s} \left( - T_s \pD{\mc{E}_s}{F_s} + \frac{k_y}{\omega} \frac{cT_s}{q_s B} \pD{\mc{Z}_s}{F_s} \right) .
\end{align}
Note the denominator of the second term, which reveals that a fluctuation with frequency $\omega$, Doppler-shifted to the frame drifting at the $-\grad\Phi_s\btimes\ex$ velocity, may be Landau- or cyclotron-resonant; the former can lead to collisionless damping, a feature of critical importance for the behaviour of convective motions. 

The next step is to carry out the integral over gyrophase angle in (\ref{eqn:sigma}). Before doing so, recall that the equilibrium distribution function $F_s$ that resides in the integrand of (\ref{eqn:sigma}) is independent of gyrophase {\em at fixed guiding centre}, not at fixed position. As explained in Section \ref{sec:orbitintegral}, we circumvent this issue by Taylor expanding $F_s(\mc{Z}_s,\mc{E}_s)$ about the particle position $z$ and the gyrophase-independent particle energy $\varepsilon_s$ (see (\ref{eqn:varepsilon})). The practical result is that the partial derivatives of $F_s$ in (\ref{eqn:sigma}) may be replaced by the expressions
\begin{subequations}\label{eqn:ZtozEtoe}
\begin{align}
\pD{\mc{Z}_s}{F_s(\mc{Z}_s,\mc{E}_s)} = \left[ 1 - \frac{v_\perp\cos\vartheta}{\Omega_s} \left( \pD{z}{} + q_s \D{z}{\Phi_s} \pD{\varepsilon_s}{} \right) \right] \pD{z}{F_s(z,\varepsilon_s)} , \\*
\pD{\mc{E}_s}{F_s(\mc{Z}_s,\mc{E}_s)} = \left[ 1 - \frac{v_\perp\cos\vartheta}{\Omega_s} \left( \pD{z}{} + q_s \D{z}{\Phi_s} \pD{\varepsilon_s}{} \right) \right] \pD{\varepsilon_s}{F_s(z,\varepsilon_s)} ,
\end{align}
\end{subequations}
which are accurate to $\mc{O}(\rho_s / H)^2$. The integrals over gyrophase angle in (\ref{eqn:sigma}) can now be performed, again with the aid of Jacobi-Anger expansions:
\begin{align}\label{eqn:uns}
\bb{v} \,{\rm e}^{\imag a_s\! \sin(\vartheta-\psi)} &= \sum_{n=-\infty}^\infty {\rm e}^{\imag n ( \vartheta - \psi )} \biggl[ \bigl( v_\parallel \ex - v_{ds} \ey \bigr) J_n \nonumber\\*
\mbox{} &\quad+ v_\perp \biggl( \cos\psi \frac{nJ_n}{a_s} + \imag\sin\psi J'_n \biggr) \ey + v_\perp \biggl( \sin\psi \frac{nJ_n}{a_s} - \imag\cos\psi J'_n \biggr) \ez \biggr] \nonumber\\*
\mbox{} &= \sum_{n=-\infty}^\infty {\rm e}^{\imag n ( \vartheta - \psi )} \,\bb{u}_{n,s} ,
\end{align}
\begin{align}\label{eqn:wns}
\bb{v} \cos\vartheta\,{\rm e}^{\imag a_s\! \sin(\vartheta-\psi)} &= \sum_{n=-\infty}^\infty {\rm e}^{\imag n ( \vartheta - \psi )} \Biggl\{ \bigl( v_\parallel \ex - v_{ds} \ey \bigr) \biggl( \cos\psi \frac{nJ_n}{a_s} + \imag\sin\psi J'_n \biggr) \nonumber\\*
\mbox{} &\quad+ \frac{v_\perp}{2} \biggl[ J_n + \cos 2\psi \bigl( J_n + 2 J''_n \bigr) + 2\imag \sin 2\psi \biggl( \frac{nJ_n}{a_s} \biggr)^{\!\!\prime} \, \biggr] \ey \nonumber\\*
\mbox{} &\quad+ \frac{v_\perp}{2} \biggl[ \sin 2\psi \bigl( J_n + 2 J''_n \bigr) - 2\imag \cos 2\psi \biggl( \frac{nJ_n}{a_s} \biggr)^{\!\!\prime} \, \biggr] \ez \Biggr\} \nonumber\\*
\mbox{} &\doteq \sum_{n=-\infty}^\infty {\rm e}^{\imag n(\vartheta - \psi)} \, \bb{w}_{n,s} .
\end{align}
The final result is given by (\ref{eqn:sigma2}).

\section{Definitions and integrals related to the Maxwell- Boltzmann equilibrium (\ref{eqn:quasimaxwell})}\label{app:integrals}

To obtain (\ref{eqn:Maxdispersion}), several $v_\parallel$-, $v_\perp$-, and Bessel-function-weighted integrals over the Maxwell-Boltzmann equilibrium distribution function (\ref{eqn:quasimaxwell}) were performed using (\ref{eqn:plasmadisp}) and (\ref{eqn:watson}). In this Appendix, we provide explicit expressions for these integrals, in addition to the tensors $\msb{U}_{n,s}$ and $\msb{W}_{n,s}$ for which they were needed.

First, we note that the Landau-type integral (\ref{eqn:plasmadisp}) with $p=0$ is just the standard plasma dispersion function $Z(\zeta)$ first tabulated by \citet{fc61}. The next six orders ($p=1\dots 6$) can be written in terms of $Z(\zeta)$ as follows \citep[e.g.][]{gdg72}:
\begin{gather}\label{eqn:Zevaluated}
Z_1(\zeta) = 1 + \zeta Z(\zeta) , \quad Z_2(\zeta) = \zeta \bigl[ 1 + \zeta Z(\zeta) \bigr] , \quad Z_3(\zeta) = \frac{1}{2} + \zeta^2 \bigl[ 1 + \zeta Z(\zeta) \bigr] , \tag{\theequation {\it a,b,c}}\\*
Z_4(\zeta) = \zeta \biggl\{ \frac{1}{2} + \zeta^2 \bigl[ 1 + \zeta Z(\zeta) \bigr] \biggr\}, \quad Z_5(\zeta) = \frac{3}{4} + \zeta^2  \biggl\{ \frac{1}{2} + \zeta^2 \bigl[ 1 + \zeta Z(\zeta) \bigr] \biggr\} ,\tag{\theequation {\it d,e}} \\*
Z_6(\zeta) = \zeta \Biggl\{ \frac{3}{4} + \zeta^2 \biggl\{ \frac{1}{2} + \zeta^2 \bigl[ 1 + \zeta Z(\zeta) \bigr] \biggr\} \Biggr\} . \tag{\theequation {\it f}}
\end{gather}
The Bessel-function-weighted integrals over $v_\perp$ are more difficult to compute, as they require multiple differentiations of the \citet{watson66} relation (\ref{eqn:watson}) with respect to its three  parameters and some creative use of the recursion relations linking different orders $n$ of the modified Bessel function $I_n$. With $a_s \doteq k_\perp v_\perp / \Omega_s$, $\alpha_s \doteq ( k_\perp \rho_s)^2/2$, and $\Gamma_n(\alpha_s) \doteq I_n(\alpha_s) \exp(-\alpha_s)$, some particularly useful integrals of this type are
\begin{subequations}\label{eqn:GammaIntegrals1}
\begin{align}
\label{eqn:gkintegral_a}
\int^\infty_0 \frac{2 v_\perp {\rm d}v_\perp}{v^2_{ths}} & \bigl[ J_n(a_s) \bigr]^2 {\rm e}^{-v^2_\perp / v^2_{ths}} = \Gamma_n(\alpha_s) ,\\
\label{eqn:gkintegral_b}
\int^\infty_0 \frac{2 v_\perp {\rm d}v_\perp}{v^2_{ths}} & \bigl[ J_n(a_s) \bigr]^2  \frac{v^2_\perp}{v^2_{ths}}  {\rm e}^{-v^2_\perp / v^2_{ths}} = \bigl[ \alpha_s \Gamma_n(\alpha_s) \bigr]' ,\\
\int^\infty_0 \frac{2 v_\perp {\rm d}v_\perp}{v^2_{ths}} & \bigl[ J_n(a_s) \bigr]^2  \frac{v^4_\perp}{v^4_{ths}}  {\rm e}^{-v^2_\perp / v^2_{ths}} =  \bigl[ \alpha^2_s \Gamma_n(\alpha_s) \bigr]'' ,\\
\int^\infty_0 \frac{2 v_\perp {\rm d}v_\perp}{v^2_{ths}} & \bigl[ J_n(a_s) \bigr]^2  \frac{v^6_\perp}{v^6_{ths}}  {\rm e}^{-v^2_\perp / v^2_{ths}} =  \bigl[ \alpha^3_s \Gamma_n(\alpha_s) \bigr]''' ,
\end{align}
\end{subequations}
\begin{subequations}\label{eqn:GammaIntegrals2}
\begin{align}
\label{eqn:gkintegral_e}
\int^\infty_0 \frac{2 v_\perp {\rm d}v_\perp}{v^2_{ths}} & \frac{2J_n(a_s) J'_n(a_s)}{a_s} \frac{v^2_\perp}{v^2_{ths}}  {\rm e}^{-v^2_\perp / v^2_{ths}} = \bigl[ \Gamma_n(\alpha_s) \bigr]' ,\\
\label{eqn:gkintegral_f}
\int^\infty_0 \frac{2 v_\perp {\rm d}v_\perp}{v^2_{ths}} & \frac{2J_n(a_s) J'_n(a_s)}{a_s} \frac{v^4_\perp}{v^4_{ths}}  {\rm e}^{-v^2_\perp / v^2_{ths}} = \bigl[ \alpha_s \Gamma_n(\alpha_s) \bigr]'' ,\\
\int^\infty_0 \frac{2 v_\perp {\rm d}v_\perp}{v^2_{ths}} & \frac{2J_n(a_s) J'_n(a_s)}{a_s} \frac{v^6_\perp}{v^6_{ths}}  {\rm e}^{-v^2_\perp / v^2_{ths}} = \bigl[ \alpha^2_s \Gamma_n(\alpha_s) \bigr]''' ,
\end{align}
\end{subequations}
\begin{subequations}\label{eqn:GammaIntegrals3}
\begin{align}
\label{eqn:gkintegral_h}
\int^\infty_0 \frac{2 v_\perp {\rm d}v_\perp}{v^2_{ths}} & \bigl[ J'_n(a_s) \bigr]^2 \frac{v^2_\perp}{v^2_{ths}} {\rm e}^{-v^2_\perp / v^2_{ths}} = \frac{n^2}{2\alpha_s} \Gamma_n(\alpha_s) - \frac{1}{\alpha_s} \bigl[ \alpha^2_s \Gamma'_n(\alpha_s) \bigr] ,\\
\label{eqn:gkintegral_i}
\int^\infty_0 \frac{2 v_\perp {\rm d}v_\perp}{v^2_{ths}} & \bigl[ J'_n(a_s) \bigr]^2 \frac{v^4_\perp}{v^4_{ths}} {\rm e}^{-v^2_\perp / v^2_{ths}} = \frac{n^2}{2\alpha_s}\bigl[ \alpha_s \Gamma_n(\alpha_s) \bigr]' -\frac{1}{\alpha_s} \bigl[ \alpha^3_s \Gamma'_n(\alpha_s) ]' ,\\
\int^\infty_0 \frac{2 v_\perp {\rm d}v_\perp}{v^2_{ths}} & \bigl[ J'_n(a_s) \bigr]^2 \frac{v^6_\perp}{v^6_{ths}} {\rm e}^{-v^2_\perp / v^2_{ths}} = \frac{n^2}{2\alpha_s}\bigl[ \alpha^2_s \Gamma_n(\alpha_s) \bigr]'' -\frac{1}{\alpha_s} \bigl[ \alpha^4_s \Gamma'_n(\alpha_s) ]'' ,
\end{align}
\end{subequations}
\begin{subequations}\label{eqn:GammaIntegrals4}
\begin{align}
\int^\infty_0 \frac{2 v_\perp {\rm d}v_\perp}{v^2_{ths}} & 2J_n(a_s) J''_n(a_s) \frac{v^2_\perp}{v^2_{ths}} {\rm e}^{-v^2_\perp / v^2_{ths}} = \frac{1}{\alpha_s} \bigl[ \alpha^2_s \Gamma''_n(\alpha_s) - \alpha_s\Gamma_n(\alpha_s) \bigr] ,\\
\int^\infty_0 \frac{2 v_\perp {\rm d}v_\perp}{v^2_{ths}} & 2J_n(a_s) J''_n(a_s) \frac{v^4_\perp}{v^4_{ths}} {\rm e}^{-v^2_\perp / v^2_{ths}} = \frac{1}{\alpha_s} \bigl[ \alpha^3_s \Gamma''_n(\alpha_s) - \alpha^2_s \Gamma_n(\alpha_s) \bigr]' ,\\
\int^\infty_0 \frac{2 v_\perp {\rm d}v_\perp}{v^2_{ths}} & 2J_n(a_s) J''_n(a_s) \frac{v^6_\perp}{v^6_{ths}} {\rm e}^{-v^2_\perp / v^2_{ths}} = \frac{1}{\alpha_s} \bigl[ \alpha^4_s \Gamma''_n(\alpha_s) - \alpha^3_s \Gamma_n(\alpha_s) \bigr]'' ,
\end{align}
\end{subequations}
\begin{subequations}\label{eqn:GammaIntegrals5}
\begin{align}
\int^\infty_0 \frac{2 v_\perp {\rm d}v_\perp}{v^2_{ths}} & 2a_s J'_n(a_s)J''_n(a_s) \frac{v^2_\perp}{v^2_{ths}} {\rm e}^{-v^2_\perp / v^2_{ths}} = \frac{1}{\alpha_s} \bigl[ \alpha^2_s \Gamma'_n(\alpha_s) \bigr] + \bigl[ \alpha^2_s \Gamma''_n(\alpha_s) \bigr]' ,\\
\int^\infty_0 \frac{2 v_\perp {\rm d}v_\perp}{v^2_{ths}} & 2a_s J'_n(a_s)J''_n(a_s) \frac{v^4_\perp}{v^4_{ths}} {\rm e}^{-v^2_\perp / v^2_{ths}} = \frac{1}{\alpha_s} \bigl[ \alpha^3_s \Gamma'_n(\alpha_s) \bigr]' + \bigl[ \alpha^3_s \Gamma''_n(\alpha_s) \bigr]'' ,\\
\int^\infty_0 \frac{2 v_\perp {\rm d}v_\perp}{v^2_{ths}} & 2a_s J'_n(a_s)J''_n(a_s) \frac{v^6_\perp}{v^6_{ths}} {\rm e}^{-v^2_\perp / v^2_{ths}} = \frac{1}{\alpha_s} \bigl[ \alpha^4_s \Gamma'_n(\alpha_s) \bigr]''+ \bigl[ \alpha^4_s \Gamma''_n(\alpha_s) \bigr]''' .
\end{align}
\end{subequations}
Using these integrals, we find that the elements of the tensor
\begin{equation}\label{eqn:Utensor}
\msb{U}_{n,s} \doteq \frac{1}{n_s} \int{\rm d}^3\bb{v} \, \frac{\bb{u}_{n,s}\bb{u}^\ast_{n,s}}{v^2_{ths}} \frac{1}{v_\parallel / v_{ths} - \zeta_{n,s}} \left[ 1 + \frac{\omega_{Ts}}{\omega} \left(  \frac{5}{2} - \frac{v^2_\parallel + v^2_\perp}{v^2_{ths}} \right) \right] f_s(z,v_\parallel,v_\perp)
\end{equation}
are
\begin{subequations}
\begin{align}
\msb{U}^{(xx)}_{n,s} &= Z_2(\zeta_{n,s}) \Gamma_n(\alpha_s) \left\{ 1 + \frac{\omega_{Ts}}{\omega} \left[ \frac{3}{2} - \frac{Z_4(\zeta_{n,s})}{Z_2(\zeta_{n,s})} - \alpha_s\frac{\Gamma'_n(\alpha_s)}{ \Gamma_n(\alpha_s)}\right] \right\} , \\
\msb{U}^{(xy)}_{n,s} &= \left( \frac{n\cos\psi}{\sqrt{2\alpha_s}} - \frac{v_{ds}}{v_{ths}} \right)  Z_1(\zeta_{n,s}) \Gamma_n(\alpha_s) \left\{ 1 + \frac{\omega_{Ts}}{\omega} \left[ \frac{3}{2} - \frac{Z_3(\zeta_{n,s})}{Z_1(\zeta_{n,s})} - \alpha_s \frac{\Gamma'_n(\alpha_s)}{\Gamma_n(\alpha_s)} \right] \right\} \nonumber\\*
\mbox{} &\quad- \imag \frac{\sin\psi}{\sqrt{2\alpha_s}} Z_1(\zeta_{n,s}) \alpha_s \Gamma'_n(\alpha_s) \left\{ 1 + \frac{\omega_{Ts}}{\omega} \left[ \frac{1}{2} - \frac{Z_3(\zeta_{n,s})}{Z_1(\zeta_{n,s})} - \alpha_s \frac{\Gamma''_n(\alpha_s)}{\Gamma'_n(\alpha_s)} \right]  \right\} , \\
\msb{U}^{(xz)}_{n,s} &= \frac{n\sin\psi}{\sqrt{2\alpha_s}} Z_1(\zeta_{n,s}) \Gamma_n(\alpha_s) \left\{ 1 + \frac{\omega_{Ts}}{\omega} \left[ \frac{3}{2} - \frac{Z_3(\zeta_{n,s})}{Z_1(\zeta_{n,s})} - \alpha_s \frac{\Gamma'_n(\alpha_s)}{\Gamma_n(\alpha_s)} \right] \right\} \nonumber\\*
\mbox{} &\quad+ \imag \frac{\cos\psi}{\sqrt{2\alpha_s}} Z_1(\zeta_{n,s}) \alpha_s\Gamma'_n(\alpha_s) \left\{ 1 + \frac{\omega_{Ts}}{\omega} \left[ \frac{1}{2} - \frac{Z_3(\zeta_{n,s})}{Z_1(\zeta_{n,s})} - \alpha_s \frac{\Gamma''_n(\alpha_s)}{\Gamma'_n(\alpha_s)} \right] \right\} ,\\
\msb{U}^{(yy)}_{n,s} &=  \left[ \left( \frac{n\cos\psi}{\sqrt{2\alpha_s}} - \frac{v_{ds}}{v_{ths}} \right)^2 + \frac{n^2\sin^2\psi}{2\alpha_s} \right] Z_0(\zeta_{n,s}) \Gamma_n(\alpha_s) \nonumber\\*
\mbox{} &\qquad \times \left\{ 1 + \frac{\omega_{Ts}}{\omega} \left[ \frac{3}{2} - \frac{Z_2(\zeta_{n,s})}{Z_0(\zeta_{n,s})} - \alpha_s \frac{\Gamma'_n(\alpha_s)}{\Gamma_n(\alpha_s)} \right] \right\} \nonumber\\*
\mbox{} &\quad- \sin^2\psi Z_0(\zeta_{n,s}) \alpha_s \Gamma'_n(\alpha_s) \left\{ 1 + \frac{\omega_{Ts}}{\omega} \left[ -\frac{1}{2} - \frac{Z_2(\zeta_{n,s})}{Z_0(\zeta_{n,s})} - \alpha_s \frac{\Gamma''_n(\alpha_s)}{\Gamma'_n(\alpha_s)} \right] \right\} , \\
\msb{U}^{(yz)}_{n,s} &= \imag \left( \frac{n}{2\alpha_s} - \frac{\cos\psi}{\sqrt{2\alpha_s}} \frac{v_{ds}}{v_{ths}} \right) Z_0(\zeta_{n,s}) \alpha_s \Gamma'_n(\alpha_s) \nonumber \\
\mbox{} &\qquad \times \left\{ 1 + \frac{\omega_{Ts}}{\omega} \left[ \frac{1}{2} - \frac{Z_2(\zeta_{n,s})}{Z_0(\zeta_{n,s})} - \alpha_s \frac{\Gamma''_n(\alpha_s)}{\Gamma'_n(\alpha_s)} \right] \right\} \nonumber\\
\mbox{} &\quad- \frac{n\sin\psi}{\sqrt{2\alpha_s}} \frac{v_{ds}}{v_{ths}} Z_0(\zeta_{n,s}) \Gamma_n(\alpha_s) \left\{ 1 + \frac{\omega_{Ts}}{\omega} \left[ \frac{3}{2} - \frac{Z_2(\zeta_{n,s})}{Z_0(\zeta_{n,s})} - \alpha_s \frac{\Gamma'_n(\alpha_s)}{\Gamma_n(\alpha_s)} \right] \right\} \nonumber\\
\mbox{} &\quad+ \frac{1}{2} \sin2\psi Z_0(\zeta_{n,s}) \alpha_s \Gamma'_n(\alpha_s) \left\{ 1 + \frac{\omega_{Ts}}{\omega} \left[ -\frac{1}{2} - \frac{Z_2(\zeta_{n,s})}{Z_0(\zeta_{n,s})} - \alpha_s \frac{\Gamma''_n(\alpha_s)}{\Gamma'_n(\alpha_s)} \right] \right\} , \\
\msb{U}^{(zz)}_{n,s} &= \frac{n^2}{2\alpha_s} Z_0(\zeta_{n,s}) \Gamma_n(\alpha_s) \left\{ 1 + \frac{\omega_{Ts}}{\omega} \left[ \frac{3}{2} - \frac{Z_2(\zeta_{n,s})}{Z_0(\zeta_{n,s})} - \alpha_s \frac{\Gamma'_n(\alpha_s)}{\Gamma_n(\alpha_s)} \right] \right\} \nonumber\\
\mbox{} &\quad- \cos^2\psi Z_0(\zeta_{n,s}) \alpha_s \Gamma'_n(\alpha_s) \left\{ 1 + \frac{\omega_{Ts}}{\omega} \left[ -\frac{1}{2} - \frac{Z_2(\zeta_{n,s})}{Z_0(\zeta_{n,s})} - \alpha_s \frac{\Gamma''_n(\alpha_s)}{\Gamma'_n(\alpha_s)} \right] \right\} .
\end{align}
\end{subequations}
The un-written components of $\msb{U}_{n,s}$ (namely, $yx$, $zx$, $zy$) are the same as their transpose counterparts ($xy$, $xz$, $yz$, respectively), but with $\imag \rightarrow -\imag$. The $\msb{W}_{n,s}$ tensor, formally one order in $\rho_s/H$ smaller than $\msb{U}_{n,s}$, is much more cumbersome. Using $q_s {\rm d}\Phi_s/{\rm d}z = -T_s {\rm d}\ln P_s/{\rm d}z = {\rm const}$ to simplify some of the derivatives of $F_s(z,\varepsilon_s)$, we find that the tensor may be written in the relatively compact form
\begin{align}\label{eqn:Wtensor}
\msb{W}_{n,s} &\equiv \frac{\rho_s}{n_s} \int{\rm d}^3\bb{v}\, \frac{v_\perp}{v_{ths}} \frac{\bb{w}_{n,s}\bb{u}^\ast_{n,s}}{v^2_{ths}}\frac{1}{v_\parallel/v_{ths} - \zeta_{n,s}}  \nonumber\\*
\mbox{} &\quad\times \left( \pD{z}{} - \D{z}{\ln T_s} \right)  \left[ 1 +  \frac{\omega_{Ts}}{\omega} \left( \frac{5}{2} - \frac{v^2_\parallel+v^2_\perp}{v^2_{ths}} \right) \right] f_s(z,v_\parallel,v_\perp) ,
\end{align}
where $\bb{w}_{n,s}$ and $\bb{u}^\ast_{n,s}$ are defined via (\ref{eqn:wns}) and (\ref{eqn:ustarns}), respectively. Once expanded with the integrals performed, (\ref{eqn:Wtensor}) contributes a {\em lot} of terms. Fortunately, none of them enters into any of the limits we consider in the main text, and so we feel no pressing need to write out the components of $\msb{W}_{n,s}$ explicitly. A precocious reader may compute them using the integrals (B\,0)--(\ref{eqn:GammaIntegrals5}), several pens, and lots of paper.

\section{Derivation of (\ref{eqn:gkDdotE}) from the gyrokinetic theory of a thermally stratified atmosphere}\label{app:gk}

In this Appendix, we show how one may instead obtain (\ref{eqn:gkDdotE}) by linearizing the nonlinear electromagnetic gyrokinetic theory of a thermally stratified atmosphere. We begin by providing a derivation of that theory. While similar derivations may be found elsewhere \citep[e.g.][]{fc82,abel13}, our restriction to slab geometry removes many of the complications introduced by those authors' applications to axisymmetric tokamaks and makes the theory rather easier grasp, which hopefully gives this Appendix some pedagogical value. Our presentation also complements the derivation of ``astrophysical gyrokinetics'' for a homogeneous plasma found in the appendix of \citet{howes06}, whose traditional approach, based on an order-by-order expansion of the Vlasov equation, we adopt. However, we make no attempt to formulate `global' gyrokinetic equations that simultaneously describe the long and short scales in our model atmosphere, nor do we investigate the transport-time evolution of a gyrokinetic stratified atmosphere. Our aim here is simply to provide the reader with an alternative path to the linear gyrokinetic theory analyzed in Section \ref{sec:gyrokinetic}.

\subsection{Gyrokinetic Ordering}

The derivation of the nonlinear gyrokinetic theory begins by adopting the ordering
\begin{equation}\label{eqn:nonlinearGKordering}
\frac{\omega}{\Omega_s} \sim \frac{k_\parallel}{k_\perp} \sim \frac{\delta\bb{B}}{B_0} \sim \frac{\delta\bb{E}}{(v_{ths}/c)B_0} \sim \frac{\rho_s}{H} \doteq \epsilon \ll 1
\end{equation}
and expanding the distribution function in powers of $\epsilon$:
\begin{equation}
f_s = F_s + \delta f_s = F_{0s} + F_{1s} + F_{2s} + \dots + \delta f_{1s} + \delta f_{2s} + \dots ,
\end{equation}
where the subscript indicates the order in $\epsilon$. The parameters $\beta_s$ and $\tau$ are taken to be order unity in this expansion. We neglect any temporal evolution of the equilibrium quantities due to interactions with the fluctuating fields; these occur on the transport timescale, much longer than any timescale of interest in this paper. Note that (\ref{eqn:nonlinearGKordering}) gives $k_\perp \rho_s \sim 1$, and thus perpendicular spatial structure in the fluctuations is allowed on Larmor scales. The difference between the ordering (\ref{eqn:nonlinearGKordering}) and that used to reduce the linearized Vlasov equation in Section \ref{sec:gyrokinetic} ({\em viz}., equation (\ref{eqn:gkordering})) is that, here, while the amplitudes of the fluctuations are small, they are not {\em infinitesimally} small (relative to $\epsilon$). As a result, certain nonlinear effects (such as the advection of the perturbed distribution function by the fluctuating $\bb{E}\btimes\bb{B}$ drift) are retained. To avoid any confusion between linear and nonlinear quantities, we retain the subscript `0' labeling the equilibrium fields. 

Following the discussion in Section \ref{sec:equilibrium}, we take the equilibrium distribution function $F_s = F_{0s} + F_{1s} + F_{2s} + \dots$ to be a function of the guiding-centre position $\bb{R}_s$ (\ref{eqn:R}), the particle energy $\mc{E}_s$ (\ref{eqn:E}), and the gyrophase angle $\vartheta$. In the presence of electromagnetic fluctuations, these variables evolve according to
\begin{gather}\label{eqn:Rdot}
\D{t}{\bb{R}_s} = v_\parallel \ex - \frac{c}{B_0} \biggl[ \D{z}{\Phi_s} \ez + \grad\chi + \left( \pD{t}{} + \bb{v}\bcdot\grad \right) \frac{\bb{A}}{c} \biggr] \btimes \ex , \\*
\D{t}{\mc{E}_s} = q_s \biggl[ \pD{t}{\chi} -  \biggl( \pD{t}{} + \bb{v}\bcdot\grad \biggr)  \varphi \biggr] ,
\end{gather}
%
%
and ${\rm d}\vartheta / {\rm d}t = -\Omega_s + \mc{O}(\epsilon \Omega_s)$, where we have used (\ref{eqn:potentials}) to write the fluctuating electromagnetic fields in terms of the scalar potential $\varphi$, the vector potential $\bb{A}$, and the combination
\begin{equation}\label{eqn:gkpotential}
\chi \doteq \varphi - \frac{\bb{v}\bcdot\bb{A}}{c} .
\end{equation}
Equation (\ref{eqn:gkpotential}) is often referred to as the {\it gyrokinetic potential}, for reasons that will soon become clear. We also use the results of Section \ref{sec:drifts} and decompose the velocity into its parallel, perpendicular, and drift parts (see (\ref{eqn:vvec})):
\begin{equation}\label{eqn:vvecapp}
\bb{v} = v_\parallel \ex - v_{ds}\ey + v_\perp \bigl( \cos\vartheta\ey + \sin\vartheta\ez \bigr),
\end{equation}
where $v_{ds} = (c/B_0) {\rm d}\Phi_s / {\rm d}z$ is the equilibrium drift speed.

The formal expansion of (\ref{eqn:vlasov})--(\ref{eqn:ampere}) that results is worked out order by order. We begin with the Vlasov equation (\ref{eqn:vlasov}), ordered relative to $\omega f_{0s}$.

\subsection{Gyrokinetic Equation}

\subsubsection{Minus-first order, $\mc{O}(1/\epsilon)$}

The largest term in (\ref{eqn:vlasov}) corresponds physically to Larmor motion of the (zeroth-order) equilibrium particle distribution about the guiding centre:
\begin{equation}\label{eqn:minusfirst}
-\Omega_s \!\left.\pD{\vartheta}{F_{0s}} \right|_{\bs{R}_s} = 0.
\end{equation}
Thus, $F_{0s}$ is independent of gyrophase (gyrotropic): $F_{0s} = F_{0s}(\bb{R}_s,\mc{E}_s)$.\footnote{At this order, there is no difference between the particle position and the guiding-centre position, since the equilibrium drifts are $\sim$$\epsilon v_{ths}$, and so $F_{0s}$ is gyrotropic at both fixed particle position and fixed guiding-centre position. At higher orders, however, there is a difference.}

\subsubsection{Zeroth order, $\mc{O}(1)$}\label{sec:gkzeroth}

Proceeding to next order and decomposing the velocity into its parallel, perpendicular, and drift parts (equation (\ref{eqn:vvecapp})), equation (\ref{eqn:vlasov}) becomes
\begin{align}\label{eqn:zeroth}
\bb{v}_\perp\bcdot\grad_\perp \delta f_{1s} &- \Omega_s \!\left. \pD{\vartheta}{\delta f_{1s}} \right|_{\bs{r}} - \Omega_s \!\left.\pD{\vartheta}{F_{1s}}\right|_{\bs{R}_s} + v_\parallel \pD{x}{F_{0s}} \nonumber\\*
\mbox{} &= q_s \pD{\mc{E}_s}{F_{0s}} \bb{v}_\perp\bcdot\grad_\perp\varphi = \bb{v}_\perp\bcdot\grad_\perp \left( q_s \varphi \pD{\mc{E}_s}{F_{0s}} \right) + \mc{O}(\epsilon^2 F_{0s}).
\end{align}
Equation (\ref{eqn:zeroth}) contains both fluctuating and mean quantities. To isolate the latter, we average over space to obtain
\begin{equation}
- \Omega_s \!\left.\pD{\vartheta}{F_{1s}}\right|_{\bs{R}_s} + v_\parallel \pD{x}{F_{0s}} = 0 ;
\end{equation}
further integrating over gyrophase and using (\ref{eqn:minusfirst}), we find that $F_{0s}$ is independent of the $x$ coordinate (as promised in \S\ref{sec:equilibrium}). As a result, $F_{1s}$, like $F_{0s}$ before it, is gyrotropic: $F_{1s} = F_{1s}(\bb{R}_s,\mc{E}_s)$.

Returning to (\ref{eqn:zeroth}) and using our newfound knowledge about the equilibrium distribution function, the first-order perturbed distribution function is related to $F_{0s}$ via
\begin{equation}\label{eqn:zeroth2}
\bb{v}_\perp\bcdot\grad_\perp \delta f_{1s} - \Omega_s \!\left. \pD{\vartheta}{\delta f_{1s}} \right|_{\bs{r}} = \bb{v}_\perp\bcdot\grad_\perp \left( q_s \varphi \pD{\mc{E}_s}{F_{0s}} \right) .
\end{equation}
Equation (\ref{eqn:zeroth2}) admits a homogeneous solution and a particular solution. The homogeneous solution $h_s$ satisfies
\begin{equation}
\bb{v}_\perp\bcdot\grad_\perp h_s - \Omega_s \!\left.\pD{\vartheta}{h_s} \right|_{\bs{r}} = -\Omega_s \!\left.\pD{\vartheta}{h_s} \right|_{\gas} = 0,
\end{equation}
where we have transformed the $\vartheta$ derivative taken at constant position $\bb{r}$ to one taken at constant guiding centre (see (\ref{eqn:R})). Thus, $h_s$ is independent of the gyrophase angle at constant guiding centre $\bb{R}_s$ (but not at constant position $\bb{r}$):
\begin{equation}\label{eqn:gkresponse}
h_s = h_s(t,\bb{R}_s,v_\parallel,v_\perp) .
\end{equation}
It represents the response of charged rings to the perturbed fields, and is thus referred to as the {\it gyrokinetic response}. The particular solution of (\ref{eqn:zeroth2}) -- the so-called adiabatic, or `Boltzmann', response -- is given by
\begin{equation}\label{eqn:boltzmannresponse}
\delta f_{1s,Boltz} = q_s \varphi \pD{\mc{E}_s}{F_{0s}} .
\end{equation}
It arises from the evolution of $F_{0s}$ under the influence of the perturbed electromagnetic fields. Indeed, 
\begin{equation}\label{eqn:newF0}
F_{0s}(\bb{R}_s,\mc{E}_s) + \delta f_{1s,Boltz} = F_{0s}(\bb{R}_s, \mc{E}_s + q_s \varphi ) + \mc{O}(\epsilon^2 F_{0s}) ;
\end{equation}
i.e.~the Boltzmann response does not change the form of the equilibrium distribution function if the latter is written as a function of sufficiently precisely conserved particle invariants.

At this point in the derivation, we {\em assume} that $F_{0s}$ is independent of the $y$ coordinate, i.e.~$F_{0s} = F_{0s}(Z_s,\mc{E}_s)$, where $Z_s$ is the position of the guiding centre projected onto the $z$ axis. At this order, we may replace $Z_s$ with $\mc{Z}_s$ (see (\ref{eqn:Z2})), and fold their $\sim$$\mc{O}(\rho_s/H)$ difference into $F_{1s}$. Combining (\ref{eqn:gkresponse}) and (\ref{eqn:newF0}), the distribution function of species $s$ may thus be written
\begin{equation}\label{eqn:fexpanded}
f_s = F_{0s}(\mc{Z}_s,\mc{E}_s+q_s\varphi) + F_{1s}(\bb{R}_s,\mc{E}_s) + F_{2s} + \dots + h_s(t,\bb{R}_s,v_\parallel,v_\perp) + \delta f_{2s} + \dots ,
\end{equation}
where we have absorbed the $\mc{O}(\epsilon^2 F_{0s})$ terms into $\delta f_{2s}$. Next we derive an evolution equation for the gyrokinetic response $h_s$.

\subsubsection{First Order, $\mc{O}(\epsilon)$}\label{sec:gkfirst}

Collecting $\mc{O}(\epsilon)$ terms in (\ref{eqn:vlasov}) and decomposing $\delta f_{1s}$ into its adiabatic (\ref{eqn:boltzmannresponse}) and non-adiabatic (\ref{eqn:gkresponse}) parts \citep[cf.][]{al80,ctb81}, we find that the gyrokinetic response evolves according to 
\begin{align}\label{eqn:first}
\pD{t}{h_s} &+ \D{t}{\bb{R}_s} \bcdot \pD{\bb{R}_s}{h_s} + q_s \pD{t}{\chi} \pD{\mc{E}_s}{F_{0s}}  + \frac{c}{B_0} \pD{y}{\chi}\pD{\mc{Z}_s}{F_{0s}} = - v_\parallel \pD{x}{F_{1s}} + \Omega_s \!\left. \pD{\vartheta}{(F_{2s} + \delta f_{2s})} \right|_{\gas} \nonumber\\*
\mbox{} & - \frac{q_s}{m_s} \left( - \grad_\perp\varphi + \frac{\bb{v}\btimes\delta\bb{B}}{c} \right)\bcdot\pD{\bb{v}}{} ( \delta f_{1s} + F_{1s} ) - \pD{\mc{Z}_s}{F_{0s}}\bb{v}_\perp\bcdot\grad_\perp \frac{A_y}{B_0},
\end{align}
where
\begin{equation}\label{eqn:R0dot}
\D{t}{\bb{R}_s} = v_\parallel \ex - v_{ds} \ey + \frac{c}{B_0} \left( - \grad_\perp\varphi + \frac{\bb{v}\btimes\delta\bb{B}}{c} \right) \btimes\ex
\end{equation}
is the velocity of the guiding centre required at this order (cf.~\ref{eqn:Rdot}). Upon performing a ring average of (\ref{eqn:first}) over $\vartheta$ at fixed $\bb{R}_s$, defined, for any function $a(t,\bb{r},\bb{v})$, as
\begin{equation}\label{eqn:ringaverage}
\langle a(t,\bb{r},\bb{v})\rangle_{\gas} \doteq \frac{1}{2\upi}\oint{\rm d}\vartheta\, a \biggl(t,\bb{R}_s - \frac{\bb{v}\btimes\ex}{\Omega_s},\bb{v}\biggr),
\end{equation}
we find that the entire right-hand side of (\ref{eqn:first}) vanishes. This follows from the periodicity of $F_{2s}$ and $\delta f_{2s}$ in $\vartheta$, the gyrotropy of $F_{0s}$ and $F_{1s}$ at fixed guiding center, and from the fact that, for any arbitrary function $a(\bb{r})$, the ring average $\langle\bb{v}_\perp\bcdot\grad_\perp a\rangle_{\gas} = 0$ (see equation (A21) of \citet{howes06}). Thus, the ring-averaged (\ref{eqn:first}) is
\begin{equation}\label{eqn:gk1}
\pD{t}{h_s} + \biggl\langle \D{t}{\bb{R}_s} \biggr\rangle_{\ggs} \!\bcdot \pD{\bb{R}_s}{h_s} = -q_s \pD{t}{\langle\chi\rangle_{\gas}} \pD{\varepsilon_s}{F_{0s}} - \frac{c}{B_0} \pD{y}{\langle\chi\rangle_{\gas}} \pD{\mc{Z}_s}{F_{0s}} .
\end{equation}
Using the decomposition $\delta\bb{B} = \grad A_\parallel \btimes \ex + \delta B_\parallel \ex$ in (\ref{eqn:Rdot}) and retaining only first-order contributions, the ring-averaged guiding-centre velocity (\ref{eqn:R0dot}) is
\begin{subequations}\label{eqn:gaRdot}
\begin{align}\label{eqn:gaRdota}
\biggl\langle \D{t}{\bb{R}_s} \biggr\rangle_{\ggs}\! &= v_\parallel \ex - v_{ds} \ey - \frac{c}{B_0} \langle\grad_\perp\varphi\rangle_{\gas} \btimes \ex + \frac{v_\parallel}{B_0} \langle\grad_\perp A_\parallel \rangle_{\gas} - \frac{1}{B_0} \langle\bb{v}_\perp\delta B_\parallel \rangle_{\gas} \\*
\label{eqn:gaRdotb}
\mbox{} &= v_\parallel \ex - v_{ds} \ey - \frac{c}{B_0} \pD{\bb{R}_s}{\langle\chi\rangle_{\gas}} \btimes \ex ,
\end{align}
\end{subequations}
where, to obtain (\ref{eqn:gaRdotb}), we have used the identity $\langle\bb{v}_\perp \delta B_\parallel\rangle_{\gas} = -\langle\grad_\perp ( \bb{v}_\perp\bcdot\bb{A}_\perp)\rangle_{\gas}$. Each of the terms in (\ref{eqn:gaRdota}) has a straightforward physical interpretation. The first term represents the streaming of guiding centres along the (uniform) background magnetic field. Combined with the fourth term in (\ref{eqn:gaRdota}), which accounts for the motion along the ring-averaged fluctuating magnetic field, we see that the guiding centres stream along the ring-averaged total (perturbed) magnetic field:
\begin{equation}
\biggl[ v_\parallel \ex + \frac{v_\parallel}{B_0} \langle \grad_\perp A_\parallel\rangle_{\gas} \biggl] \bcdot \pD{\bb{R}_s}{h_s} = \left\langle v_\parallel \frac{\bb{B}}{B_0} \right\rangle_{\ggs} \!\bcdot\pD{\bb{R}_s}{h_s} .
\end{equation}
The second term in (\ref{eqn:gaRdota}) captures the (slow) $-\grad\Phi_s\btimes\bb{B}$ drift across the magnetic-field lines. The third term is due to the ring-averaged fluctuating $\bb{E}\btimes\bb{B}$ drift. Finally, the fourth term represents the $\grad B$ drift; if we expand the ring average to lowest order in $\bb{v}\btimes\ex/\Omega_s$, we obtain the familiar drift velocity
\begin{equation}
- \frac{1}{B_0} \langle\bb{v}_\perp \delta B_\parallel \rangle_{\gas} \simeq - \frac{\mu_s \grad B \btimes \ex}{m_s \Omega_s} ,
\end{equation}
where $\grad B = \grad \delta B_\parallel$ is evaluated at the center of the ring, $\bb{r} = \bb{R}_s$.

Substituting (\ref{eqn:gaRdot}) into (\ref{eqn:gk1}), we finally obtain the {\it gyrokinetic equation}
\begin{equation}\label{eqn:gkeqn}
\left( \pD{t}{} + v_\parallel \pD{x}{} - v_{ds} \pD{y}{} \right) h_s = - q_s \pD{t}{\langle\chi\rangle_{\gas}} \pD{\mc{E}_s}{F_{0s}} - \frac{c}{B_0} \{ \langle\chi\rangle_{\gas} , F_{0s} + h_s \} ,
\end{equation}
where the Poisson bracket is defined by $\{ U,V \} = \ex \bcdot \{ (\partial U/\partial \bb{R}_s)\btimes(\partial V/\partial \bb{R}_s) \}$. In order to solve (\ref{eqn:gkeqn}), we must relate the electromagnetic potentials to the gyrokinetic response. This is accomplished via the field equations.

\subsection{Field equations}

To obtain equations governing the electromagnetic potentials, we substitute (\ref{eqn:fexpanded}) into the leading-order expansions of the quasineutrality constraint (\ref{eqn:quasineutrality}) and Amp\`{e}re's law (equation (\ref{eqn:ampere})). To first order, $\mc{O}(\varepsilon)$, (\ref{eqn:quasineutrality}) becomes
\begin{equation}\label{eqn:gkqn1}
0 = \sum_s q_s \delta n_s = \sum_s q_s \left[ \int{\rm d}^3\bb{v} \, h_s \biggl( t, \bb{r} + \frac{\bb{v}\btimes\ex}{\Omega_s}, v_\parallel, v_\perp \biggr) - n_s \frac{q_s \varphi}{T_s}  \right] .
\end{equation}
Because the electrostatic potential $\varphi$ is a function of the spatial variable $\bb{r}$, the velocity integral of $h_s$ must be performed at constant location $\bb{r}$ of the charges rather than at constant guiding centre $\bb{R}_s$. This introduces a gyro-averaging operation dual to the ring averaged defined in (\ref{eqn:ringaverage}):
\begin{equation}\label{eqn:gyroaverage}
\langle h_s (t,\bb{R}_s,v_\parallel,v_\perp) \rangle_{\bs{r}} \doteq \frac{1}{2\upi} \oint{\rm d}\vartheta \, h_s \biggl(t,\bb{r}+\frac{\bb{v}\btimes\ex}{\Omega_s},v_\parallel,v_\perp\biggr).
\end{equation}
Equation (\ref{eqn:gkqn1}) may then be written as
\begin{equation}\label{eqn:gkqn}
0 = \sum_s q_s \left[ \int{\rm d}^3\bb{v} \, \langle h_s\rangle_{\bs{r}} - n_s \frac{q_s \varphi}{T_s}  \right] .
\end{equation}
Likewise, the parallel and perpendicular components of Amp\`{e}re's law become, respectively,
\begin{gather}\label{eqn:prlampere}
\nabla^2_\perp A_\parallel = - \frac{4\upi}{c} j_\parallel = - \frac{4\upi}{c} \sum_s q_s \int{\rm d}^3\bb{v}\, v_\parallel \langle h_s \rangle_{\bs{r}} , \\*
\label{eqn:prpampere}
\nabla^2_\perp \delta B_\parallel = - \frac{4\upi}{c} \ex\bcdot(\grad_\perp\btimes\bb{j}_\perp) = -\frac{4\upi}{c}\ex\bcdot \left[ \grad_\perp\btimes\sum_s q_s \int{\rm d}^3\bb{v}\, \langle\bb{v}_\perp h_s \rangle_{\bs{r}} \right] .
\end{gather}
Together with the gyrokinetic equation (\ref{eqn:gkeqn}), the field equations (\ref{eqn:gkqn})--(\ref{eqn:prpampere}) provide a closed system that describes the nonlinear evolution of a gyrokinetic stratified atmosphere. 

\subsection{Fourier space}\label{sec:gkfourier}

Our next task is to linearize (\ref{eqn:gkeqn}) and substitute the resulting expression for the gyrokinetic response into the field equations. Before doing so, let us calculate the Fourier representation of the ring-averaging (\ref{eqn:ringaverage}) and gyro-averaging (\ref{eqn:gyroaverage}) operations. Decomposing the gyrokinetic potential into plane waves, $\chi(t,\bb{r},\bb{v}) = \sum_{\bs{k}} \chi_{\bs{k}} \exp(\imag\bb{k}\bcdot\bb{r})$, the ring average of its Fourier coefficient is
\begin{align}\label{eqn:gkpotentialk}
\langle\chi_{\bs{k}}(t,\bb{v})\rangle_{\gas} &= \frac{1}{2\upi} \oint{\rm d}\vartheta \left( \varphi_{\bs{k}} - \frac{v_\parallel A_{\parallel\bs{k}}}{c} - \frac{\bb{v}_\perp\bcdot\bb{A}_{\perp\bs{k}}}{c} \right) \exp\left( -\imag\bb{k}\bcdot \frac{\bb{v}_\perp\btimes\ex}{\Omega_s} \right) \nonumber\\*
\mbox{} &= {\rm J}_0(a_s) \left( \varphi_{\bs{k}} - \frac{v_\parallel A_{\parallel\bs{k}}}{c} \right) + \frac{T_s}{q_s} \frac{2v^2_\perp}{v^2_{ths}} \frac{{\rm J}_1(a_s)}{a_s} \frac{\delta B_{\parallel\bs{k}}}{B_0} ,
\end{align}
where $a_s \doteq k_\perp v_\perp / \Omega_s$ is the argument of the zeroth- (${\rm J}_0$) and first-order (${\rm J}_1$) Bessel functions. Similarly, we can Fourier-decompose $h_s(t,\bb{R}_s,v_\parallel,v_\perp) = \sum_{\bs{k}} h_{s\bs{k}}(t,v_\parallel,v_\perp) \exp(\imag\bb{k}\bcdot\bb{R}_s)$ and perform its gyro-averages at constant $\bb{r}$, {\it viz.}
\begin{align}
\langle h_{s\bs{k}} (t,v_\parallel,v_\perp)\rangle_{\bs{r}} &= \frac{1}{2\upi} \oint{\rm d}\vartheta \, h_{s\bs{k}} \exp\left( \imag \bb{k}\bcdot \frac{\bb{v}_\perp\btimes\ex}{\Omega_s} \right) = {\rm J}_0(a_s) h_{s\bs{k}} , \\*
\langle \bb{v}_\perp h_{s\bs{k}} (t,v_\parallel,v_\perp) \rangle_{\bs{r}} &= \frac{1}{2\upi} \oint{\rm d}\vartheta \, \bb{v}_\perp h_{s\bs{k}} \exp\left( \imag \bb{k}\bcdot \frac{\bb{v}_\perp\btimes\ex}{\Omega_s} \right) \nonumber\\*
\mbox{} &= -\imag\bb{k}\btimes\ex \frac{v^2_\perp}{\Omega_s} \frac{{\rm J}_1(a_s)}{a_s} h_{s\bs{k}}
\end{align}
Using these formulae, the field equations (\ref{eqn:gkqn})--(\ref{eqn:prpampere}) expressed in Fourier space are
\begin{gather}\label{eqn:gkqnk}
\sum_s \frac{q^2_s n_{0s}}{T_{0s}} \varphi_{\bs{k}} = \sum_s q_s \int{\rm d}^3\bb{v}\, {\rm J}_0(a_s) h_{s\bs{k}} , \\*
\frac{k^2_\perp c^2}{4\upi} \frac{A_{\parallel\bs{k}}}{c} = \sum_s q_s \int{\rm d}^3\bb{v}\, v_\parallel {\rm J}_0(a_s)  h_{s\bs{k}} , \\*
\label{eqn:gkampk}
- \frac{\delta B_{\parallel\bs{k}}}{B_0} = \sum_s \frac{\beta_s}{n_{0s}} \int{\rm d}^3\bb{v} \, \frac{v^2_\perp}{v^2_{ths}} \frac{{\rm J}_1(a_s)}{a_s} h_{s\bs{k}} .
\end{gather}
We now derive the linear gyrokinetic theory.

\subsection{Linear theory}

Returning to the gyrokinetic equation (\ref{eqn:gkeqn}), we drop the nonlinear term $\propto$$\{\langle\chi\rangle_{\gas},h_s\}$ and adopt plane-wave solutions to find
\begin{equation}\label{eqn:hsk}
h_{s\bs{k}} = \frac{\omega}{\omega + k_y v_{ds} - k_\parallel v_\parallel }  \frac{q_s\langle\chi_{\bs{k}}\rangle_{\gas}}{T_s} \left( - T_s \pD{\mc{E}_s}{}  + \frac{k_y}{\omega} \frac{cT_s}{q_sB} \pD{\mc{Z}_s}{} \right) F_s(\mc{Z}_s,\mc{E}_s) ,
\end{equation}
where we have safely omitted the `0' subscript adorning the equilibrium quantities. (This is essentially equation 24 of \citealt{al80}.) At this order in the gyrokinetic expansion, we may replace the phase-space variables $(\mc{Z}_s,\mc{E}_s)$ by their leading-order counterparts $(z,\varepsilon_s)$ (see equation (\ref{eqn:varepsilon}) and related discussion in \S\ref{sec:orbitintegral}) and use (\ref{eqn:dFM}), valid for a Maxwell-Boltzmann equilibrium distribution, to evaluate the phase-space derivatives of $F_s$ in (\ref{eqn:hsk}). Substituting the resulting expression for $h_{s\bs{k}}$ into the field equations (\ref{eqn:gkqnk})--(\ref{eqn:gkampk}) and recalling the definitions of the temperature-gradient and diamagnetic drift frequencies,
\[
\omega_{Ts} \doteq -k_y \frac{cT_s}{q_s B} \D{z}{\ln T_s} \quad{\rm and}\quad \omega_{Ps} \doteq - k_y \frac{cT_s}{q_s B} \D{z}{\ln P_s} ,
\]
we have
\begin{gather}\label{eqn:gkfieldqn}
\sum_s \frac{q^2_s n_s}{T_s} \varphi_{\bs{k}} = \sum_s \frac{q^2_s}{T_s} \int{\rm d}^3\bb{v}\, \frac{\omega  {\rm J}_0(a_s) \langle\chi_{\bs{k}}\rangle_{\gas}}{\omega + k_y v_{ds} - k_\parallel v_\parallel } \left[ 1 + \frac{\omega_{Ts}}{\omega} \left( \frac{5}{2} - \frac{v^2_\parallel + v^2_\perp}{v^2_{ths}} \right) \right] f_{M,s} , \\*
\frac{k^2_\perp c^2}{4\upi} \frac{A_{\parallel\bs{k}}}{c} = \sum_s \frac{q^2_s}{T_s} \int{\rm d}^3\bb{v}\, v_\parallel  \frac{\omega {\rm J}_0(a_s) \langle\chi_{\bs{k}}\rangle_{\gas}}{\omega + k_y v_{ds} - k_\parallel v_\parallel } \left[ 1 + \frac{\omega_{Ts}}{\omega} \left( \frac{5}{2} - \frac{v^2_\parallel + v^2_\perp}{v^2_{ths}} \right) \right] f_{M,s} , \\*
\label{eqn:gkfieldamp}
- \frac{k^2_\perp c^2}{4\upi} \frac{\delta B_{\parallel\bs{k}}}{k_\perp c} = \sum_s  \frac{q^2_s}{T_s} \int{\rm d}^3\bb{v} \, v_\perp \frac{\omega {\rm J}_1(a_s) \langle\chi_{\bs{k}}\rangle_{\gas}}{\omega + k_y v_{ds} - k_\parallel v_\parallel }  \left[ 1 + \frac{\omega_{Ts}}{\omega} \left( \frac{5}{2} - \frac{v^2_\parallel + v^2_\perp}{v^2_{ths}} \right) \right] f_{M,s} ,
\end{gather}
with $f_{M,s} = f_{M,s}(z,v_\parallel,v_\perp)$ given by (\ref{eqn:fMs}). The final step in the derivation of the linear gyrokinetic theory is to replace the ring-averaged gyrokinetic potential in (\ref{eqn:gkfieldqn})--(\ref{eqn:gkfieldamp}) by (\ref{eqn:gkpotentialk}) and perform the resulting integrals using (\ref{eqn:plasmadisp}), (\ref{eqn:Zevaluated}{\it ,d}), and (\ref{eqn:GammaIntegrals1}{\it a,b}), (\ref{eqn:GammaIntegrals2}{\it a,b}), (\ref{eqn:GammaIntegrals3}{\it a}). This leads identically to (\ref{eqn:gkDdotE}), as promised.

\bibliographystyle{jpp}
\bibliography{XK16_postreview}

\begin{thebibliography}{71}
\expandafter\ifx\csname natexlab\endcsname\relax\def\natexlab#1{#1}\fi

\bibitem[{Abel} {\em et~al.\/}(2013){Abel}, {Plunk}, {Wang}, {Barnes},
  {Cowley}, {Dorland} \& {Schekochihin}]{abel13}
{\sc {Abel}, I.~G., {Plunk}, G.~G., {Wang}, E., {Barnes}, M., {Cowley}, S.~C.,
  {Dorland}, W. \& {Schekochihin}, A.~A.} 2013 {Multiscale gyrokinetics for
  rotating tokamak plasmas: fluctuations, transport and energy flows}. {\em
  Reports on Progress in Physics\/} {\bf 76}~(11), 116201.

\bibitem[{Antonsen} \& {Lane}(1980)]{al80}
{\sc {Antonsen}, Jr., T.~M. \& {Lane}, B.} 1980 {Kinetic equations for low
  frequency instabilities in inhomogeneous plasmas}. {\em Physics of Fluids\/}
  {\bf 23}, 1205--1214.

\bibitem[{Avara} {\em et~al.\/}(2013){Avara}, {Reynolds} \&
  {Bogdanovi{\'c}}]{avara13}
{\sc {Avara}, M.~J., {Reynolds}, C.~S. \& {Bogdanovi{\'c}}, T.} 2013 {Role of
  Magnetic Field Strength and Numerical Resolution in Simulations of the
  Heat-flux-driven Buoyancy Instability}. {\em \apj\/} {\bf 773}, 171.

\bibitem[{Balbus}(2000)]{balbus00}
{\sc {Balbus}, S.~A.} 2000 {Stability, Instability, and ``Backward'' Transport
  in Stratified Fluids}. {\em \apj\/} {\bf 534}, 420--427.

\bibitem[{Balbus}(2001)]{balbus01}
{\sc {Balbus}, S.~A.} 2001 {Convective and Rotational Stability of a Dilute
  Plasma}. {\em \apj\/} {\bf 562}, 909--917.

\bibitem[{Barnes}(1966)]{barnes66}
{\sc {Barnes}, A.} 1966 {Collisionless Damping of Hydromagnetic Waves}. {\em
  \pof\/} {\bf 9}, 1483--1495.

\bibitem[{Bogdanovi{\'c}} {\em et~al.\/}(2009){Bogdanovi{\'c}}, {Reynolds},
  {Balbus} \& {Parrish}]{bogdanovic09}
{\sc {Bogdanovi{\'c}}, T., {Reynolds}, C.~S., {Balbus}, S.~A. \& {Parrish},
  I.~J.} 2009 {Simulations of Magnetohydrodynamics Instabilities in
  Intracluster Medium Including Anisotropic Thermal Conduction}. {\em \apj\/}
  {\bf 704}, 211--225.

\bibitem[{Braginskii}(1965)]{braginskii65}
{\sc {Braginskii}, S.~I.} 1965 {Transport Processes in a Plasma}. {\em Reviews
  of Plasma Physics\/} {\bf 1}, 205.

\bibitem[{Brunt}(1927)]{brunt27}
{\sc {Brunt}, D.} 1927 {The period of simple vertical oscillations in the
  atmosphere}. {\em Quart.~J.~Roy.~Meteor.~Soc.\/} {\bf 53}, 30.

\bibitem[{Carilli} \& {Taylor}(2002)]{ct02}
{\sc {Carilli}, C.~L. \& {Taylor}, G.~B.} 2002 {Cluster Magnetic Fields}. {\em
  \araa\/} {\bf 40}, 319--348.

\bibitem[{Catto} \& {Simakov}(2004)]{cs04}
{\sc {Catto}, P.~J. \& {Simakov}, A.~N.} 2004 {A drift ordered short mean free
  path description for magnetized plasma allowing strong spatial anisotropy}.
  {\em Physics of Plasmas\/} {\bf 11}, 90--102.

\bibitem[{Catto} {\em et~al.\/}(1981){Catto}, {Tang} \& {Baldwin}]{ctb81}
{\sc {Catto}, P.~J., {Tang}, W.~M. \& {Baldwin}, D.~E.} 1981 {Generalized
  gyrokinetics}. {\em Plasma Physics\/} {\bf 23}, 639--650.

\bibitem[{Chew} {\em et~al.\/}(1956){Chew}, {Goldberger} \& {Low}]{cgl56}
{\sc {Chew}, G.~F., {Goldberger}, M.~L. \& {Low}, F.~E.} 1956 {The Boltzmann
  Equation and the One-Fluid Hydromagnetic Equations in the Absence of Particle
  Collisions}. {\em Proceedings of the Royal Society of London Series A\/} {\bf
  236}, 112--118.

\bibitem[{Coppi} {\em et~al.\/}(1967){Coppi}, {Rosenbluth} \& {Sagdeev}]{crs67}
{\sc {Coppi}, B., {Rosenbluth}, M.~N. \& {Sagdeev}, R.~Z.} 1967 {Instabilities
  due to Temperature Gradients in Complex Magnetic Field Configurations}. {\em
  Physics of Fluids\/} {\bf 10}, 582--587.

\bibitem[{Cowley} {\em et~al.\/}(1991){Cowley}, {Kulsrud} \& {Sudan}]{cks91}
{\sc {Cowley}, S.~C., {Kulsrud}, R.~M. \& {Sudan}, R.} 1991 {Considerations of
  ion-temperature-gradient-driven turbulence}. {\em Physics of Fluids B\/} {\bf
  3}, 2767--2782.

\bibitem[{Dimits} {\em et~al.\/}(2000){Dimits}, {Bateman}, {Beer}, {Cohen},
  {Dorland}, {Hammett}, {Kim}, {Kinsey}, {Kotschenreuther}, {Kritz}, {Lao},
  {Mandrekas}, {Nevins}, {Parker}, {Redd}, {Shumaker}, {Sydora} \&
  {Weiland}]{dimits00}
{\sc {Dimits}, A.~M., {Bateman}, G., {Beer}, M.~A., {Cohen}, B.~I., {Dorland},
  W., {Hammett}, G.~W., {Kim}, C., {Kinsey}, J.~E., {Kotschenreuther}, M.,
  {Kritz}, A.~H., {Lao}, L.~L., {Mandrekas}, J., {Nevins}, W.~M., {Parker},
  S.~E., {Redd}, A.~J., {Shumaker}, D.~E., {Sydora}, R. \& {Weiland}, J.} 2000
  {Comparisons and physics basis of tokamak transport models and turbulence
  simulations}. {\em Physics of Plasmas\/} {\bf 7}, 969--983.

\bibitem[{Dorland} {\em et~al.\/}(2000){Dorland}, {Jenko}, {Kotschenreuther} \&
  {Rogers}]{dorland00}
{\sc {Dorland}, W., {Jenko}, F., {Kotschenreuther}, M. \& {Rogers}, B.~N.} 2000
  {Electron Temperature Gradient Turbulence}. {\em Physical Review Letters\/}
  {\bf 85}, 5579--5582.

\bibitem[{Fried} \& {Conte}(1961)]{fc61}
{\sc {Fried}, B.~D. \& {Conte}, S.~D.} 1961 {\em {The Plasma Dispersion
  Function}\/}. New York: Academic Press, 1961.

\bibitem[{Frieman} \& {Chen}(1982)]{fc82}
{\sc {Frieman}, E.~A. \& {Chen}, L.} 1982 {Nonlinear gyrokinetic equations for
  low-frequency electromagnetic waves in general plasma equilibria}. {\em
  Physics of Fluids\/} {\bf 25}, 502--508.

\bibitem[{Goodrich} {\em et~al.\/}(1972){Goodrich}, {Durney} \& {Grow}]{gdg72}
{\sc {Goodrich}, L.~C., {Durney}, C.~H. \& {Grow}, R.~W.} 1972 {General
  Recurrence Relation for Use in Evaluating Moments of the Integrand of the
  Plasma Dispersion Function}. {\em Physics of Fluids\/} {\bf 15}, 715--716.

\bibitem[{Heinemann} \& {Quataert}(2014)]{hq14}
{\sc {Heinemann}, T. \& {Quataert}, E.} 2014 {Linear Vlasov Theory in the
  Shearing Sheet Approximation with Application to the Magneto-rotational
  Instability}. {\em \apj\/} {\bf 792}, 70.

\bibitem[{Hellinger} \& {Tr{\'a}vn{\'{\i}}{\v c}ek}(2008)]{ht08}
{\sc {Hellinger}, P. \& {Tr{\'a}vn{\'{\i}}{\v c}ek}, P.~M.} 2008 {Oblique
  proton fire hose instability in the expanding solar wind: Hybrid
  simulations}. {\em Journal of Geophysical Research (Space Physics)\/} {\bf
  113}~(A12), A10109.

\bibitem[{Hellinger} \& {Tr{\'a}vn{\'{\i}}{\v c}ek}(2015)]{ht15}
{\sc {Hellinger}, P. \& {Tr{\'a}vn{\'{\i}}{\v c}ek}, P.~M.} 2015 {Proton
  temperature-anisotropy-driven instabilities in weakly collisional plasmas:
  Hybrid simulations}. {\em Journal of Plasma Physics\/} {\bf 81}~(1),
  305810103.

\bibitem[{Horton}(1999)]{horton99}
{\sc {Horton}, W.} 1999 {Drift waves and transport}. {\em Reviews of Modern
  Physics\/} {\bf 71}, 735--778.

\bibitem[{Howes} {\em et~al.\/}(2006){Howes}, {Cowley}, {Dorland}, {Hammett},
  {Quataert} \& {Schekochihin}]{howes06}
{\sc {Howes}, G.~G., {Cowley}, S.~C., {Dorland}, W., {Hammett}, G.~W.,
  {Quataert}, E. \& {Schekochihin}, A.~A.} 2006 {Astrophysical Gyrokinetics:
  Basic Equations and Linear Theory}. {\em \apj\/} {\bf 651}, 590--614.

\bibitem[{Islam}(2014)]{islam14}
{\sc {Islam}, T.} 2014 {The Collisionless Magnetoviscous-thermal Instability}.
  {\em \apj\/} {\bf 787}, 53.

\bibitem[{Kaufman}(1960)]{kaufman60}
{\sc {Kaufman}, A.~N.} 1960 {Plasma Viscosity in a Magnetic Field}. {\em
  Physics of Fluids\/} {\bf 3}, 610--616.

\bibitem[{Kim} {\em et~al.\/}(1993){Kim}, {Horton} \& {Dong}]{khd93}
{\sc {Kim}, J.~Y., {Horton}, W. \& {Dong}, J.~Q.} 1993 {Electromagnetic effect
  on the toroidal ion temperature gradient mode}. {\em Physics of Fluids B\/}
  {\bf 5}, 4030--4039.

\bibitem[{Kingsep} {\em et~al.\/}(1990){Kingsep}, {Chukbar} \&
  {Yan'kov}]{kingsep90}
{\sc {Kingsep}, A.~S., {Chukbar}, K.~V. \& {Yan'kov}, V.~V.} 1990 {}. In {\em
  Reviews of Plasma Physics\/}, , vol.~16, p. 243. {Consultants Bureau, New
  York}.

\bibitem[{Komarov} {\em et~al.\/}(2016){Komarov}, {Churazov}, {Kunz} \&
  {Schekochihin}]{komarov16}
{\sc {Komarov}, S.~V., {Churazov}, E.~M., {Kunz}, M.~W. \& {Schekochihin},
  A.~A.} 2016 {Thermal conduction in a mirror-unstable plasma}. {\em \mnras\/}
  {\bf 460}, 467--477.

\bibitem[{Kulsrud}(1964)]{kulsrud64}
{\sc {Kulsrud}, R.~M.} 1964 {}. In {\em Teoria dei plasmi\/} (ed.
  {M.~N.~Rosenbluth}), p.~54. {Academic Press}.

\bibitem[{Kulsrud}(1983)]{kulsrud83}
{\sc {Kulsrud}, R.~M.} 1983 {MHD description of plasma}. In {\em Basic Plasma
  Physics: Selected Chapters, Handbook of Plasma Physics, Volume 1\/} (ed.
  {A.~A.~Galeev \& R.~N.~Sudan}), p.~1.

\bibitem[{Kunz}(2011)]{kunz11}
{\sc {Kunz}, M.~W.} 2011 {Dynamical stability of a thermally stratified
  intracluster medium with anisotropic momentum and heat transport}. {\em
  \mnras\/} {\bf 417}, 602--616.

\bibitem[{Kunz} {\em et~al.\/}(2012){Kunz}, {Bogdanovi{\'c}}, {Reynolds} \&
  {Stone}]{kunz12}
{\sc {Kunz}, M.~W., {Bogdanovi{\'c}}, T., {Reynolds}, C.~S. \& {Stone}, J.~M.}
  2012 {Buoyancy Instabilities in a Weakly Collisional Intracluster Medium}.
  {\em \apj\/} {\bf 754}, 122.

\bibitem[{Kunz} {\em et~al.\/}(2014){Kunz}, {Schekochihin} \& {Stone}]{kss14}
{\sc {Kunz}, M.~W., {Schekochihin}, A.~A. \& {Stone}, J.~M.} 2014 {Firehose and
  Mirror Instabilities in a Collisionless Shearing Plasma}. {\em Physical
  Review Letters\/} {\bf 112}~(20), 205003.

\bibitem[{Landau}(1946)]{landau46}
{\sc {Landau}, L.} 1946 {On the Vibrations of the Electronic Plasma}. {\em
  Zh.~Exp.~Teor.~Fiz. (English translation: 1946, J.~Phys.~U.S.S.R., 10, 25)\/}
  {\bf 16}, 574.

\bibitem[{Latter} \& {Kunz}(2012)]{lk12}
{\sc {Latter}, H.~N. \& {Kunz}, M.~W.} 2012 {The HBI in a quasi-global model of
  the intracluster medium}. {\em \mnras\/} {\bf 423}, 1964--1972.

\bibitem[{McCourt} {\em et~al.\/}(2011){McCourt}, {Parrish}, {Sharma} \&
  {Quataert}]{mccourt11}
{\sc {McCourt}, M., {Parrish}, I.~J., {Sharma}, P. \& {Quataert}, E.} 2011 {Can
  conduction induce convection? On the non-linear saturation of buoyancy
  instabilities in dilute plasmas}. {\em \mnras\/} {\bf 413}, 1295--1310.

\bibitem[{Melville} {\em et~al.\/}(2016){Melville}, {Schekochihin} \&
  {Kunz}]{melville16}
{\sc {Melville}, S., {Schekochihin}, A.~A. \& {Kunz}, M.~W.} 2016
  {Pressure-anisotropy-driven microturbulence and magnetic-field evolution in
  shearing, collisionless plasma}. {\em \mnras\/} {\bf 459}, 2701--2720.

\bibitem[{Mikellides} {\em et~al.\/}(2011){Mikellides}, {Tassis} \&
  {Yorke}]{mikellides11}
{\sc {Mikellides}, I.~G., {Tassis}, K. \& {Yorke}, H.~W.} 2011 {2D
  Magnetohydrodynamics simulations of induced plasma dynamics in the near-core
  region of a galaxy cluster}. {\em \mnras\/} {\bf 410}, 2602--2616.

\bibitem[Mikhailovskii(1962)]{mikhailovskii62}
{\sc Mikhailovskii, A.B.} 1962 Dielectric properties of an inhomogeneous
  plasma. {\em Nuclear Fusion\/} {\bf 2}~(3-4), 162.

\bibitem[{Mikhailovskii}(1967)]{mikhailovskii67}
{\sc {Mikhailovskii}, A.~B.} 1967 {Oscillations of an Inhomogeneous Plasma}.
  {\em Reviews of Plasma Physics\/} {\bf 3}, 159.

\bibitem[{Mikhailovskii}(1974)]{mikhailovskii74}
{\sc {Mikhailovskii}, A.~B.} 1974 {\em {Theory of Plasma Instabilities, Vol.~2,
  Instabilities of an Inhomogeneous Plasma}\/}. New York: Consultants Bureau.

\bibitem[{Mikhailovskii}(1992)]{mikhailovskii92}
{\sc {Mikhailovskii}, A.~B.} 1992 {\em {Electromagnetic instabilities in an
  inhomogeneous plasma}\/}. Bristol: IOP Publishing Press.

\bibitem[{Mikhailovskii} \& {Tsypin}(1971)]{mt71}
{\sc {Mikhailovskii}, A.~B. \& {Tsypin}, V.~S.} 1971 {Transport equations and
  gradient instabilities in a high pressure collisional plasma}. {\em Plasma
  Physics\/} {\bf 13}, 785--798.

\bibitem[{Mikhailovskii} \& {Tsypin}(1984)]{mt84}
{\sc {Mikhailovskii}, A.~B. \& {Tsypin}, V.~S.} 1984 {Transport equations of
  plasma in curvilinear magnetic field}. {\em Beitr\"{a}ge aus der
  Plasmaphysik\/} {\bf 24}, 335--354.

\bibitem[{Neronov} \& {Vovk}(2010)]{neronov10}
{\sc {Neronov}, A. \& {Vovk}, I.} 2010 {Evidence for Strong Extragalactic
  Magnetic Fields from Fermi Observations of TeV Blazars}. {\em Science\/} {\bf
  328}, 73.

\bibitem[{Parrish} {\em et~al.\/}(2012){Parrish}, {McCourt}, {Quataert} \&
  {Sharma}]{parrish12}
{\sc {Parrish}, I.~J., {McCourt}, M., {Quataert}, E. \& {Sharma}, P.} 2012 {The
  effects of anisotropic viscosity on turbulence and heat transport in the
  intracluster medium}. {\em \mnras\/} {\bf 422}, 704--718.

\bibitem[{Parrish} \& {Quataert}(2008)]{pq08}
{\sc {Parrish}, I.~J. \& {Quataert}, E.} 2008 {Nonlinear Simulations of the
  Heat-Flux-driven Buoyancy Instability and Its Implications for Galaxy
  Clusters}. {\em \apjl\/} {\bf 677}, L9.

\bibitem[{Parrish} {\em et~al.\/}(2009){Parrish}, {Quataert} \&
  {Sharma}]{parrish09}
{\sc {Parrish}, I.~J., {Quataert}, E. \& {Sharma}, P.} 2009 {Anisotropic
  Thermal Conduction and the Cooling Flow Problem in Galaxy Clusters}. {\em
  \apj\/} {\bf 703}, 96--108.

\bibitem[{Parrish} \& {Stone}(2007)]{ps07}
{\sc {Parrish}, I.~J. \& {Stone}, J.~M.} 2007 {Saturation of the Magnetothermal
  Instability in Three Dimensions}. {\em \apj\/} {\bf 664}, 135--148.

\bibitem[{Parrish} {\em et~al.\/}(2008){Parrish}, {Stone} \& {Lemaster}]{psl08}
{\sc {Parrish}, I.~J., {Stone}, J.~M. \& {Lemaster}, N.} 2008 {The
  Magnetothermal Instability in the Intracluster Medium}. {\em \apj\/} {\bf
  688}, 905--917.

\bibitem[{Planck Collaboration} {\em et~al.\/}(2015){Planck Collaboration},
  {Ade}, {Aghanim}, {Arnaud}, {Arroja}, {Ashdown}, {Aumont}, {Baccigalupi},
  {Ballardini}, {Banday} \& et~al.]{planck15}
{\sc {Planck Collaboration}, {Ade}, P.~A.~R., {Aghanim}, N., {Arnaud}, M.,
  {Arroja}, F., {Ashdown}, M., {Aumont}, J., {Baccigalupi}, C., {Ballardini},
  M., {Banday}, A.~J. \& et~al.} 2015 {Planck 2015 results. XIX. Constraints on
  primordial magnetic fields}. {\em ArXiv e-prints\/} .

\bibitem[{Quataert}(2008)]{quataert08}
{\sc {Quataert}, E.} 2008 {Buoyancy Instabilities in Weakly Magnetized
  Low-Collisionality Plasmas}. {\em \apj\/} {\bf 673}, 758--762.

\bibitem[{Quataert} {\em et~al.\/}(2002){Quataert}, {Dorland} \&
  {Hammett}]{qdh02}
{\sc {Quataert}, E., {Dorland}, W. \& {Hammett}, G.~W.} 2002 {The
  Magnetorotational Instability in a Collisionless Plasma}. {\em \apj\/} {\bf
  577}, 524--533.

\bibitem[{Quataert} {\em et~al.\/}(2015){Quataert}, {Heinemann} \&
  {Spitkovsky}]{qht15}
{\sc {Quataert}, E., {Heinemann}, T. \& {Spitkovsky}, A.} 2015 {Linear
  instabilities driven by differential rotation in very weakly magnetized
  plasmas}. {\em \mnras\/} {\bf 447}, 3328--3341.

\bibitem[{Ramos}(2005)]{ramos05}
{\sc {Ramos}, J.~J.} 2005 {General expression of the gyroviscous force}. {\em
  Physics of Plasmas\/} {\bf 12}~(11), 112301.

\bibitem[{Reynders}(1994)]{reynders94}
{\sc {Reynders}, J.~V.~W.} 1994 {Finite-{$\beta$} modification of the
  ion-temperature-gradient-driven instability in a sheared slab geometry}. {\em
  Physics of Plasmas\/} {\bf 1}, 1953--1961.

\bibitem[{Riquelme} {\em et~al.\/}(2016){Riquelme}, {Quataert} \&
  {Verscharen}]{riquelme16}
{\sc {Riquelme}, M., {Quataert}, E. \& {Verscharen}, D.} 2016 {PIC Simulations
  of the Effect of Velocity Space Instabilities on Electron Viscosity and
  Thermal Conduction}. {\em ArXiv e-prints\/} .

\bibitem[{Riquelme} {\em et~al.\/}(2015){Riquelme}, {Quataert} \&
  {Verscharen}]{riquelme15}
{\sc {Riquelme}, M.~A., {Quataert}, E. \& {Verscharen}, D.} 2015
  {Particle-in-cell Simulations of Continuously Driven Mirror and Ion Cyclotron
  Instabilities in High Beta Astrophysical and Heliospheric Plasmas}. {\em
  \apj\/} {\bf 800}, 27.

\bibitem[{Rudakov} \& {Sagdeev}(1961)]{rs61}
{\sc {Rudakov}, L.~I. \& {Sagdeev}, R.~Z.} 1961 {On the instability of a
  nonuniform rarefied plasma in a strong magnetic field}. {\em
  Dokl.~Akad.~Nauk.~SSSR\/} {\bf 138}, 581.

\bibitem[{Rutherford} \& {Frieman}(1968)]{rf68}
{\sc {Rutherford}, P.~H. \& {Frieman}, E.~A.} 1968 {Drift Instabilities in
  General Magnetic Field Configurations}. {\em Physics of Fluids\/} {\bf 11},
  569--585.

\bibitem[{Schekochihin} {\em et~al.\/}(2010){Schekochihin}, {Cowley}, {Rincon}
  \& {Rosin}]{schekochihin10}
{\sc {Schekochihin}, A.~A., {Cowley}, S.~C., {Rincon}, F. \& {Rosin}, M.~S.}
  2010 {Magnetofluid dynamics of magnetized cosmic plasma: firehose and
  gyrothermal instabilities}. {\em \mnras\/} {\bf 405}, 291--300.

\bibitem[{Schwarzschild}(1906)]{schwarzschild06}
{\sc {Schwarzschild}, K.} 1906 {}. {\em Nach.~Gesell.~Wiss.~G\"{o}ttingen\/}
  {\bf 41}.

\bibitem[{Sharma} {\em et~al.\/}(2003){Sharma}, {Hammett} \& {Quataert}]{shq03}
{\sc {Sharma}, P., {Hammett}, G.~W. \& {Quataert}, E.} 2003 {Transition from
  Collisionless to Collisional Magnetorotational Instability}. {\em \apj\/}
  {\bf 596}, 1121--1130.

\bibitem[{Sironi} \& {Narayan}(2015)]{sn15}
{\sc {Sironi}, L. \& {Narayan}, R.} 2015 {Electron Heating by the Ion Cyclotron
  Instability in Collisionless Accretion Flows. I. Compression-driven
  Instabilities and the Electron Heating Mechanism}. {\em \apj\/} {\bf 800},
  88.

\bibitem[{Snyder} \& {Hammett}(2001)]{sh01}
{\sc {Snyder}, P.~B. \& {Hammett}, G.~W.} 2001 {Electromagnetic effects on
  plasma microturbulence and transport}. {\em Physics of Plasmas\/} {\bf 8},
  744--749.

\bibitem[{Stix}(1962)]{stix62}
{\sc {Stix}, T.~H.} 1962 {\em {The Theory of Plasma Waves}\/}. New York:
  McGraw-Hill.

\bibitem[{Taylor} \& {Hastie}(1968)]{th68}
{\sc {Taylor}, J.~B. \& {Hastie}, R.~J.} 1968 {Stability of general plasma
  equilibria - I formal theory}. {\em Plasma Physics\/} {\bf 10}, 479--494.

\bibitem[{V\"{a}is\"{a}l\"{a}}(1925)]{vaisala25}
{\sc {V\"{a}is\"{a}l\"{a}}, V.} 1925 {\"{U}ber die Wirkung der Windschwankungen
  auf die pilot Beobachtungen}. {\em Soc.~Sci.~Fennica, Commentaliones
  Phys.~Math II\/} {\bf 2}, 46.

\bibitem[{Watson}(1966)]{watson66}
{\sc {Watson}, G.~N.} 1966 {\em {A Treatise on the Theory of Bessel
  Functions}\/}. Cambridge: Cambridge Univ.~Press.

\end{thebibliography}

\end{document}